\documentclass[a4paper,11pt]{article}
\pdfoutput=1

\usepackage{jheppub}

\usepackage{graphicx}
\usepackage{amsmath}
\usepackage{amssymb,amsfonts}
\usepackage{color}
\usepackage{xspace}
\usepackage{orcidlink}
\usepackage[small]{subfigure}
\usepackage{marginnote}

\newcommand{\eq}[1]{Eq.~\eqref{eq:#1}}

\renewcommand{\sec}[1]{Sec.~\ref{sec:#1}}

\newcommand{\app}[1]{App.~\ref{app:#1}}

\newcommand{\fig}[1]{Fig.~\ref{fig:#1}}

\newcommand{\OMIT}[1]{}

\newcommand{\ord}[1]{\mathcal{O}(#1)}

\newcommand{\df}{\mathrm{d}}
\newcommand{\nn}{\nonumber}

\newcommand{\taus}{{\tau_s}}

\newcommand{\beq}{\begin{equation}}
\newcommand{\eeq}{\end{equation}}
\newcommand{\beqa}{\begin{eqnarray}}
\newcommand{\eeqa}{\end{eqnarray}}

\newcommand{\lqcd}{\Lambda_{\rm QCD}}

\newcommand{\msbar}{\overline{\textrm{MS}}}

\newcommand{\muj}{\mu_J}
\newcommand{\muh}{\mu_H}
\newcommand{\mus}{\mu_s}

\newcount\hour \newcount\minute
\hour=\time \divide \hour by 60
\minute=\time
\newcommand{\mydate}{\ \today \ - \number\hour :\ifnum \minute<10 0\fi
\number\minute}

\def\wt{\widetilde}

\newcommand{\rLO}{ {L_1} }
\newcommand{\rLT}{ {L_2} }

\newcommand{\rL}{ {{ L}} }

\newcommand\tf{\widetilde{S}_f}

\newcommand{\rnO}{ {y_1} }
\newcommand{\rnT}{ {y_2} }

\newcommand{\rnOp}{ {y_1^\prime} }
\newcommand{\rnTp}{ {y_2^\prime} }

\newcommand{\im}{\text{Im}\,}
\newcommand{\re}{\text{Re}\,}

\usepackage{xcolor}
\usepackage[normalem]{ulem}

\title{\boldmath Precision $e^+e^-$ Hemisphere Masses in the Dijet Region with Power Corrections}

\preprint{\begin{flushright}
UWThPh-2025-5,
MIT-CTP 5028
\end{flushright}\vspace*{-0.75cm}}

\author[a]{Andr\'e H. Hoang\orcidlink{0000-0002-8424-9334},}
\author[b]{Vicent Mateu\orcidlink{0000-0003-0902-5012},}
\author[c]{Matthew D.\ Schwartz\orcidlink{0000-0001-6344-693X},}
\author[a,d]{Iain W.\ Stewart\orcidlink{0000-0003-0248-0979}}

\affiliation[a]{University of Vienna, Faculty of Physics, Boltzmanngasse 5, A-1090 Wien, Austria}
\affiliation[b]{Departamento de F\'isica Fundamental and IUFFyM, Universidad de Salamanca, Plaza de la Merced S/N, E-37008 Salamanca, Spain}
\affiliation[c]{Department of Physics, Harvard University, Cambridge, MA~02138, U.S.A.}
\affiliation[d]{Center for Theoretical Physics -- a Leinweber Institute, Massachusetts Institute of Technology, Cambridge, MA 02139, USA}

\emailAdd{andre.hoang@univie.ac.at}
\emailAdd{vmateu@usal.es}
\emailAdd{schwartz@g.harvard.edu}
\emailAdd{iains@mit.edu}

\abstract{We derive high-precision results for the $e^+e^-$ heavy jet mass (HJM) $\df \sigma/\df \rho$ and dihemisphere mass (DHM) $\df^2\sigma/(\df s_1\df s_2)$ distributions, for
$s_1\sim s_2$, in the dijet region. New results include: i)~the N$^3$LL resummation for HJM of large logarithms $\ln^n(\rho)$ at small $\rho$ including the exact two-loop non-global hemisphere soft function, the 4-loop cusp anomalous dimension and the 3-loop hard and jet functions, ii)~N$^3$LL results for DHM with resummation of logarithms $\ln(s_{1,2}/Q^2)$ when there is no large separation between $s_1$ and $s_2$, iii)~profile functions for HJM to give results simultaneously valid in the peak and tail regions, iv)~a complete two-dimensional basis of non-perturbative functions which can be used for double differential observables, that are needed for both HJM and DHM in the peak region, and v)~an implementation of renormalon subtractions for large-angle soft radiation to ${\cal O}(\alpha_s^3)$ together with a resummation of the additional large $\ln(Q\rho/\Lambda_{\rm QCD})$ logarithms.
Here $Q$ is the $e^+e^-$ center-of-mass energy.
Our resummation results are combined with known fixed-order ${\cal O}(\alpha_s^3)$ results and we discuss the convergence and remaining perturbative uncertainty in the cross section.
We also prove that, at order $1/Q$, the first moment of the HJM distribution involves an additional non-perturbative parameter compared to the power correction that shifts the tail of the spectrum (where $1\gg \rho\gg \Lambda_{\rm QCD}/Q$). This differs from thrust where a single non-perturbative parameter at order $1/Q$ describes both the first moment and the tail, and it disfavors models of power corrections employing a single non-perturbative parameter, such as the low-scale effective coupling model.
In this paper we focus only on the dijet region, not the far-tail distribution for $\rho \gtrsim 0.2$ beyond which the trijet factorization and resummation become important.}

\begin{document}
\maketitle
\flushbottom
\newpage
\section{Introduction}
\label{sec:intro}
The study of event-shape distributions has attracted renewed interest in the past few years. Although little new experimental data has become available, there has been considerable theoretical progress. On the one hand, for a large selection of $e^+e^-$ event shapes, fixed-order perturbative predictions are now available numerically at ${\cal O}(\alpha_s^3)$~\cite{GehrmannDeRidder:2007bj,GehrmannDeRidder:2007hr,Weinzierl:2008iv,Gehrmann-DeRidder:2014hxk, Weinzierl:2009ms,DelDuca:2016ily}, building on earlier analytic results at ${\cal O}(\alpha_s)$~\cite{Ellis:1980wv}, and numerical~\cite{Catani:1996jh, Catani:1996vz} and analytical~\cite{Dixon:2018qgp} results at ${\cal O}(\alpha_s^2)$. On the other hand, factorization theorems~\cite{Fleming:2007qr,Schwartz:2007ib,Bauer:2008dt} derived in Soft-Collinear Effective Theory (SCET)~\cite{Bauer:2000ew, Bauer:2000yr, Bauer:2001ct, Bauer:2001yt, Bauer:2002nz} have enabled a resummation of large perturbative logs at next-to-next-to-next-to-leading log (N$^3$LL) accuracy for thrust~\cite{Becher:2008cf}, C-parameter~\cite{Hoang:2014wka}, Energy-Energy correlator (EEC)~\cite{Moult:2018jzp} and heavy jet mass (HJM)~\cite{Chien:2010kc}, as well as next-to-next-to-leading-log (N$^2$LL) precision in angularities~\cite{Hornig:2009vb, Bell:2018gce} and Jet Broadening~\cite{Becher:2012qc} (see Refs.~\cite{Becher:2011pf,Chiu:2011qc, Chiu:2012ir} for the derivation of the factorization theorem). SCET generalizes the factorization-based approach of Collins, Soper and Sterman (CSS)~\cite{Collins:1981uk,Korchemsky:1998ev, Korchemsky:1999kt,Korchemsky:2000kp,Berger:2003iw} to the language of the effective field theories. This extends the resummation at the next-to-leading-log (NLL) level obtained much earlier with the classical approach of Catani, Trentadue, Turnock and Webber (CTTW)~\cite{Catani:1992ua} (see Ref.~\cite{Banfi:2004yd} for an automated, numerical implementation of resummation at this order), and N$^2$LL level analytically for the EEC in Ref.~\cite{deFlorian:2004mp}, and through a numerical procedure in~\cite{Banfi:2014sua} for other event shapes. The computation of the 2-loop soft function for generic event shapes has been automated numerically in Refs.~\cite{Bell:2018vaa,Bell:2018oqa,Bell:2020yzz}, while progress in the same direction for the jet function has been achieved in Ref.~\cite{Bell:2021dpb}.
Recently, factorization theorems and resummation have also been studied for thrust and HJM near the trijet limit~\cite{Bhattacharya:2022dtm,Bhattacharya:2023qet} where Sudakov shoulder logarithms~\cite{Catani:1997xc} become important. A renormalon-based approach has also been used to study power corrections away from the dijet limit~\cite{Luisoni:2020efy,Caola:2021kzt,Caola:2022vea}.
These fixed-order computations and factorization theorems have been generalized to event shapes in which the angle between beam and thrust axes is measured in Refs.~\cite{Hagiwara:2010cd,Mateu:2013gya,Bris:2022cdr}, to massive quark production in Refs.~\cite{Fleming:2007qr,Fleming:2007xt,Lepenik:2019jjk,Bachu:2020nqn,Bris:2020uyb}, to account for the effects of massive quark radiation in Refs.~\cite{Gritschacher:2013pha,Pietrulewicz:2014qza,Bris:2024bcq}, and to include an infrared resolution cutoff scale analogous to that of a parton shower in Ref.~\cite{Hoang:2018zrp,Hoang:2024nqi}.

An advantage of the effective field theory approach is that it naturally allows for the incorporation of power corrections. These corrections can be both perturbative (referred to as non-singular here),
such as those suppressed by powers of an event shape variable that vanishes in the dijet limit,
and non-perturbative, associated with power-law dependence on $\Lambda_{\rm QCD}$.
Subleading non-singular power corrections for event shapes have been studied in Refs.~\cite{Freedman:2013vya,Freedman:2014uta,Boughezal:2016zws,Moult:2016fqy,Feige:2017zci,Moult:2017rpl,Moult:2017jsg,Goerke:2017lei,Beneke:2017ztn} in SCET. Finally in Refs.~\cite{Moult:2018jjd,Moult:2019uhz,Beneke:2022obx,Agarwal:2023fdk} the leading logarithmic resummation for subleading power terms was achieved for thrust.
The leading non-perturbative effects arise from hadronization, where the physical particles observed are color-neutral massive hadrons, not massless colored partons. Methods in SCET have been developed for $e^+e^-\to$~dijets to include these non-perturbative power corrections in a systematically improvable fashion~\cite{Hoang:2007vb,Ligeti:2008ac,Abbate:2010xh,Agarwal:2020uxi}, building on early work in~\cite{Manohar:1994kq,Webber:1994cp,Dokshitzer:1995zt,Akhoury:1995sp,Nason:1995hd,Korchemsky:1994is,MovillaFernandez:2001ed,Korchemsky:1999kt,Gardi:2001ny}.
All those theoretical developments have been used to determine the strong coupling constant with high precision comparing to thrust and C-parameter experimental data in Refs.~\cite{Gehrmann:2009eh,Gehrmann:2012sc,Abbate:2010xh,Abbate:2012jh,Hoang:2015hka,Bell:2023dqs,Nason:2023asn}.

The non-perturbative power corrections must in general also include hadron mass effects, which were first investigated for event shapes in Ref.~\cite{Salam:2001bd}. In Ref.~\cite{Mateu:2012nk} a quantum-field-theory-based formalism for these corrections was derived. For HJM the use of massless kinematics suggest a universality relation $2\Omega_1^\rho = \Omega_1^\tau$, however the hadron mass corrections are quite important and violate this universality relation. Thus, the dominant power correction parameter differs from that of thrust, $2\Omega_1^\rho \ne \Omega_1^\tau$.
In analyses of multi-purpose Monte Carlo simulations the inclusion of hadron mass effects suggests a relation closer to $\Omega_1^\rho \simeq \Omega_1^\tau$~\cite{Mateu:2012nk}.
The analysis of hadron masses shows that the leading power correction also carries a non-trivial anomalous dimension.\footnote{In Ref.~\cite{Hoang:2015hka} the effect of the hadron-mass anomalous dimension in the thrust and C-parameter cross sections was investigated in the context of tail fits, and it was concluded that their phenomenological impact was rather small. Therefore, although implementing them is a straightforward task, to avoid further complications we will not discuss them further in this article.} In general, these corrections also depend on whether energy or momentum variables are used for the measurements (e.g.~E-scheme or P-scheme).

There are also perturbative differences
between HJM and thrust, in particular that soft and collinear effects from the trijet region act very differently on the two observables. One way this difference is manifested is that the heavy jet mass has a Sudakov shoulder both to the left and to the right of the 3-particle threshold at $\rho=1/3$, while thrust only has a Sudakov shoulder to the right of $\tau=1/3$~\cite{Catani:1997xc,Bhattacharya:2022dtm,Bhattacharya:2023qet}.

Although understanding this effect will be important for $\alpha_s$ fits, our focus here is solely on the dijet peak and tail regions, leaving the inclusion of the trijet region $\rho \sim 1/3$ and fits to $\alpha_s$ for a separate dedicated study. In this paper, when we consider the trijet region, our focus is only to ensure that it does not get contaminated by our dijet resummation, providing the initial conditions needed for a proper treatment of this region.

In this paper, we focus on formulating a high-precision theoretical expression for the dihemisphere mass (DHM) and HJM distributions. Our ultimate goal is to carry out global fits for $\alpha_s(m_Z)$ comparing to HJM experimental data. Our work builds on the N$^3$LL results for HJM of Ref.~\cite{Chien:2010kc}, which incorporate the ${\cal O}(\alpha_s^3)$ results of Refs.~\cite{GehrmannDeRidder:2007bj,Weinzierl:2008iv}. We extend these results by obtaining an N$^3$LL$^\prime$\,+\,${\cal O}(\alpha_s^3)$ cross section that is simultaneously valid in the peak and tail by implementing a rigorous field-theory-based method to deal with power corrections, and by updating the non-singular distribution at ${\cal O}(\alpha_s^3)$ with numerical output from {\sc CoLoRFulNNLO}~\cite{DelDuca:2016ily}. For massless QCD we thereby achieve the same level of precision for HJM as was achieved for thrust in Refs.~\cite{Abbate:2010xh,Abbate:2012jh} or C-parameter in Refs.~\cite{Hoang:2014wka,Hoang:2015hka}.

In \sec{pert} we discuss factorization-based results for the DHM distribution $\df ^2\sigma/(\df s_1\df s_2)$ \cite{Fleming:2007qr,Fleming:2007xt}. Here, phase space is divided into left (L) and right (R) hemispheres by the plane perpendicular to the thrust axis, and the hemisphere jet invariant masses are
\begin{equation}
s_1 =\biggl( \sum_{i\in L} p_i^\mu \biggr)^{\!\!2} \,,\qquad s_2 =\biggl( \sum_{i\in R} p_i^\mu \biggr)^{\!\!2} \,.
\end{equation}
The thrust axis is defined as the direction $\hat n$ which maximizes the sum of the absolute values of the three-momenta projections. The thrust event shape is defined as~\cite{Farhi:1977sg}
\begin{align}\label{eq:thrustDef}
\tau = 1 - \dfrac{1}{\sum_i |\vec{p}_i|}\max_{\hat n}\sum_i|{\hat n}\cdot {\vec p}_i|\,.
\end{align}
The HJM event shape is defined~\cite{Clavelli:1979md, Chandramohan:1980ry, Clavelli:1981yh} as $\rho = \max(s_1,s_2)/Q^2$. The distribution is characterized by a dijet peak region where $s_i\sim Q\lqcd$, a dijet tail region where $Q\lqcd \ll s_i \ll Q^2$, and a (trijet) far-tail region where $s_i\sim Q^2$ (which includes the Sudakov shoulder region where $s_i\sim Q^2/3$).

Both thrust and HJM are single-variable event shapes which are free of non-global logs. In the DHM distribution, which is doubly differential, there are non-global logarithms of $s_1/s_2$, which become large for $s_1\gg s_2$, which are not summed by the type of factorization theorem we consider here.\footnote{See Refs.~\cite{Dasgupta:2001sh,Banfi:2002hw} for the leading logarithmic summation of non-global logarithms with large $N_c$, Ref.~\cite{Khelifa-Kerfa:2015mma} for LL summation beyond large $N_c$, and Refs.~\cite{Larkoski:2015zka,Becher:2015hka,Becher:2016mmh,Larkoski:2016zzc} for methods to sum up these non-global logarithms within SCET. The coefficients of the leading non-global large logarithmic terms have also been computed analytically to 5-loops in~\cite{Schwartz:2014wha}.}
In our analysis of the doubly differential distribution, we take $s_1\sim s_2$ which avoids large non-global logarithms. We also discuss the mixed distribution-cumulative cross-section, $\Xi$, and its diagonal projection which is the heavy jet mass distribution, $\df \sigma/\df \rho$,
\begin{align} \label{eq:Xirhodefn}
\Xi(s_1^c,s_2) \equiv \int_0^{s_1^c} \df s_1 \frac{\df ^2\sigma}{\df s_1\df s_2} \,,\qquad\qquad
\frac{\df \sigma}{\df \rho} = 2 Q^2\Xi(Q^2\rho,Q^2\rho)\,.
\end{align}
The second equation is equivalent to the direct projection from the DHM onto HJM which uses the projection
$Q^2[\delta(Q^2\,\rho - s_1)\theta(s_1 - s_2) + \delta(Q^2\,\rho - s_2)\theta(s_2 - s_1)]$.
We will also discuss the double cumulative cross-section, $\Sigma(s_1^c,s_2^c)$ whose diagonal
projection is the cumulative heavy jet mass distribution $\Sigma^\rho(\rho^c)$:
\begin{equation}
\Sigma(s_1^c,s_2^c) \equiv \!\int_0^{s_1^c} \df s_1\! \int_0^{s_2^c}\! \df s_2
\frac{\df ^2\sigma}{\df s_1\df s_2} \,,\qquad
\Sigma^\rho(\rho^c)
\equiv \!\int_0^{\rho^c}\!\df \rho\frac{\df\sigma}{\df\rho}\,
= \Sigma(Q^2\rho^c,Q^2\rho^c)\,.
\end{equation}

The content of this paper is as follows.
In Sec.~\ref{sec:pert} we review the factorization theorem that sums logarithms to all orders in perturbation theory for the most singular terms in these cross sections. This is achieved by factorizing the cross section into hard, jet, and soft functions, $\df^2 \sigma/(\df s_1\df s_2)= H(\mu) \times J(\mu) \otimes J(\mu) \otimes S(\mu)$. Hard, jet, and soft renormalization scales, $\mu_H$, $\mu_J$, and $\mu_s$ are introduced so that $H(\mu_H)$, $J(\mu_J)$, and $S(\mu_s)$ have fixed-order perturbative expansions in $\alpha_s$ with small logarithms. The argument of the large logs depends
on ratios of the three $\mu_i$ scales and are resummed by renormalization-group (RG) evolving the corresponding function. For $\df \sigma/\df \rho$ and $\Sigma^\rho(\rho^c)$ in the peak and tail regions, our results have a perturbative level of precision of
N$^3$LL$^\prime$\,+\,${\cal O} (\alpha_s^3)$
and include non-perturbative corrections that scale as ${\cal O}([Q\lqcd /s_i]^j)$ for any $j$ (encoded in a shape function),
but do not include power corrections that start at $\lqcd^2/s_i$.

In Sec.~\ref{sec:double} we discuss how we efficiently implement the dihemisphere mass soft function $S(\ell_1,\ell_2)$ at ${\cal O}(\alpha_s^2)$~\cite{Kelley:2011ng,Hornig:2011iu,Monni:2011gb}, within a resummed cross section at N$^3$LL$^\prime$ order. This function implements in fixed-order perturbation theory global logs, the non-global double logarithm~\cite{Dasgupta:2001sh}, as well as non-global single logarithms and non-logarithmic terms.

In Sec.~\ref{sec:nonsing} we discuss the extraction of the non-singular cross section for $\df \sigma/\df \rho$,
which must be included to bring the level of precision to ${\cal O}(\alpha_s^3)$. This cross section is implemented as a fixed-order expansion in $\alpha_s(\mu_{\rm ns})$, and has residual $\mu_{\rm ns}$ scale dependence from beyond the order of the truncation. The non-singular terms play a key role in the multi-jet region where there are large cancellations between them and the singular terms, order-by-order in $\alpha_s$.
Our non-singular extraction at ${\cal O}(\alpha_s^2)$ and ${\cal O}(\alpha_s^3)$ makes use of the numerical Monte Carlos {\sc event2}~\cite{Catani:1996jh,Catani:1996vz} and {\sc CoLoRFulNNLO}~\cite{DelDuca:2016ily}.
The analytic results for the $\alpha_s^2\,\delta(\rho)$ term in the $\df\sigma/\df\rho$ cross section~\cite{Kelley:2011ng,Monni:2011gb} improve the accuracy of these non-singular extractions.

We turn to the discussion of power corrections in the double differential DHM and HJM spectra in Sec.~\ref{sec:power}. In the peak region these distributions depend on a two-dimensional non-perturbative shape function $F(k_1,k_2) = F(k_2,k_1)$, which falls off exponentially when either $k_1\gg \lqcd $ or $k_2 \gg \lqcd $, and has support for positive values of $k_i$. We construct a complete basis of functions for $F$ by extending the one-dimensional basis of Ref.~\cite{Ligeti:2008ac} in Sec.~\ref{sec:shape}. Perturbative renormalon subtractions are implemented to define this function $F$ in a scheme that stabilizes the first moment $\Omega_{1,0}$, see Sec.~\ref{sec:Rgap}.
In an Operator Product Expansion (OPE) which is valid in the tail region of the distributions for thrust, HJM, and DHM, positive-definite power corrections $\Omega_{i,j}$ appear, where
\begin{equation}\label{eq:Omega-ij}
\Omega_{i,j}\,=\,\Omega_{j,i}\,=\!\int\!\df k_1\,\df k_2\,k_1^i\,k_2^j\, F(k_1,k_2)\,.
\end{equation}
The moments of the DHM distribution depends on additional parameters $\Upsilon_{i,j}$ which are defined as
\begin{equation}\label{eq:Upsilon-ij}
\Upsilon_{i,j}
=\!\int_0^\infty\!\df k_1\df k_2\,( k_1- k_2)
\theta( k_1- k_2) k_1^i k_2^j F( k_1, k_2)
= \Upsilon_{j,i}+\Omega_{i+1,j}-\Omega_{i,j+1}
\,,
\end{equation}
and are also positive definite, as described in \sec{power}. Unlike for thrust or C-parameter, the first moment of the HJM distribution does not have the usual form of an OPE, with a partonic first moment plus a term involving a non-perturbative parameter divided by $Q$. This also applies to higher moments. Due to the complicated dynamics between the two hemispheres, perturbative and non-perturbative physics do not factor from each other except for the $0$-th moment, that is, for the total cross section.

In Sec.~\ref{sec:profiles} we discuss how we simultaneously obtain valid results in the dijet peak and tail regions for HJM. Resummation requires the scales $\mu_J$ and $\mu_s$ to depend on kinematic variables, and the transition between these three regions puts constraints on the form of the resulting profile functions for $\mu_J$ and $\mu_s$. We demonstrate that the same profile functions used for thrust and C-parameter in Ref.~\cite{Hoang:2014wka}, which are a refinement of those used in Refs.~\cite{Abbate:2010xh,Abbate:2012jh}, can be used for HJM with only a minimal modification.

In Sec.~\ref{sec:results} we provide numerical analyses of our predictions. We vary theoretical parameters as a method of estimating the perturbative uncertainties in the cross section related to higher-order effects from resummation, fixed-order corrections, and from the transition regions. We show that these uncertainties decrease as we increase the order from NLL$^\prime$, to N$^2$LL$^\prime$, and to N$^3$LL$^\prime$. On top of those, we vary parameters related to the parametrization of non-singular corrections beyond 1-loop. Section~\ref{sec:conclusion} contains our conclusions and outlook.

\section{Formalism}
\label{sec:pert}

\subsection{Double and Single Differential Cross Sections and Shape Functions}\label{sec:factorization}
Here we consider the double differential DHM distribution $\df^2\sigma/(\df s_1\df s_2)$, as well as its cumulative analogs, and two single differential distributions that can be derived from it. From the doubly differential distribution one can obtain the single differential distribution for HJM using Eq.~\eqref{eq:Xirhodefn}, and for the sum of hemisphere masses (SHM) $\taus$ using the projection \mbox{$Q^2\,\delta(Q^2\,\taus - s_1 - s_2)$}. This yields
\begin{align}\label{eq:thrust-projection}
\frac{\df \sigma}{\df \taus} = \,&\, Q^2\! \!\int_0^{Q^2\taus}\!\! \df s\,
\frac{\df^2 \sigma}{\df s_1 \df s_2}(s_1=s,s_2=Q^2\,\taus - s)\,,\\
\frac{\df \sigma}{\df \rho} \,= \,&\, 2Q^2\!\! \int_0^{Q^2\rho}\!\!\df s\,
\frac{\df^2 \sigma}{\df s_1 \df s_2}(s_1=Q^2\rho,s_2=s)=\, 2 Q^2 \Xi(Q^2\rho,Q^2\rho)\,,\nonumber
\end{align}
where we have used that the double differential distribution and projection operators are symmetric under the exchange $s_1\leftrightarrow s_2$. For the singular partonic cross sections, the distributions for $\taus$ and thrust are identical. The exact relation between thrust and $s_{1,2}$ computed for massless particles can be found e.g.\ in Ref.~\cite{Mateu:2013gya}. In order to keep the same simple relationship in the non-singular region, here we use the sum of hemisphere masses $\taus$ rather than thrust.

It is generally easier to work with cumulative distributions (whose perturbative parts are functions) rather than the differential cross sections (whose perturbative parts are distributions).
For the SHM and HJM they take the form:
\begin{align}\label{eq:cumulative-projection}
\Sigma^{\taus}(\taus) &= \int_0^{\taus}\df \tau_s^\prime \frac{\df\sigma}{\df\tau_s^\prime}\, =\!
\int_0^{Q^2\taus}\!\! \df s_1\!\int_0^{Q^2\taus-s_1\!}\df s_2 \frac{\df^2 \sigma}{\df s_1 \df s_2}\,, \\
\Sigma^{\rho}(\rho) &= \int_0^{\rho}\df \rho^\prime \frac{\df\sigma}{\df\rho^\prime}\,=
\int_0^{Q^2\rho}\!\!\df s_1\!\int_0^{Q^2\rho}\!\df s_2 \frac{\df^2 \sigma}{\df s_1 \df s_2}\,
=\Sigma(Q^2\rho,Q^2\rho)\,,\nonumber
\end{align}
where we have used $\theta(Q^2\rho - s_1)\,\theta(s_1 - s_2) + \theta(Q^2\rho - s_2)\,\theta(s_2 - s_1) =
\theta(Q^2\rho - s_1)\,\theta(Q^2\rho - s_2)$.

The DHM distribution $\df^2\sigma/(\df s_1\df s_2)$ has a perturbative contribution $\df^2\hat \sigma/(\df s_1\df s_2)$ as well as important non-perturbative corrections related to hadronization, which in the dijet region is described by a two-dimensional shape function $F( k_1, k_2)$. The partonic distribution can be split into singular and non-singular terms,
\begin{align}\label{eq:partonic-hemisphere-total}
\frac{\df^2 \hat\sigma}{\df s_1\df s_2}
& =\frac{\df \hat\sigma_{\rm sing}}{\df s_1\df s_2}+
\frac{\df^2 \hat\sigma_{\rm ns}}{\df s_1\df s_2}
\,,
\end{align}
and has support only for positive values of $s_i$. The singular part can be computed with a resummation of large logarithms using SCET and the non-singular part can be obtained from fixed-order perturbative QCD predictions. The SCET factorization theorem also provides a rigorous operator definition for the shape function $F(k_1,k_2)$, and yields the precise manner in which $F(k_1,k_2)$ must be convolved with the singular partonic distribution $\df \hat\sigma_{\rm sing}/(\df s_1\df s_2)$. To the level of precision that we are working and as long as one stays within the dijet region, the shape function can also be convolved with the non-singular distribution (even though technically in an even higher precision treatment some components of the non-singular distribution would need to be convolved with different non-perturbative shape functions). Corrections to this approximation enter at ${\cal O}[\alpha_s\,\Lambda_{\rm QCD}/(Q\,s_i)]$ and are beyond the order we are working.\footnote{The DHM distribution starts at ${\cal O}(1/s_i)$, so the corrections are suppressed by $\alpha_s\,\Lambda_{\rm QCD}/Q$ respect to the leading
term.}
Within this setup, the non-perturbative double differential distribution can be factorized as
\begin{equation}\label{eq:double-tot}
\frac{\df \sigma}{\df s_1\df s_2}\,=\!\int \!\df k_1\df k_2\frac{\df \hat\sigma}{\df s_1\df s_2}
(s_1-Q\, k_1,s_2-Q\, k_2)F(k_1, k_2)\,,
\end{equation}
where this factorization theorem captures only the soft hadronization effects through $F$, which dominate over collinear hadronization, which enters at ${\cal O}(\Lambda_{\rm QCD}^2/s_i^2)$, beyond the order we are working.

For the mixed differential-cumulative and for the double cumulative distributions the corresponding factorization theorems are
\begin{subequations} \label{eq:double-mixed-sigma}
\begin{align}
\Xi(s_1^c,s_2)\,
&=\!\int\! \df k_1\df k_2\,\widehat\Xi(s_1^c-Q\, k_1,s_2-Q\, k_2) F( k_1, k_2)
\label{eq:double-mixed-sigmaa}
\\
&=Q \!\int \!\df k_1\df k_2\,\frac{\df \hat\sigma}{\df s_1\df s_2}
(s^c_1-Q\, k_1,s_2-Q\, k_2) F^{\Xi}( k_1, k_2)
\,,\label{eq:double-mixed-sigmab}
\\
\Sigma(s_1^c,s_2^c)\,
&=\!\int\!\df k_1\df k_2\,\widehat\Sigma(s_1^c-Q\, k_1,s^c_2-Q\, k_2) F( k_1, k_2)
\label{eq:double-mixed-sigmac}
\\
&= Q^2\!\int \!\df k_1\df k_2\,\frac{\df \hat\sigma}{\df s_1\df s_2}
(s^c_1-Q\, k_1,s^c_2-Q\, k_2)F^\Sigma( k_1, k_2)
\,,\label{eq:double-mixed-sigmad}
\end{align}
\end{subequations}
where the results are either written with the same $F$ as Eq.~\eqref{eq:double-tot}, or equivalently with the single- and double-cumulative versions of the shape function
\begin{equation}\label{eq:Shape-cumulative}
F^\Xi( k_1^c, k_2) \,\equiv\! \int_0^{ k_1^c}\! \df k_2 F( k_1, k_2)\,,\qquad
F^\Sigma( k_1^c, k_2^c) \,\equiv\! \int_0^{ k_1^c}\! \df k_1\!\int_0^{ k_2^c}\! \df k_2
F( k_1, k_2)\,.
\end{equation}
Note that only the second function is symmetric under the exchange of its two arguments. Each of the two ways in which the shape function for $\Xi$ and $\Sigma$ can be taken into account has numerical advantages, and our code uses Eqs.~(\ref{eq:double-mixed-sigmaa},\ref{eq:double-mixed-sigmac}) and Eqs.~(\ref{eq:double-mixed-sigmab},\ref{eq:double-mixed-sigmad}) for the singular and non-singular
terms, respectively.

Let us next project Eqs.~\eqref{eq:double-mixed-sigma} into the sum of hemisphere masses and heavy jet mass using Eqs.~\eqref{eq:thrust-projection} and \eqref{eq:cumulative-projection}. For the SHM, the projection collapses the double integration into the convolution of the partonic $\taus$ distribution with a one-dimensional shape function:%
\begin{align}\label{eq:thrust-hadronic}
\frac{\df \sigma}{\df \taus}
=& \int\! \df k \frac{\df\hat \sigma}{\df \taus}\biggl(\!\taus - \frac{k}{Q}\biggr) F_\tau(k)\,,\\[0.1cm]
\Sigma^\taus(\tau_s^c)
=& \int\! \frac{\df k}{Q} \frac{\df\hat \sigma}{\df \taus}\biggl(\!\tau_s^c - \frac{k}{Q}\biggr)
F^\Sigma_\tau(k)
=\! \int\! \df k\,\widehat\Sigma^\taus\biggl(\!\tau_s^c - \frac{k}{Q}\biggr) F_\tau(k)\,, \nonumber\\
F_\tau(k) \equiv&\! \int_0^k\! \df k^\prime F(k^\prime,k - k^\prime)
\,,\qquad
F^\Sigma_\tau(k) \equiv\! \int_0^k \!\df k^\prime F_\tau(k^\prime )\,.\nonumber
\end{align}
The SHM distribution has the simplification that both the singular and non-singular partonic distributions are one dimensional, and the non-singular can be determined directly from the fixed-order $\taus$ distribution.
In contrast, for the projection onto HJM the formulae combining perturbative singular and non-singular functions and the non-perturbative shape function have two-dimensional convolutions:
\begin{subequations}\label{eq:HJM-hadronic}\begin{align}
\frac{\df \sigma}{\df \rho} =&\,\, 2Q^2\!\!\int \!\df k_1\df k_2\,
\widehat\Xi(Q^2\rho-Qk_1,Q^2\rho-Qk_2)
F(k_1,k_2)\\
\,=&\,\, 2Q^2\!\!\int \!\df k_1\df k_2
\frac{\df\hat\sigma}{\df s_1\df s_2}(Q^2\rho-Qk_1,Q^2\rho-Qk_2)
F^\Xi(k_1,k_2)\,,\\
\Sigma^\rho(\rho^c) =&\! \int \!\df k_1\df k_2\,
\widehat\Sigma(Q^2\rho^c-Qk_1,Q^2\rho^c-Qk_2)
F(k_1,k_2)\label{eq:HJMCumF}\\
\,=&\,Q^2\!\! \int \!\df k_1\df k_2\,
\frac{\df\hat\sigma}{\df s_1\df s_2}(Q^2\rho^c-Q\,k_1,Q^2\rho^c-Qk_2)
F^\Sigma(k_1,k_2)\\
\,=&\,Q\!\! \int \!\df k_1\df k_2\,\widehat\Xi(Q^2\rho^c-Qk_1,Q^2\rho^c-Qk_2)
F^\Xi(k_2,k_1)\,.
\end{align}\end{subequations}
The two-dimensional shape function is also needed for DHM as seen in \eq{double-tot}.
For HJM this observation was first discussed in Ref.~\cite{Korchemsky:1999kt} in a form analogous to our $\Sigma^\rho = \widehat\Sigma \otimes F$ equation.
In principle, this implies that the two dimension non-singular DHM is needed for these distributions.
At ${\cal O}(\alpha_s)$ it is still straightforward to obtain since the fixed-order DHM is known analytically. However, for orders beyond ${\cal O}(\alpha_s)$ only numerical results are available, and the numerical determination of a two dimensional non-singular distribution is more subtle and less precise than a one dimensional distribution.
Since the convolution of the non-singular distribution with $F$ was anyway an approximation, it turns out to suffice to use an effectively one dimensional non-singular distribution at the level of precision that we are aiming for,
see Sec.~\ref{sec:power}.

\subsection{Factorization Theorems for Singular Contributions} \label{sec:singular}
For the partonic singular terms, the factorization theorem in SCET implies~\cite{Fleming:2007qr,Fleming:2007xt,Hoang:2007vb}
\begin{align} \label{eq:hemidist}
\frac{1}{\sigma_0} \frac{\df^2 \hat{\sigma}_{\rm sing}}{\df s_1 \df s_2}
= &\, H(Q, \mu)\! \int \!
\df \ell_1 \df \ell_2 \, J (s_1 - Q\, \ell_1 - Q\,\bar\Delta(R,\mu), \mu) \\
&\times J (s_2 - Q\, \ell_2- Q\,\bar\Delta(R,\mu), \mu)\,
e^{\delta(R,\mu)\bigl( \frac{\partial}{\partial \ell_1}+ \frac{\partial}{\partial \ell_2}\bigr)}
\widehat S(\ell_1, \ell_2, \mu)\nn\\
= & \,H(Q, \mu_H)\, U_H(Q,\mu_H,\mu_J) \! \int \!
\df \ell_1 \df \ell_2 \,
J (s_1 - Q\, \ell_1- Q\,\bar\Delta(R,\mu_s), \mu_J)\nn \\
& J (s_2 - Q\, \ell_2- Q\,\bar\Delta(R,\mu_s), \mu_J)
\!\int \! \df \ell_1^\prime \df \ell_2^\prime \,
U_S(\ell_1-\ell_1^\prime,\mu_J,\mu_s) \nn\\
&\times U_S(\ell_2-\ell_2^\prime,\mu_J,\mu_s)\,
e^{\delta(R,\mu_s)\bigl( \frac{\partial}{\partial \ell_1^\prime} + \frac{\partial}{\partial \ell_2^\prime}\bigr)}
\widehat S(\ell_1^\prime, \ell_2^\prime, \mu_s) \,, \nn
\end{align}
where $H(Q,\mu)$ is the hard function, $J(s,\mu)$ is the single-hemisphere jet function, and $\widehat S(\ell_1,\ell_2,\mu)$ is the partonic dihemisphere soft function.
The functions $\delta(R,\mu)$ and $\bar\Delta(R,\mu)$, defined precisely in Eqs.~\eqref{eq:delta-scheme} and \eqref{eq:DeltaRevolution} below, implement the subtraction of the $\ord{\lqcd}$ renormalon~\cite{Hoang:2007vb}. In the first line, these functions are evaluated at a common factorization scale $\mu$.
To resum logarithms, one computes in fixed-order perturbation theory $H$, $J$, $\widehat S$ at the scales $\mu_H$, $\mu_J$ and $\mu_s$ respectively, such that they do not contain large logarithms. These functions are then evolved to the common scale $\mu$ using the renormalization group. Eq.~\eqref{eq:hemidist} displays the hard evolution factor $U_H$, whose expression is given in Eq.~\eqref{eq:UH}, and the momentum-space soft evolution factor $U_S$, an expression for which can be found e.g.\ in Eq.~(79) of Ref.~\cite{Fleming:2007xt}. For convenience, we choose $\mu=\mu_J$. Details on how to perform the resummation in Laplace space are given in Sec.~\ref{sec:double}. Details on this renormalon subtraction procedure are given in Sec.~\ref{sec:Rgap}.

The fixed-order hard function $H$ can be obtained from the on-shell quark (vector-current) form factor, which is known to ${\cal O}(\alpha_s)$ \cite{Matsuura:1987wt,Matsuura:1988sm}, ${\cal O}(\alpha_s^2)$~\cite{Gehrmann:2005pd,Moch:2005id}, and ${\cal O}(\alpha_s^3)$~\cite{Lee:2010cg,Baikov:2009bg}. The fixed-order inclusive quark jet function $J$ is known at $O(\alpha_s)$~\cite{Lunghi:2002ju,Bauer:2003pi}, ${\cal O}(\alpha_s^2)$~\cite{Becher:2006qw}, and ${\cal O}(\alpha_s^3)$~\cite{Bruser:2018rad,Banerjee:2018ozf}. These two functions have been computed to all orders in the large-$\beta_0$ limit in Ref.~\cite{Gracia:2021nut}. The four-loop cusp anomalous dimension is also needed and was computed analytically in Refs.~\cite{vonManteuffel:2020vjv,Henn:2019swt,Henn:2019rmi,Bruser:2019auj,Moch:2018wjh,Moch:2017uml}.
The hemisphere soft function $\widehat S$ has been analytically computed at ${\cal O}(\alpha_s)$ \cite{Schwartz:2007ib,Fleming:2007qr} and at ${\cal O}(\alpha_s^2)$ in \cite{Kelley:2011ng,Hornig:2011iu,Monni:2011gb}. At ${\cal O}(\alpha_s^3)$ the $\mu$-dependent logs of the soft function are known through the three-loop anomalous dimensions, and given that at this order the hard and jet functions are fully known, together with the fixed-order distribution the $\mu$-independent pieces can in principle be extracted numerically, but so far these results are not robust given current fixed-order uncertainties, see for example~\cite{Bruser:2018rad}. The three-loop non-logarithmic piece for the thrust projection of the soft function has been computed in Ref.~\cite{Baranowski:2024vxg,Baranowski:2024ysi}.

The hard function involves only the c.m.\ energy $Q$ and does not participate in the convolution. Each of the two identical jet functions depends only on a single-hemisphere invariant mass. The soft function is more complicated and
depends on two light-cone momenta, one for each hemisphere. The two-dimensional soft function involves global effects (logarithms of a single invariant mass over the renormalization scale) as well as non-global effects (that are a function of the ratio of the two invariant masses)~\cite{Fleming:2007xt,Hoang:2008fs,Chien:2010kc,Kelley:2011ng,Hornig:2011iu}. More details and formulas are given in Sec.~\ref{sec:double}.

For the sum of hemisphere masses $\taus$ the partonic factorization theorem can be rewritten in term of single convolutions, yielding the simpler factorization structure known from thrust:
\begin{align} \label{eq:thrust-dist}
\frac{1}{\sigma_0} \frac{\df \hat{\sigma}_{\rm sing}}{\df \taus} =\,& Q^2H(Q, \mu)\! \int \!
\df \ell \, J_\tau (Q^2\taus - Q\, \ell, \mu) \widehat S_\tau(\ell, \mu) \,,\\
\frac{1}{\sigma_0} \widehat \Sigma_{\rm sing}^{\taus}(\taus)=\,& QH(Q, \mu)\! \int \!
\df \ell \, J_\tau (Q^2\taus - Q\, \ell, \mu) \widehat S^{\,\Sigma}_\tau(\ell, \mu) \,,
\nn
\end{align}
with
\begin{equation}\label{eq:thrust-jet-soft}
J_\tau (s,\mu) \equiv \!\int_0^s\! \df s^\prime J (s^\prime,\mu)J (s\,-\,s^\prime,\mu)\,,\qquad
\widehat S_\tau(\ell, \mu) \equiv \! \int_0^\ell\! \df \ell^\prime\, \widehat S(\ell^\prime, \ell - \ell^\prime, \mu)\,,
\end{equation}
and
\begin{equation}
\widehat S^{\,\Sigma}_\tau(\ell)\equiv\int_0^\ell\df\ell^\prime \widehat S_\tau(\ell^\prime)\,.
\end{equation}
When projecting Eq.~\eqref{eq:hemidist} into HJM the outcome is not as simple as for $\taus$ and, much as it happened with the convolution of the partonic cross section with the shape function in Eq.~\eqref{eq:HJM-hadronic},
it is impossible to avoid genuine two-dimensional integrations:
\begin{align} \label{eq:HJM-dist}
& \frac{1}{\sigma_0} \frac{\df \hat{\sigma}_{\rm sing}}{\df \rho} =\; 2Q^3 H(Q, \mu)\! \int \!
\df \ell_1 \df \ell_2 \, J (Q^2\rho - Q\, \ell_1, \mu) J (Q^2\rho - Q\, \ell_2, \mu)
\widehat S^{\,\Xi}(\ell_1, \ell_2, \mu) \,,\\
& \frac{1}{\sigma_0} \widehat\Sigma_{\rm sing}^\rho(\rho^c)=\;Q^2 H(Q, \mu)\! \int \!
\df \ell_1 \df \ell_2 \, J (Q^2\rho^c - Q\, \ell_1, \mu) J (Q^2\rho^c - Q\, \ell_2, \mu)
\widehat S^{\,\Sigma}(\ell_1, \ell_2, \mu) \,,\nn
\end{align}
with
\begin{align}
& \widehat S^{\,\Xi}(k_1^c, k_2)= \int_0^{ k_1^c}\! \df k_1\, \widehat S( k_1, k_2)\,,
& \widehat S^{\,\Sigma}(k_1^c, k_2^c) &= \int_0^{ k_1^c}\!\! \df k_1\int_0^{ k_2^c}\!\! \df k_2\,
\widehat S( k_1, k_2)\,.
\end{align}

\subsection{Factorization Theorems in Laplace Space}\label{sec:fourier}
In this section we write the SCET factorization theorem for the DHM distribution in Laplace space as well as its projections into SHM and HJM. Let us start by defining the Laplace transforms of the various cross sections (the equivalent Fourier transforms are obtained by the replacements $y_j\to iy_j$ and $\xi_j\to i\xi_j$ below)
\begin{align}
\tilde\sigma_\taus(y) \,=& \!\int \! \df\tau\, e^{-\,y\taus}\frac{\df \sigma}{\df \taus}\,,\\\nonumber
\tilde\sigma(\xi_1,\xi_2)\,=& \!\int \! \df s_1\df s_2\,e^{-(\xi_1 s_1+\xi_2 s_2)}
\frac{\df^2 \sigma}{\df s_1\df s_2}\,,
&\tilde\sigma_\rho(y) \,=& \!\int \! \df\rho\, e^{-\,y\rho}\,\frac{\df \sigma}{\df \rho}\,,\\\nonumber
\widetilde\Xi(\xi_1,\xi_2)\,=& \!\int \! \df s_1\df s_2\,e^{-(\xi_1 s_1+\xi_2 s_2)}\Xi(s_1,s_2)\,,
&\widetilde\Sigma^\taus(y) \,=& \!\int \! \df\taus\, e^{-\,y\taus}\,\Sigma^{\taus}(\taus)\,,\\\nonumber
\widetilde\Sigma(\xi_1,\xi_2)\,=& \!\int \! \df s_1\df s_2\,e^{-(\xi_1 s_1+\xi_2s_2)}\Sigma(s_1,s_2)\,,
&\widetilde\Sigma^\rho(y) \,=& \!\int \! \df\rho\, e^{-\,y\rho}\,\Sigma^\rho(\rho)\,,
\end{align}
where the integrations proceed from zero to infinity along the real axis. With a few simple manipulations it is straightforward to prove the following relations:
\begin{align}
\widetilde\Xi(\xi_1,\xi_2) \,= \,&\, \frac{\tilde\sigma(\xi_1,\xi_2)}{\xi_1}\,,
&\widetilde\Sigma(\xi_1,\xi_2) \,=\,&\, \frac{\tilde\sigma(\xi_1,\xi_2)}{\xi_1\,\xi_2}\,,\\\nonumber
\tilde\sigma_\tau(y) \,=\,&\,\tilde\sigma\biggl(\!\frac{y}{Q^2},\frac{y}{Q^2}\!\biggr)\,,
&\widetilde\Sigma_\tau(y) \,=\, &\,\frac{\tilde\sigma_\tau(y)}{y}\,.
\end{align}
The HJM differential and cumulative distributions are simply obtained via inverse Laplace transforms
\begin{align}
\frac{\df \sigma}{\df \rho} & = 2Q^2\!\!\int_{0^+-\,i\,\infty}^{0^+ +\,i\,\infty}\!
\frac{\df \xi_1\df \xi_2}{(2\pi)^2}\,
e^{Q^2\rho(\xi_1+\xi_2)}\,\widetilde\Xi(\xi_1,\xi_2)\,,\\\nonumber
\Sigma^\rho(\rho) &= \!\!\int_{0^+-\,i\,\infty}^{0^+ +\,i\,\infty}\!\frac{\df \xi_1\df \xi_2}{(2\pi)^2}\,
e^{Q^2\rho(\xi_1+\xi_2)}\,\widetilde\Sigma(\xi_1,\xi_2)\,.
\end{align}
The contours for these inverse Laplace integrals\footnote{The integration to compute the inverse Fourier transform runs from $-\infty$ to $\infty$ on the real axis with a small shift upwards into the complex plane. In addition, the arguments in exponentials appear multiplied by $i$.} are slightly shifted to the right of the imaginary
axis, indicated by the $0^+$ above. For simplicity we will omit $0^+$ when quoting the integration boundaries from now on. Finally, the Laplace transforms of the HJM differential and cumulative distributions read
\begin{align}
\tilde\sigma_\rho(y) \,=\,\frac{2}{Q^2}\!\!\int\!\frac{\df x}{2\pi}\:
\widetilde\Xi\biggl(\!\frac{x}{Q^2},\frac{y-x}{Q^2}\!\biggr)\,,\qquad
\widetilde\Sigma^\rho(y) \,=\,\frac{1}{Q^4}\!\!\int\!\frac{\df x}{2\pi}\:
\widetilde\Sigma\biggl(\!\frac{x}{Q^2},\frac{y-x}{Q^2}\!\biggr)\,.
\end{align}
From Eq.~\eqref{eq:hemidist} one can show that in Laplace space the convolutions turn into regular products, yielding
\begin{align}\label{eq:doubleLaplace}
\frac{1}{\sigma_0}\tilde\sigma(\xi_1,\xi_2) = \,& H(Q, \mu)\widetilde{J}(L^{\mu^2}_{\xi_1},\alpha_s(\mu))
\widetilde{J}(L^{\mu^2}_{\xi_2},\alpha_s(\mu))\,
e^{-\,Q(\xi_1+\xi_2)\,[\,\delta(R,\mu)+\bar\Delta(R,\mu)\,]}
\\
&\times \widetilde{S}(L^\mu_{Q\,\xi_1}, L^\mu_{Q\,\xi_1},\alpha_s(\mu))
\widetilde{F}(Q\,\xi_1, Q\,\xi_2)
\nonumber \\[4pt]
= \,& H(Q, \mu_H)U_H(Q,\mu_H,\mu_J)\: \widetilde{J}(L^{\mu^2_J}_{\xi_1},\alpha_s(\mu_J))\:
\widetilde{J}(L^{\mu^2_J}_{\xi_2},\alpha_s(\mu_J))\:
e^{2K(-\Gamma_{\rm cusp},\gamma_S,\mu_J,\mu_s)}
\nonumber\\[4pt]
&\times(\mu_s\,e^{\gamma_E}Q\,\xi_1)^{-\omega(\Gamma_{\rm cusp},\mu_J,\mu_s)}
\: (\mu_s\,e^{\gamma_E}Q\,\xi_2)^{-\omega(\Gamma_{\rm cusp},\mu_J,\mu_s)}
\nonumber \\[4pt]
&\times
e^{-\,Q(\xi_1+\xi_2)\,[\,\delta(R,\mu_s)+\bar\Delta(R,\mu_s)\,]}\:
\widetilde{S}(L^{\mu_s}_{Q\,\xi_1}, L^{\mu_s}_{Q\,\xi_1},\alpha_s(\mu_s))\: \widetilde{F}(Q\,\xi_1, Q\,\xi_2)
\,. \nn
\end{align}
The Laplace transforms of the jet, partonic soft, and shape functions read
\begin{align}\label{eq:lapdef}
\widetilde{S}(L^\mu_{y_1}, L^\mu_{y_2},\alpha_s(\mu)) \,=&\! \int \! \df \ell_1 \df \ell_2\,
e^{-(y_1 \ell_1 \,+\,y_2 \ell_2)}\widehat S(\ell_1, \ell_2,\mu)\,,\\\nonumber
\widetilde{F}(y_1, y_2) \,=&\! \int \! \df \ell_1 \df \ell_2\,
e^{-(y_1 \ell_1 \,+\,y_2 \ell_2)}F(\ell_1, \ell_2)\,,\\\nonumber
\widetilde{J}(L^{\mu^2}_\xi,\alpha_s(\mu)) \,=& \!\int \! \df s\,e^{-\,\xi s} J(s, \mu)\,,
\end{align}
where we have defined
\begin{align} \label{eq:Ldefn}
L^\mu_x\equiv \ln(\mu\, x\, e^{\gamma_E}) \,,
\end{align}
with $\gamma_E$ being the Euler-Mascheroni constant.\footnote{Note that in Refs.~\cite{Becher:2008cf,Chien:2010kc} $L^\mu_x$ is defined with a minus sign relative to the notation in this article.} The evolution kernels have the following form:
\begin{align}\label{eq:UH}
U_H(Q,\mu_H,\mu_J) & = e^{2K(-\Gamma_{\rm cusp},\gamma_H,\mu_H,\mu_J)}
\biggl(\frac{Q}{\mu_H}\biggr)^{\!\!2\omega(\Gamma_{\rm cusp},\mu_H,\mu_J)}\,,\\\nonumber
K(\Gamma,\gamma,\mu_1,\mu_2) &= K(\Gamma,0,\mu_1,\mu_2) + \frac{1}{2}\omega(\gamma,\mu_1,\mu_2)\,,\\\nonumber
\omega(\Gamma,\mu,\mu_0) &\,=\, 2\!\int_{\alpha_s(\mu_0)}^{\alpha_s(\mu)}\frac{\df \alpha}{\beta(\alpha)}\,\Gamma(\alpha)\,,
\\\nonumber
K(\Gamma,0,\mu,\mu_0) &\,=\, 2\!\int_{\alpha_s(\mu)}^{\mu_0}\frac{\df \alpha}{\beta(\alpha)}\,\Gamma(\alpha)\!\!
\int_{\alpha_s(\mu_0)}^{\alpha^\prime}\frac{\df \alpha^\prime}{\beta(\alpha^\prime)}\,.
\end{align}
Explicit expressions for $K$ and $\omega$ up to N$^3$LL can be found in Eqs.~(A19) and (A20) of Ref.~\cite{Hoang:2014wka}, and a result for $\omega$ valid to any order in Eq.~(A19) of Ref.~\cite{Hoang:2021fhn}. To compute the Laplace transform of the partonic DHM distribution one simply sets $\widetilde{F}\to 1$ in Eq.~\eqref{eq:doubleLaplace}.

We emphasize that the exponential of $\delta(R,\mu_s)$ has to be expanded strictly in $\alpha_s(\mu_s)$ together with the rest of the matrix elements to subtract the $\mathcal{O}(\lqcd)$ renormalon in the partonic soft function, as explained in Sec.~\ref{sec:power}.
This scale-dependent renormalon subtraction implies the appearance of the gap subtraction constant $\bar\Delta(R,\mu_s)$ that leads to a shift in the hemisphere masses variables $s_i\to s_i - Q\bar\Delta$. To compute the cross section with power corrections in the $\msbar$ scheme for the soft function (i.e.\ without any renormalon subtraction) one simply sets $\delta = \bar\Delta = 0$.

\section{DHM and HJM Resummation with a 2-Dimensional Soft Function}
\label{sec:double}
In this section we discuss the resummation of large perturbative logarithms for DHM and HJM in the context of SCET.
We therefore focus on the partonic distribution, ignoring non-perturbative effects and renormalon subtractions. In the factorization formula~(\ref{eq:doubleLaplace})
this is
achieved by setting the shape function to a Dirac delta function, and the $\delta$ subtractions and finite shift $\bar\Delta$ to zero. In the case of thrust or the sum of hemisphere masses one does not loose generality, since, as we have seen in Eq.~\eqref{eq:thrust-hadronic}, the full (resummed) partonic distribution exactly factors from the shape function.

For HJM and DHM, hadronization effects do not factor from resummation in the same simple way, and therefore one needs to figure out a clever and efficient way of including both simultaneously. To that end, in \sec{resummation} we derive an exact expression to sum up large logarithms for these event shapes when the non-global part of the soft function is taken into account. In \sec{dihemiS} we show that the exact resummation admits a convergent expansion for the non-global part of the soft function that can be carried out to arbitrary precision. This expansion can be easily combined with a two-dimensional shape function to take into account hadronization effects in an efficient way. The expansion also justifies parametrizing the unknown non-global 3-loop soft function in terms of only three parameters.

\subsection[Resummation for DHM with $s_1\sim s_2$ and HJM]
{Resummation for DHM with $\boldsymbol{s_1\sim s_2}$ and HJM}\label{sec:resummation}
The RG evolution is easiest to compute in Laplace space, where it does not involve convolutions. For the dihemisphere soft function we know from its RG equation (RGE) that its dependence on $\mu$ must take the form of a power series of logarithms whose coefficients are set by the soft function cusp and non-cusp anomalous dimensions. Thus it is helpful to explicitly separate out the $\mu$ dependence. Defining
\begin{align}
L_1 \equiv L_{y_1}^\mu\,,
\qquad\qquad
L_2 \equiv L_{y_2}^\mu\,,
\end{align}
with \eq{Ldefn},
\begin{equation}\label{eq:softsplit}
\widetilde S({\rLO},{\rLT},\alpha_s(\mu))
= \widetilde U_s\bigl({\rLO}, \alpha_s(\mu)\bigr)\,
\widetilde U_s\bigl({\rLT},\alpha_s(\mu)\bigr)\,
\widetilde S_f\bigl( {\rLO} - {\rLT}, \alpha_s(\mu_{12}) \bigr) \,,
\end{equation}
where $\mu_{12} \equiv \mu\, e^{-(L_1+L_2)/2}$. Here, the function $\widetilde S_f$ contains constants and the non-global structure which
is a function of $L=L_1-L_2=\ln(y_1/y_2)$. It can be written as
\begin{equation}\label{eq:Ttilde}
\widetilde S_f\bigl( {\rLO} - {\rLT}, \alpha_s(\mu_{12}) \bigr)
= \exp\Big[{\widetilde T(y_1,y_2)} \Big] \,,
\end{equation}
where the form of $\widetilde T$ is constrained due to the non-Abelian exponentiation theorem. We note that $L_1-L_2$ and $\mu_{12}$ are $\mu$ independent. However, within our code we need to expand $\alpha_s(\mu_{12})$ back in terms of $\alpha_s(\mu)$ and $L_1+L_2$:
\begin{align}
\tilde T(y_1,y_2)={}&2\sum_{i = 1} \left[ \frac{\alpha_s (\mu)}{4 \pi} \right]^i \sum_{j = 0}^{i -1} T_{i j} (L) (L_1 + L_2)^j \,,\\
T_{n j}(L) ={}& \frac{1}{j} \sum_{i = j + 1}^{n - 1} i \beta_{n - i - 1}T_{i, j - 1}(L)\,,\nn
\end{align}
where $T_{ij}(L)=T_{ij}(-L)$.
The factor $\widetilde U_s(L_i,\alpha_s(\mu))$ is the RG evolution kernel in Laplace space which evolves from $\mu=e^{-\gamma_E}/y_i$ to $\mu$, and is known to N$^3$LL accuracy. It has a perturbative expansion in terms of the cusp and non-cusp anomalous dimensions that reads:
\begin{align} \label{eq:smuexp}
\widetilde U_s({\rL_i},\alpha_s)
&= \exp\Big[
{K(-\Gamma_{\rm cusp},\gamma_S,\mu,\mu\,e^{-L_i})\,-\,L_i\,\omega(\Gamma_{\rm cusp},\mu,\mu\,e^{-L_i})}
\Big]\,,
\end{align}
with $K$ and $\omega$ defined in Eqs.~\eqref{eq:UH}.
The function $\widetilde S_f(\rL,\alpha_s)$, which contains the non-global structure, is known to 2-loops~\cite{Kelley:2011ng,Hornig:2011iu}.

In terms of the Laplace-transformed jet and partonic soft functions, the $\tau_s$ cumulative distribution takes the form
\begin{align} \label{eq:scetdist}
\frac{1}{\sigma_0}\widehat\Sigma^{\taus}(\taus) \,=\, &\,
H(Q,\mu_H)\,U_H(Q,\mu_H,\mu_J) \,e^{2K(-\Gamma_{\rm cusp},\gamma_S,\mu_J,\mu_s)}\,
\widetilde S_T(-\partial_{\eta},\mu_s) \\
&\times
\biggl[\widetilde J\Bigl( \ln\frac{\mu_J^2}{\mu_s\, Q}-\partial_\eta,\mu_J\Bigr)\biggr]^2
\biggl[\biggl(\frac{Q\,\taus}{\mu_s} \biggr)^{\!\!\eta} \!\frac{e^{-\gamma_E\, \eta}}{\Gamma(\eta+1)}\biggr]
\,,\nonumber
\end{align}
where $\eta = 2\omega(\Gamma_{\rm cusp},\mu_J, \mu_s)$, and with the Laplace-transformed thrust soft function $\widetilde{S}_T(\ell,\mu)$ given by taking $L_1=L_2=\ell$,
\begin{equation}
\widetilde S_T(\ell,\alpha_s(\mu)) \equiv \widetilde{S}(\ell,\ell,\alpha_s(\mu)) =
\bigl[\,\widetilde U_s\bigl(\ell,\alpha_s(\mu)\bigr)\bigr]^2 \,\widetilde S_f \bigl(0,\alpha_s\bigl(\mu\, e^{-\ell}\bigr)\bigr) \,,
\end{equation}
as can be easily derived from Eq.~\eqref{eq:thrust-jet-soft}. The
fixed-order hard and jet functions, $H(Q,\mu)$ and $\wt{J}(\ell,\mu)$, and their respective anomalous dimensions can be found in an analytical form in Ref.~\cite{Becher:2008cf}, or numerically for $n_f=5$ in Ref.~\cite{Abbate:2010xh}. Equation~\eqref{eq:scetdist} can be easily modified to compute the differential partonic SHM distribution simply setting $\eta\to\eta\, - \,1$ after taking the $\partial_\eta$ derivatives and multiplying by the global factor $Q\,e^{\gamma_E}/\mu_s$.

Note that the $\tau_s$ distribution only depends on
\begin{equation}
\widetilde S_f(0,\alpha_s) =
\exp\biggl[2\sum_i \Bigl(\frac{\alpha_s}{4\pi}\Bigr)^{\!i} s_i\biggr],
\end{equation}
where $\alpha_s$ is evaluated at $\mu \exp(-L)$ and $s_i=T_{i0}(0)$ with~\cite{Monni:2011gb,Kelley:2011ng,Baranowski:2024vxg,Baranowski:2024ysi}
\begin{align}\label{eq:ctwoeq}
s_1 ={}& \!-\!C_F \frac{\pi^2}{2}
\,,\\
s_2={}& C_F\biggl[\biggl(\frac{40}{81} +\frac{77 \pi ^2}{27} -\frac{52 \zeta_3}{9} \biggr) T_{F\,} n_f +\!
\biggl(\frac{7 \pi^4}{15} - \frac{1070}{81} - \frac{871 \pi ^2}{108} + \frac{143 \zeta_3}{9} \biggr)\biggr]
C_F C_A\,,
\nn\\
s_3 ={}& C_F\biggl[34.47129\,C_Fn_fT_F + 419.8619\,C_A n_fT_F-376.88789\,C_A^2\nn\\
&-\!\biggl(\frac{132704}{6561}-\frac{25952 \zeta_3}{243}-\frac{200 \pi ^2}{243}+\frac{164 \pi ^4}{1215}\biggr)n_f^2T_F^2\biggr]
\,.\nonumber
\end{align}
Note that the color factor $C_F^2$ does not show up in $s_2$, while $C_F^3$ and $C_AC_F^2$ are absent in $s_3$, as required by the non-Abelian exponentiation theorem. The same feature was found in the explicit three-loop computation carried out in Ref.~\cite{Clavero:2024yav} and is present in the three-loop soft and cusp anomalous dimensions as well.
$\widehat\Sigma^{\taus}(\taus)$ depends on $\widetilde S_f(0,\alpha_s(\mu e^{-\ell}))$, so after expanding in powers of $\alpha_s(\mu)$ only depend on powers of $\ell$, which we can evaluate by acting with a finite number of derivatives $\partial_\eta$.

The partonic resummed DHM doubly cumulative distribution can be written as
\begin{align} \label{eq:mass-dist}
\frac{1}{\sigma_0} \widehat\Sigma(s_1,s_2)
=\,& H(Q,\mu_H)U_H(Q,\mu_H,\mu_J) e^{2K(-\Gamma_{\rm cusp},\gamma_S,\mu_J,\mu_s)}
\widetilde J\Bigl( \ln\frac{\mu_J^2}{\mu_s\, Q}-\partial_{\eta_1},\mu_J\Bigr)\\
\times \,&
\widetilde J\Bigl( \ln\frac{\mu_J^2}{\mu_s\, Q}-\partial_{\eta_2},\mu_J \Bigr)
\widetilde U_s\bigl(-\partial_{\eta_1},\alpha_s(\mu_s)\bigr)
\widetilde U_s\bigr(-\partial_{\eta_2},\alpha_s(\mu_s)\bigr)
\nonumber \\
\times \,&
\widetilde S_f\bigl(\partial_{\eta_1}-\partial_{\eta_2},
\alpha_s\bigl(\mu_s\, e^{(\partial_{\eta_1}+\partial_{\eta_2})/2}\bigr)\bigr)
\biggl(\frac{s_1}{Q\,\mu_s} \biggr)^{\!\!{\eta_1}}
\biggl(\frac{s_2}{Q\,\mu_s} \biggr)^{\!\!{\eta_2}}\!\!
\frac{e^{-\gamma_E\, \eta_1}}{\Gamma(\eta_1+1)}\,
\frac{e^{-\gamma_E\, \eta_2}}{\Gamma(\eta_2+1)}\,,\nonumber
\end{align}
evaluated at $\eta_1 =\eta_2= \omega(\Gamma_{\rm cusp}, \mu_J, \mu_s)$. This equation can be employed to compute the partonic HJM cumulative cross section $\hat\Sigma^\rho$ by setting $s_1=s_2=Q^2\rho$. To compute the partonic differential HJM distribution one makes in addition the replacement $\eta_1\to \eta_1 - 1$ after taking the derivatives and multiplies the whole expression by $2Qe^{\gamma_E}/\mu_s$, where the factor of $2$ comes about because of the symmetries of Eq.~\eqref{eq:mass-dist}.

\begin{figure}[t]\centering
\includegraphics[width=0.6\textwidth]{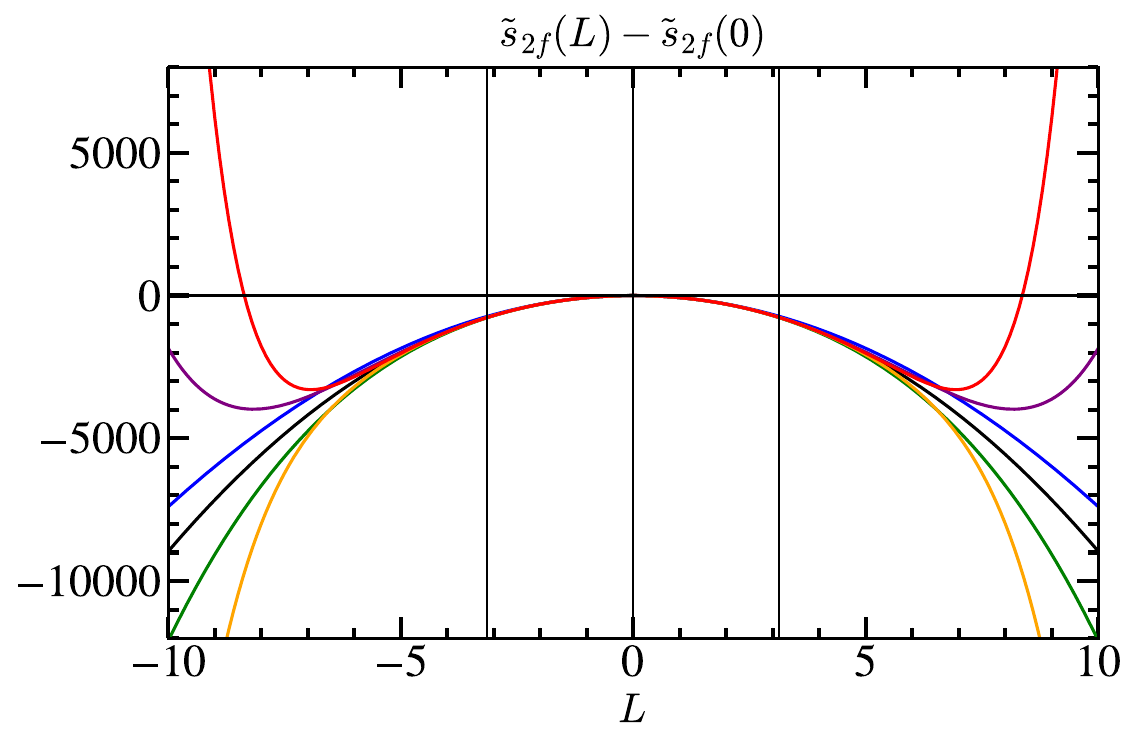}
\caption{The non-global part of the 2-loop soft function, $\tilde s_{2f}(\rL) - \tilde s_{2f}(0)$ is shown
as a function of \mbox{$\rL = \ln(\rnO/\rnT)$} (black curve). The blue, green, purple, orange and red lines correspond to expansions of $\tilde s_{2f}(\rL) - \tilde s_{2f}(0)$ around $\rL=0$ to second, fourth, sixth, eight and tenth order in $\rL$, respectively. As explained in the text, only agreement between $-\pi < \rL <\pi$ (marked with vertical lines) is needed.\label{fig:sftplot}}
\end{figure}

The operator $\tf(\partial_{\eta_1}-\partial_{\eta_2},\alpha_s)$ acting on a function of $\eta_1$ and $\eta_2$ is not straightforward to evaluate unless $\tf(\rL,\alpha_s)$ is (or can be approximated by) a polynomial. In Ref.~\cite{Hoang:2008fs}, Hoang and Kluth conjectured that this function is a second-order even polynomial in $\rL$. In Ref.~\cite{Chien:2010kc}, the possibility of a more general form was considered by Chien and Schwartz. However, since the non-global structure was not yet known, Ref.~\cite{Chien:2010kc} also used a polynomial approximation to $\tf(\rL,\alpha_s)$, including uncertainties which were meant to account for the ignorance in $\tf(\rL,\alpha_s)$. This function was eventually computed exactly in momentum space in Refs.~\cite{Kelley:2011ng,Hornig:2011iu}, and in Laplace space in \cite{Hornig:2011iu}, and was found to involve 3rd-order polylogarithms at \mbox{2-loops}. In our notation
\begin{equation}
\widetilde S_f(\rL,\alpha_s) =\exp\Bigl[-\Bigl(\frac{\alpha_s}{4\pi}\Bigr) C_F \pi^2
+\Bigl(\frac{\alpha_s}{4\pi}\Bigr)^{\!2} \tilde{s}_{2f}(\rL)+ {\mathcal{O}}(\alpha_s^3)\Bigr],
\end{equation}
where
\begin{equation} \label{eq:s2fdefn}
\tilde s_{2f}(\rL) = 2T_{20}(L)=
2 t_2(e^\rL)\,,
\end{equation}
with $t_2(x)$ given in Eq.~(3.31) of Ref.~\cite{Hornig:2011iu}. The function $\tilde s_{2f}(\rL)$ is shown in Fig.~\ref{fig:sftplot} along with
Taylor expansions around $L=0$ truncated at different orders.
While the partonic DHM with this soft function in Eq.~\eqref{eq:mass-dist} can be evaluated exactly, the hadronic DHM with the additional shape function integrations becomes very difficult without additional approximations.
Below, we justify using the expansion of $\tilde s_{2f}(\rL)$ around $\rL=0$
for numerical analyses of the hadronic DHM.

\subsection[Polar Expansion of the 2-Dimensional Soft Function for $s_1\sim s_2$] {Polar Expansion of the 2-Dimensional Soft Function for $\boldsymbol{s_1\sim s_2}$}\label{sec:dihemiS}
If $\tilde s_{2f}$ in \eq{s2fdefn} could be written as the sum of products of functions depending only on $L_1$ or $L_2$, the convolution could be organized as the sum of products of one-dimensional integrals. This is unfortunately not the case.

Before the ${\cal O}(\alpha_s^2)$ piece of the soft function was computed~\cite{Kelley:2011ng,Hornig:2011iu}, a simple ansatz was proposed by Hoang and Kluth \cite{Hoang:2008fs}, who fitted the unknown constants to the output of the {\sc event2} generator. The ansatz reads
\begin{equation}
\tilde s_{2f}(L)\,=\,2\bigl(s_2+s_{2,2}\,L^2\bigr)\,,\label{eq:HK}
\end{equation}
where $s_2$ directly corresponds to the non-log term of the thrust soft function. This ansatz falls into the category of being the sum of products of one-variable functions.
Consider the expansion of the known exact $\tilde s_{2f}$ function in powers of $L$:\,\footnote{For simplicity, we will take $n_f=5$ active flavors in all our numerical analyses.}
\begin{align}
\label{eq:small-theta-numeric}
\tilde s_{2f}(L) =2\sum_{k = 0}s_{2,2k}L^{2k} ={} &2\,[
-40.6804 - 18.4928L^2 - 0.117284\, L^4
+ 0.00255967L^6 \nn \\
&\quad-\!0.0000599137 L^8 + 1.44569\times 10^{-6}L^{10}+{\cal O}(L^{12})\,]
\,,
\end{align}
where we identify $s_2 \equiv s_{2,0}$. Interestingly, the $s_{2,2}$ in \eq{small-theta-numeric} is numerically very similar to the fitted value for $s_{2,2}$ from Ref.~\cite{Hoang:2008fs}, which quotes $s_{2,2}=-15.4\pm2.5$. Furthermore, the coefficients of the $L^4$ and higher terms in \eq{small-theta-numeric} are numerically much smaller. When performing resummation for HJM, we integrate over the variables in $L$. The rapid convergence of the expansion hints that an accurate resummed result may be obtained by keeping only some number of terms in the expansion for small $L$. We confirm this expectation below.

To evaluate the resumed DHM cumulative, we transform to Laplace space where the RG equation does not involve convolutions, solve the RG, and transform back. The Laplace-transformed soft function is defined in Eq.~\eqref{eq:lapdef}. The contribution of the soft function to the DHM cumulative distribution [\,i.e.\ setting $H=1$ and $J(s)=\delta(s)$\,] in fixed-order perturbation theory, with all renormalization scales set equal to $\mu$, is given by
\begin{equation}
\widehat\Sigma_{\rm soft}(s_1,s_2)=\! \int_0^{s_1/Q} \!\!\df \ell_1\! \int_0^{s_2/Q}\!\! \df \ell_2\!
\int_{-i\infty}^{+i\infty} \frac{\df \rnO}{2\pi i}
\int_{-i\infty}^{+i\infty} \frac{\df \rnT}{2\pi i}\,
e^{\rnO \ell_1+\rnT \ell_2} \,\widetilde{S}(L_1, L_1,\alpha_s(\mu))\,.
\label{eq:doublemasscont1}
\end{equation}
Performing the $\ell_1$ and $\ell_2$ integrals, this simplifies to~\cite{Hornig:2011iu}
\begin{equation}
\widehat\Sigma_\text{soft}(s_1,s_2)= \frac{1}{(2\pi i)^2}\!
\int\! \frac{\df \rnO}{\rnO}\!
\int\! \frac{\df \rnT}{\rnT}\, e^{(\rnO s_1\,+\,\rnT s_2)/Q}\, \widetilde{S}(L_1, L_2,\alpha_s(\mu))\,.
\label{eq:doublemasscont2}
\end{equation}
Using the factorized expression in Eq.~\eqref{eq:softsplit}, the resummed Laplace-transformed partonic soft function is
\begin{align}\label{eq:FO-NGL}
\widetilde{S}(L_1, L_2,\alpha_s(\mu)) \,=\,&\, e^{K+\,2 A_\Gamma (\rLO + \rLT)}
\widetilde U_s\bigl(L_1^{s}, \alpha_s(\mu_s)\bigr)
\widetilde U_s\bigl(L_2^{s},\alpha_s(\mu_s)\bigr)\\\nonumber
&\times\widetilde S_f\Bigl( L_1-L_2, \alpha_s\bigl(\mu_s\,e^{-\frac{L^{s}_1+L^{s}_2}{2}}\bigr)\Bigr) \,,
\end{align}
with $K = -K(\Gamma_{\rm cusp},\gamma_S,\mu,\mu_s)$, $L_i^s\equiv L_i^{\mu_s}$, and $2A_\Gamma = -\,\omega(\Gamma_{\rm cusp},\mu,\mu_s)$. Here $\mu$ corresponds to the common renormalization scale, which takes the value $\mu=\mu_J$ in Eq.~\eqref{eq:mass-dist}.
\begin{figure*}[t!]
\subfigure[]
{\includegraphics[width=0.48\textwidth]{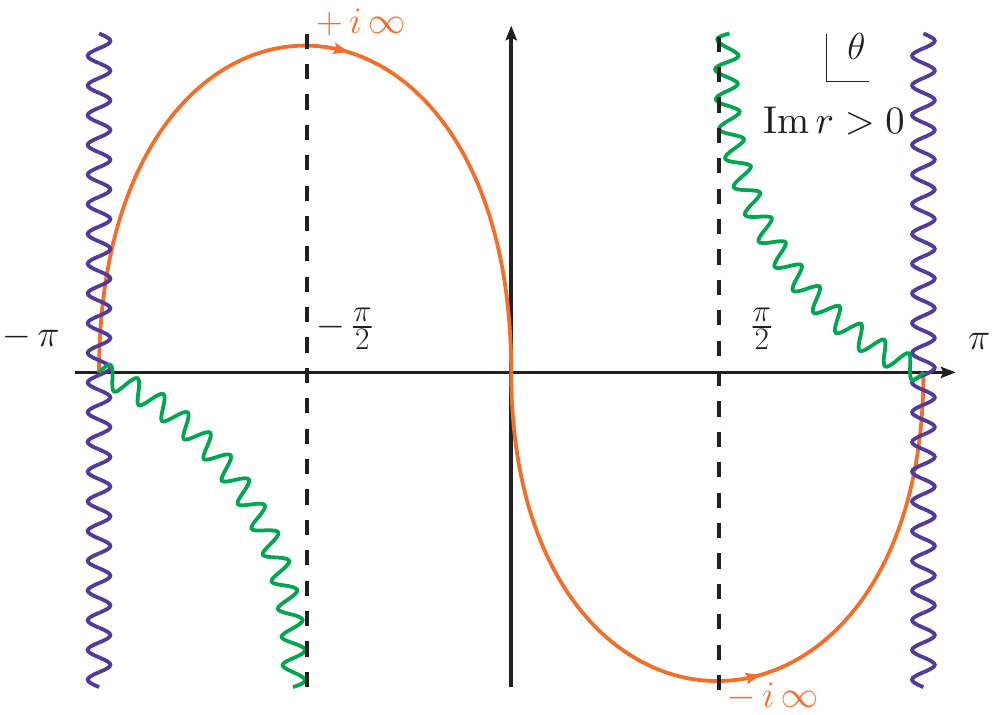}
\label{fig:cuts1}}
\subfigure[]{\includegraphics[width=0.48\textwidth]{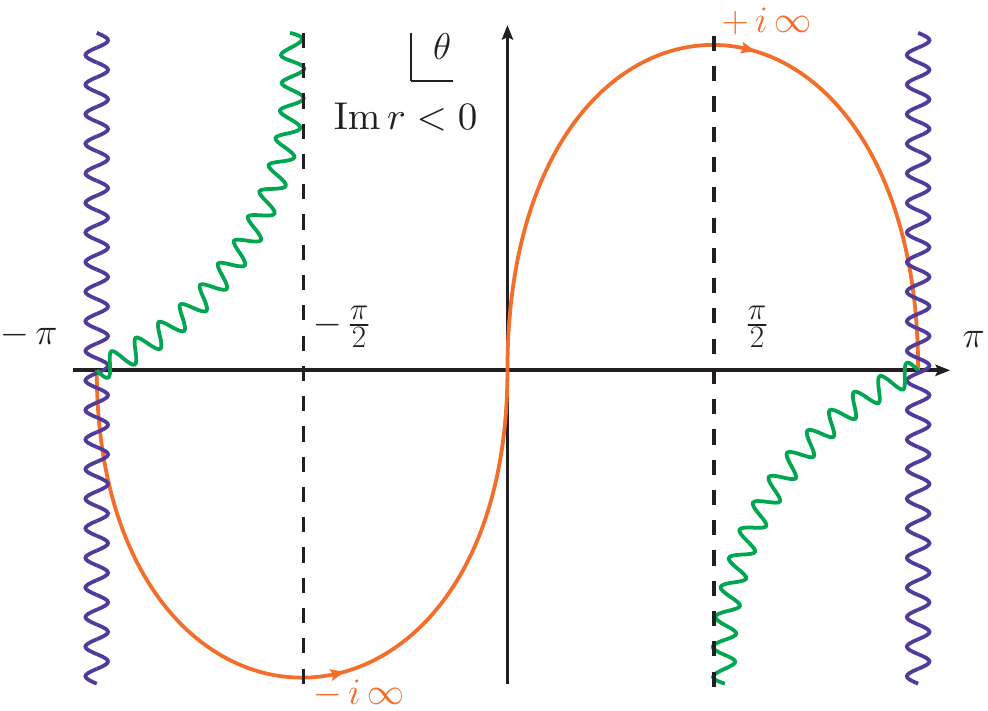}
\label{fig:cuts2}}
\caption{Contour integration and branch cuts in the complex $\theta$ plane. The orange solid line shows the original contour for the $\theta$ integration at fixed $r$ with $\im r \ne 0$ and $\re r=0^+$. The blue wiggly line corresponds to the branch cuts induced by $\tilde s_{2f}$, and the green wiggly lines correspond to the cuts induced by $\widetilde U_s$. The left panel corresponds to the situation $\im r >0$, and the two green branch cuts are defined by the functions $\im \theta=\ln[-\cos (\re \theta)]$ for $-\pi<\re \theta<0$ and $\im \theta= -\ln[-\cos(\re \theta)]$ for $0<\re \theta <\pi$. The right panel corresponds to $\im r <0$, and the branch cuts are defined by $\im \theta=-\ln[-\cos(\re \theta)]$ for $-\pi<\re \theta<0$ and $\im \theta=\ln[-\cos (\re \theta)]$ for $0<\re \theta<\pi$.}
\label{fig:cuts}%
\end{figure*}
Next, let us rescale $y_i= Q^2y_i'/s_i$ and change variables to
\begin{equation}\label{eq:change-variables}
i \theta=\ln \biggl( \frac{\rnOp}{\rnTp}\biggr)\,, \qquad r \equiv \rnOp + \rnTp\,,
\end{equation}
so that $\df \rnO \df \rnT/(\rnO \rnT) =\df \rnOp \df \rnTp/(\rnOp \rnTp) = \df r \df \theta/r $. We also define
\begin{equation}
\kappa_1 \equiv 1+ e^{-i\theta}\,, \qquad \kappa_2 \equiv 1+ e^{i\theta}\,,
\end{equation}
which let us write
\begin{equation}
L_i^s = \ln \biggl(\frac{\mu_s e^{\gamma_E} Q^2r }{\kappa_is_i}\biggr)\,.
\end{equation}
Note that
\begin{align}
L_1^s - L_2^s = L_1-L_2=\ln(s_2/s_1) + i \theta\,,
\end{align}
and $\kappa_1 \kappa_2 =[\, 2\cos(\theta/2)\,]^2$. We then have
\begin{align}
\label{eq:RpreL}
\widehat\Sigma_{\rm soft}(s_1,s_2)=\,&\, e^K\!\!\! \int_C \frac{\df \theta}{2\pi i}
\widetilde S_f\Bigl( \ln(s_2/s_1) + i\theta, \alpha_s\bigl(\mu_s\,e^{-\frac{L_1^s+L_2^s}{2}}\bigr)\Bigr)\!\!
\int_{-i\infty}^{+i\infty} \frac{\df r}{2\pi i}
\frac{e^{r}}{r}\\\nonumber
&\times
\widetilde U_s\bigl(L_1^s, \alpha_s(\mu_s)\bigr) \widetilde U_s\bigl(L_2^s,\alpha_s(\mu_s)\bigr) \!
\biggl( \frac{\mu_s^2\, r^2\, Q^4\,e^{2 \gamma_E} }{\kappa_1 \kappa_2s_1s_2}\biggr)^{\!\!2 A_\Gamma}.
\end{align}
The $r$ contour corresponds to the ordinary inverse Laplace transform (along the imaginary axis). The $\theta$ integration goes from $-\,\pi$ to $\pi$ along a contour which depends on the sign of the imaginary part of $r$, as shown in Fig.~\ref{fig:cuts}. In Ref.~\cite{Hornig:2011iu} the simple case of convolution with no resummation of Eq.~\eqref{eq:FO-NGL} was worked out, and it was shown that the contour in the complex $\theta$ plane can be flattened down to the real axis because no branch cut prevents the deformation. We generalize the proof showing that in the case in which resummation is turned on, the same contour deformation is still possible. On the one hand, $\widetilde S_f\bigl(\,\ln(s_2/s_1) + i\theta,\alpha_s\bigr)$ has no singularities for \mbox{$-\,\pi<\text{Re}(\theta)<\pi$}, at least up to two loops; on the other, $\rLO$ and $\rLT$ depend on $\ln[\,r/(1+e^{\pm i\,\theta})\,]$ yielding branch cuts in the integrand of Eq.~\eqref{eq:RpreL} as shown in Fig~\ref{fig:cuts}. We can thus again deform the $\theta$ integral to the real axis, between $-\pi$ and $\pi$. The DHM cumulative distribution then simplifies to
\begin{align} \label{eq:Rwithdel}
\widehat\Sigma_{\rm soft}(s_1,s_2)={}& e^K\, \widetilde U_s\bigl(-\partial_{\eta_1}, \alpha_s(\mu_s)\bigr) \,
\widetilde U_s\bigl(-\partial_{\eta_2},\alpha_s(\mu_s)\bigr)\int_{-\pi}^\pi\frac{\df \theta}{2\pi}\\
&
\times\tf\bigl( \ln(s_2/s_1) + i \theta,\alpha_s\bigl(\mu_s\, e^{(\partial_{\eta_1}+\partial_{\eta_2})/2}\bigr)\bigr)
\frac{e^{-\gamma_E (\eta_1 + \eta_2)}}{\Gamma ( 1 + \eta_1 + \eta_2)}\,,\nonumber
\end{align}
evaluated at $\eta_1 = \eta_2 = \omega(\Gamma_{\rm cusp},\mu,\mu_s)$. This form of the cumulative is amenable to a direct, numerical, integration. Of course, the derivatives with respect to $\eta_{1,2}$ should be performed before the integration is carried out. The dependence on $\partial_{\eta_i}$ is always polynomial, since $\alpha_s\bigl(\mu_s\, e^{(\partial_{\eta_1}+\partial_{\eta_2})/2}\bigr)\bigr)$ has to be expanded in terms of $\alpha_s(\mu_s)$ and powers of $\partial_{\eta_1}+\partial_{\eta_2}$.

To explore this expression, let us consider its dependence on the two-loop non-global structure $\tilde s_{2f}(i\theta)$.
First consider the HJM cumulative distribution (that is, $s_1=s_2=Q^2\rho$), turn off resummation (setting $K= \omega = 0$ and $\mu_s = \mu$) and put in the fixed-order expansion of $\widetilde S_f(\rL,\alpha_s)$. For $\Sigma^\rho_{\rm soft}(\rho)=\Sigma_{\rm soft}(Q^2\rho, Q^2\rho)$ at order $\alpha_s^2$, for $\tf(\rL,\alpha_s)$ (which starts at order $\alpha_s^2$) to contribute, we must take the $\alpha_s^0$ part of the evolution factors.
Then the contribution to the HJM cumulative from the non-global structure is just a number
\begin{equation}\label{eq:c2rho}
c_2^\rho \,=\, \frac{1}{2\pi} \!\int_{-\pi}^\pi\! \df\theta\,\tilde s_{2f}(i \theta)\,.
\end{equation}
At order $\alpha_s^3$, the two-loop non-global structure contributes to two terms: one with $\rho$ dependence
\begin{equation}
c_2^\zeta\, \ln\rho
= \ln \rho\, \frac{1}{2\pi} \! \int_{-\pi}^\pi \! \df\theta\, \tilde s_{2f}(i\theta)
\ln \biggl(\!2\cos\frac{\theta}{2}\biggr)\,,
\end{equation}
and one contributing to the 3-loop constant
\begin{equation}
c_3^\rho \,=\,
\frac{1}{2\pi} \! \int_{-\pi}^\pi \! \df\theta\, \tilde s_{2f}(i\theta)
\biggl[\ln^2 \biggl(\!2\cos\frac{\theta}{2}\biggr) -\frac{\theta^2}{4}\biggr]\,.
\end{equation}
These terms are generated e.g.\ by products of the one-loop jet function and $\tilde s_{2f}$. The integral forms for $c_2^\rho$ and $c_2^\zeta$ and their numerical extractions from {\sc event2} data were presented in Ref.~\cite{Chien:2010kc}. An exact expression for $c_2^\rho$ was computed in~\cite{Kelley:2011ng}.

\begin{figure*}[t!]
\subfigure[]
{\includegraphics[width=0.49\textwidth]{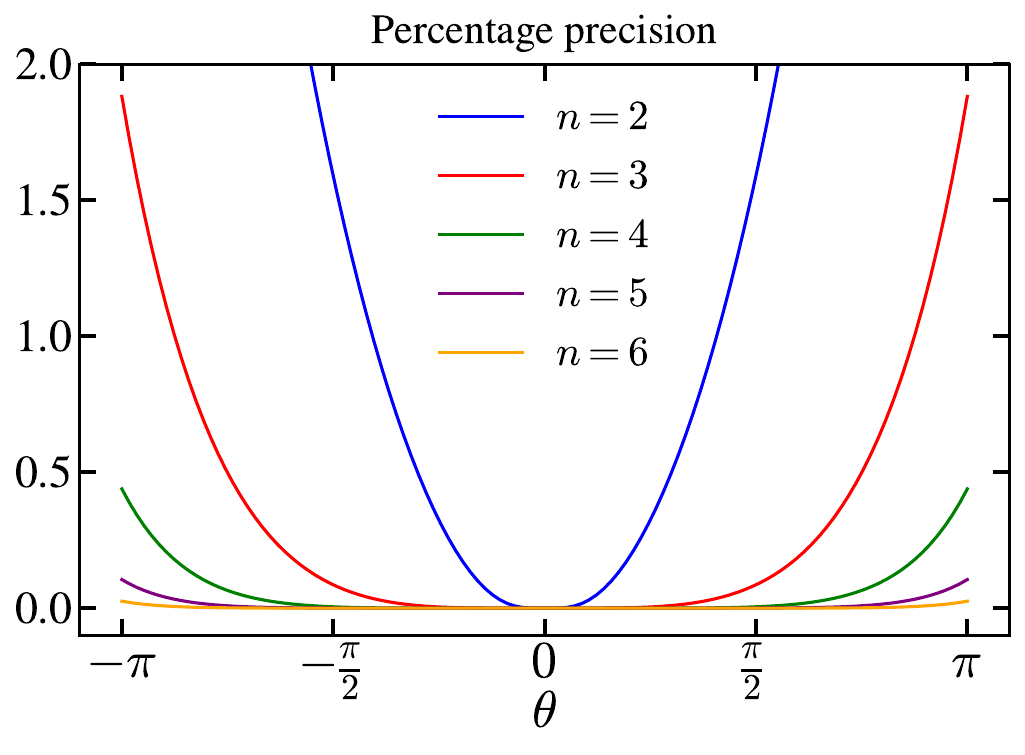}
\label{fig:Convergence-theta}}
\subfigure[]{\includegraphics[width=0.5\textwidth]{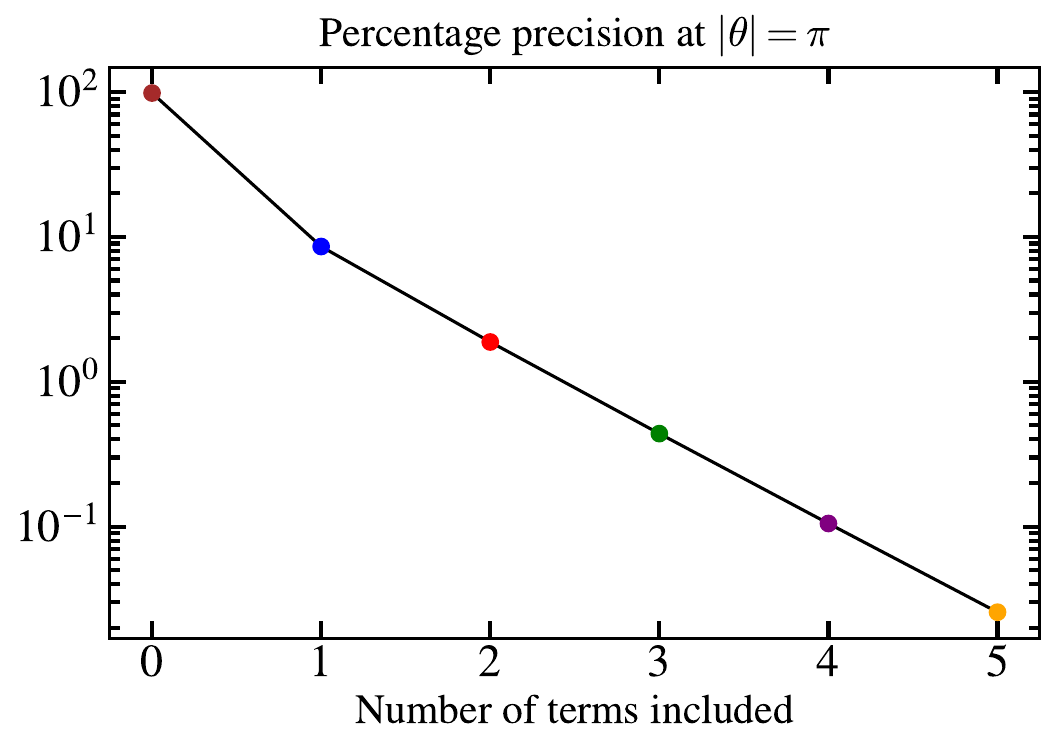}
\label{fig:Convergence-edge}}
\caption{Convergence of the polar expansion. The left panel shows the percent accuracy of the expansion of $\tilde s_{2f}(i\theta)$ when more terms are included: blue, red, green and purple lines correspond to including up to the $1$st, $2$nd, $3$rd, and $4$th terms in the expansion, respectively.
The right panel shows the percent accuracy of the small angle expansion at the point where it is least accurate ($\theta=\pm\,\pi$), as a function of the number of terms included in the expansion.}
\label{fig:convergence}%
\end{figure*}

\begin{figure}\centering
\includegraphics[width=0.6\textwidth]{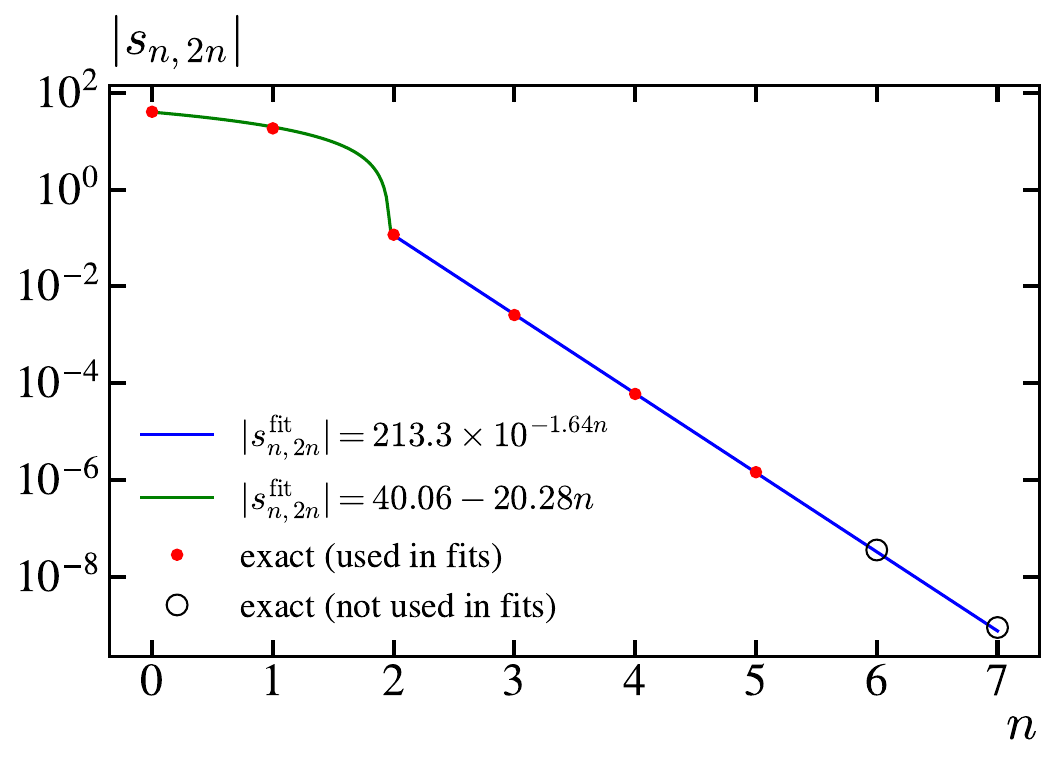}
\caption{\label{fig:sncoef}
Asymptotic behavior for $n\ge 2$ of the polar expansion coefficients for $\tilde s_{2f}$ in \eq{small-theta-numeric}. Exact numerical values of the coefficients are represented by red dots. The blue line shows a fit to the $n=2$ to $n=5$ coefficients, and is in excellent agreement with the numerical values even for larger $n$.
We also show how well a linear fit (green curve) works for the first three coefficients. }
\end{figure}

A simplifying observation is that on the interval $-\,\pi \le \theta \le \pi$ the function \mbox{$\tilde s_{2f}(L= i \theta)$} is well-behaved: it is analytic, symmetric, and bounded. In fact, as can be seen in Fig.~\ref{fig:sftplot}, it looks very much like a parabola in $L$. Thus, we expect that replacing $\tilde s_{2f}(\rL = i\theta)$ by its Taylor expansion around \mbox{$\theta=0$} should be rapidly convergent, as illustrated for the first few terms in~\eq{HK}. We call this series the {\it polar expansion} (the polar and small $L = \ln(z)$ expansions are trivially related by an alternating sign).
The excellent convergence of this expansion on the $-\,\pi \le \theta \le \pi$ interval is shown in \fig{sftplot}. Another way of quantifying its
convergence is by looking into the asymptotic behavior of its coefficients. We find that starting from the third term all coefficients have a negative sign and follow an exponential fall off with $n$ to a very good approximation. Fitting for the coefficients we find \mbox{$s_{2,2n}\simeq -\,213.3\times 10^{-1.64\,n}$} for $n\geq 2$.%
\footnote{The first three coefficients do not follow this pattern, but however we find that to a good approximation they satisfy the linear equation $|s_{2,2n} | \simeq 40.06-20.28\, n = 40.06 (1- 0.5064\, n)$ for $0\leq n \leq 2$.}
This can be seen graphically in Fig.~\ref{fig:sncoef}
and implies that the series is convergent for $|L|<10^{0.82}\approx 6.6$, which can also be seen in Fig.~\ref{fig:sftplot}.
The deformation of the $\theta$ contour together with the convergence of the expansion on the $|\theta|<\pi$ interval justifies expanding $\widetilde S_f(\rL,\alpha_s)$ around $\rL=0$.
The advantage of using this expansion, to finite order, is that
for a polynomial in $\rL$ it becomes straightforward to evaluate the HJM cumulative by taking the integral of the
partonic $\hat \Sigma(s_1,s_2)$ in Eq.~\eqref{eq:mass-dist} with the two dimensional shape function.
This is a remarkable result.

\begin{figure}[t]\centering
\includegraphics[width=0.55\textwidth]{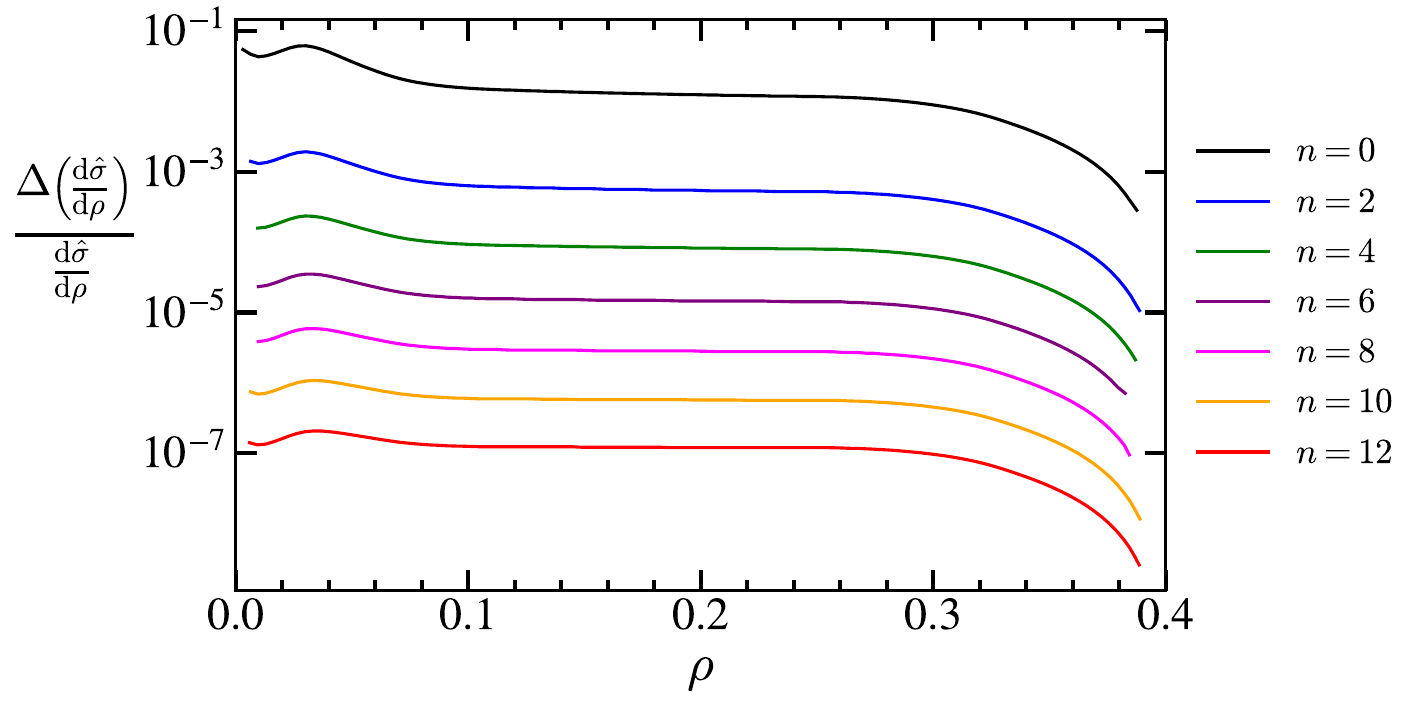}
\caption{Convergence of the full-spectrum cross section when including more terms in the polar expansion. The plot shows $\Delta(\df\hat\sigma/\df\rho)/(\df\hat\sigma/\df\rho)$, a difference obtained by taking the exact result (using the unexpanded non-global term in the soft function) and subtracting various approximations, normalized to the exact result $\df\hat\sigma/\df\rho$. The black, blue, green, purple, magenta, orange and red lines correspond to approximations obtained by expansions that include from zero to six terms, respectively (i.e.\ from $\theta^0$ to $\theta^{12}$). The curves use the singular partonic differential distribution at N$^3$LL (that is, including two-loop matrix elements, three- and four-loop non-cusp and cusp anomalous dimensions, respectively). }\label{fig:fracdiff}
\end{figure}

To check the quality of this approximation, let us examine the perturbative HJM distribution $\df\sigma/\df\rho$ obtained from Eq.~\eqref{eq:mass-dist}, when approximating $\tilde s_{2f}(\rL)$ by a polynomial versus computed exactly.%
\footnote{If one is computing the double hemisphere mass cumulative in \eq{massdistfull}, the polar expansion can be performed directly on $\tilde s_{2f}(L)$ around $L$ = 0 with a substitution $L\to L + \ln(s_2/s_1)$ after the expansion is performed, cf. Eq.~\eqref{eq:Rwithdel}.}
Including the hard, jet and soft functions, the exact result for the DHM cumulative is
\begin{align} \label{eq:massdistfull}
&\frac{1}{\sigma_0} \widehat\Sigma(s_1,s_2) =
H(Q,\mu_H)U_H(Q,\mu_H,\mu_J) e^{K(-\Gamma_{\rm cusp},\gamma_S,\mu_J,\mu_s)}
\widetilde J\Bigl( \ln\frac{\mu_J^2}{Q\,\mu_s}-\partial_{\eta_1},\mu_J\Bigr) \\
&\quad\qquad\times\!
\widetilde J\Bigl( \ln\frac{\mu_J^2}{Q\,\mu_s}-\partial_{\eta_2},\mu_J \Bigr)
\widetilde U_s\bigl(-\partial_{\eta_1}, \alpha_s(\mu_s)\bigr)
\widetilde U_s\bigl(-\partial_{\eta_2},\alpha_s(\mu_s)\bigr)
\nonumber\\
&\quad\qquad\times\!\! \int_{-\pi}^\pi\frac{\df \theta}{2\pi}
\widetilde S_f\bigl( \ln(s_1/s_2)+ i \theta,\alpha_s\bigl(\mu_s\, e^{(\partial_{\eta_1}+\partial_{\eta_2})/2}\bigr)\bigr)
\biggl( \frac{\kappa_1s_1}{Q\mu_s}\biggr)^{\!\!\eta_1}\biggl( \frac{\kappa_2s_2}{Q\mu_s}\biggr)^{\!\!\eta_2}
\frac{e^{-\gamma_E (\eta_1 + \eta_2)}}{\Gamma ( 1 + \eta_1 + \eta_2)}\,.\nonumber
\end{align}
The relative difference between the exact result and the expansion with up to seven terms is shown in Fig.~\ref{fig:fracdiff}.
For our numerical analyses later on we choose to keep the first six terms. This results in a difference less than $10^{-6}$ everywhere, which is much smaller than our perturbative uncertainty at N$^3$LL$^\prime$.

Note that if a convolution with a general two-dimensional shape function is to be performed, there is no way to reduce the general result to a single integral, like in Eq.~\eqref{eq:massdistfull}, since resummation and non-perturbative corrections get entangled. Since the shape function depends on one non-perturbative dimensionful parameter $\lqcd$ (implicitly), one cannot simply assume that the non-global part of $F$ can only depend on the ratio $k_1/k_2$. Hence, the change of variables in Eq.~\eqref{eq:change-variables} does not imply any simplification. Therefore one is left with a (numerically very intense) two-dimension integration. Instead, if one uses the polynomial expansion to $\tilde s_{2f}(\rL)$, then the analysis becomes numerically tractable. In particular, our use of a complete basis for the shape function in two dimensions, detailed in Sec.~\ref{sec:power}, makes it simple to convolve with the partonic cross section. We therefore conclude that the most convenient approach to resummation for DHM or HJM is to perform a small angle expansion of $\tilde s_{2f}$ to some desired accuracy permitting an analytic computation of resummation even when taking into account the shape function.

\subsection{Non-global Soft Function at Three Loops}\label{sec:NG3L}

While the hard and jet functions are fully known at ${\cal O}(\alpha_s^3)$, we do not know the two dimensional partonic soft function at this order. For the soft function at ${\cal O}(\alpha_s^3)$, we know the logarithmic terms predicted by RGE evolution (global logs), and the constant required for the complete ${\cal O}(\alpha_s^3)$ thrust projection, which was recently computed in Ref.~\cite{Baranowski:2024vxg,Baranowski:2024ysi}, cf.~\eq{ctwoeq}, and which for $n_f=5$ and $N_c=3$ takes the value
\begin{equation}\label{eq:s3}
s_3=-1030.2166\,.
\end{equation}
However, HJM mass requires the full dihemisphere three-loop function, and hence the unknown piece $\tilde s_{3f}$ is a function of $z=e^{L}=y_1/y_2$.
The only known properties for this function are its value at the particular value $z=1$ which defines $s_3$, the location of its cuts, its symmetry under the replacement $z\to 1/z$ and its exponentiation properties~\cite{Hoang:2008fs}.

However, even if the full functional form of $\tilde s_{3f}$ is unknown, we only need its projection onto the unit circle of the complex variable $z$. This projected function is real (otherwise the cross section would be complex) and analytic on the real axis. Moreover it again admits the polar expansion around $L=0$, which we expect to converge within the relevant range of $|L|<\pi$. With a few terms of this expansion we already have an excellent representation of the full projected function, and so the problem gets reduced to the determination of a few unknown constants. The question is of course how many terms should we retain and how to determine them.
Guided by our experience with the two-loop soft function $\tilde s_{2f}$,
we will estimate uncertainties by keeping the first three terms in the expansion at three-loops,
\begin{equation}
\frac{\tilde s_{3f}(L)}{2} = t_3(e^L) =s_3+s_{3,2} L^2 +s_{3,4} L^4 + \ldots \,,
\label{eq:HK-3loop}
\end{equation}
where the constant $s_3=s_{3,0}$ is the thrust soft function non-log piece at three loops, see Eq.~\eqref{eq:s3}. To determine $s_{3,2}$ and $s_{3,4}$ one could fit Eq.~(\ref{eq:HK-3loop}) to the {\sc eerad3} or {\sc CoLoRFulNNLO} results (for instance to asymmetric thrust and HJM). This is in practice not possible because the output of these numerical programs is too noisy in the dijet region to get a reliable extraction. What we will do instead is varying these constants as part of our estimate of perturbative uncertainties. A conservative choice would be to vary $s_{3,2}$ and $s_{3,4}$ as much as $s_3$ was varied in Ref.~\cite{Abbate:2010xh} before this constant was known, that is between $-\,500$ and $500$. Numerical studies reveal that these variations are too strong since they induce an unnaturally large error band in the peak region. In particular, it translates into a norm uncertainty three times larger than at NLL$^\prime$, indicating that this crude estimate overshoots the actual size of these constants. We have explored three less extreme possibilities: a)~tuning the $s_{3,2}$ and $s_{3,4}$ variation such that they have the same effect on the norm as varying $s_3$ in the range $[-500,500]$, b)~assuming that the asymptotic behavior sets in from the very first coefficient,
with the same exponential fall off observed at two loops $|s_{3,2n}|=|s_{3}|\times10^{-1.64\,n}$, c)~assuming a linear fall off analogous to that observed for the first three coefficients of the two loop function $s_{3,2n}\simeq s_{3}(1-0.5\,n)$. We find that the three choices give very similar results. For our numerical analyses we will use the truncation in \eq{HK-3loop} together with option c), which translates into the following variations:
$|s_{3,2}|\leq 250$ and $|s_{3,4}|\leq 7$. These constants mainly affect the peak of the distribution and have a moderate impact in the tail.

\section{Non-singular Distributions for heavy jet mass}
\label{sec:nonsing}
We include perturbative corrections beyond the leading dijet contributions discussed so far through the non-singular distribution. For this non-singular contribution, which is power-suppressed in the dijet limit, we employ fixed-order perturbation theory, i.e.\ we do not carry out any resummation of logarithms.

\subsection{Non-singular heavy jet mass}

The non-singular distribution for HJM $\df \hat{\sigma}_{\rm ns}/\df\rho$ is computed using
\begin{align} \label{eq:nssubt}
\frac{\df \hat{\sigma}_{\rm ns}}{\df \rho}(Q,\mu_{\rm ns}) \,=\,
\frac{\df \hat{\sigma}_{\text{full}}^{\text{FO}}}{\df \rho}(Q,\mu_{\rm ns}) \,-\,
\frac{\df \hat{\sigma}_{s}^{\text{FO}}}{\df \rho}(Q,\mu_{\rm ns})\,,
\end{align}
where $\mu_{\rm ns}$ is the renormalization scale of the strong coupling $\alpha_s(\mu_{\rm ns})$ in the fixed-order series. Even though the natural choice appears to be $\mu_{\rm ns}=Q$, in order to estimate missing resummation in next-to-leading power terms, we employ a $\rho$-dependent parametrization displayed in Eq.~\eqref{eq:muNSprofile}. The expression for the fixed-order singular distribution $\df \hat{\sigma}_{s}^{\text{FO}}/\df \rho$ on the RHS of Eq.~\eqref{eq:HJM-dist}, in which no resummation is carried out, is obtained from strictly expanding the factorized expression in Eq.~\eqref{eq:HJM-dist} in $\alpha_s(\mu_{\rm ns})$.
We can parametrize the non-singular distribution in the form
\begin{equation} \label{eq:nonsingular-expansion}
\frac{1}{\sigma_0} \frac{\df \hat{\sigma}_{\rm ns}}{\df \rho} = \frac{\alpha_s(Q)}{2\pi}
f_1(\rho)+\biggl[\frac{\alpha_s(Q)}{2\pi}\biggr]^{2} f_2(\rho)
+\biggl[\frac{\alpha_s(Q)}{2\pi}\biggr]^{3} f_3(\rho) +\,\ldots\,,
\end{equation}
and recover the expansion in powers of $\alpha_s(\mu_{\rm ns})$ from RG-invariance using Eq.~(3.5) of Ref.~\cite{Mateu:2017hlz}. The functions $f_i(\rho)$ are obtained by subtracting the fixed-order singular
distribution from the complete fixed-order distribution, as indicated in Eq.~\eqref{eq:nssubt}

At tree-level and $\mathcal{O}(\alpha_s)$ (where the thrust and HJM fixed-order distributions coincide), the HJM distribution is given by
\begin{align}\label{eq:OasNonSing}
\frac{\df \hat\sigma_{\rm ns}^{(0+1)}}{\df\rho} = \,&\delta(\rho)
+ \frac{\alpha_s(\mu)}{2\pi} \biggl\{\frac{4}{3 \rho }\biggl[9 \rho ^2+6 \rho+ 4\ln(\rho)+2 \biggl(3 \rho -\frac{2}{1-\rho }\biggr)\! \ln \biggl(\frac{\rho }{1-2 \rho }\biggr)
\\
&-4\biggr(\frac{1}{\rho}\biggl)_{\!\!+} -\frac{16}{3} \biggr(\frac{\ln(\rho)}{\rho}\biggl)_{\!\!+}\, \biggr]\theta(1-3\rho)
+ \delta(\rho) \biggl(\frac{4 \pi ^2}{9}-\frac{4}{3} \biggr) \biggr\}
\,. \nn
\end{align}
The singular expression is just the leading term of the asymptotic expansion for $\rho\to 0$.
For the cumulative up to $\mathcal{O}(\alpha_s)$ we have
\begin{align}
\label{eq:OasNonSingcumulat}
\Sigma_\rho^{(0+1)}(\rho_c)= \,& \int_0^{\rho_c}\df\rho \frac{\df \hat\sigma^{(0+1)}}{\df\rho}=\theta(\rho_c) + \frac{\alpha_s(\mu)}{2\pi}\Sigma_1^{\rm FO}(\rho_c) \,,
\end{align}
with
\begin{align} \label{eq:Sigma1}
\Sigma_1^{\rm FO}(\rho_c<1/3) =\,& \frac{2}{3} \biggl[16 \text{Li}_2(1-2 \rho_c )-4 \text{Li}_2\bigl[(1-2 \rho_c )^2\bigr]+8\text{Li}_2\!\biggl(1-\frac{1}{\rho_c }\biggr)
-2\\
&+3
\rho_c (3 \rho_c +4)+6 (1-2 \rho_c ) \ln \biggl(\frac{1}{\rho_c }-2\biggr)-8 \ln \biggl(\frac{1}{\rho_c }-1\biggr)\! \ln (1-2 \rho_c )\biggr]\nonumber\\
& \stackrel{\rho_c\sim 0}{=} \frac{4 \pi ^2}{9}-\frac{4}{3}-\frac{8}{3} \!\ln ^2(\rho_c )-4\! \ln (\rho_c )+\!\mathcal{O}(\rho_c)\!
\nn\\
& \stackrel{\rho_c\sim 1/3}{=} 2-48\biggr(\!\rho_c-\frac{1}{3}\biggl)^{\!\!2}+\mathcal{O}\Biggr[\!\biggr(\!\rho_c-\frac{1}{3}\biggl)^{\!\!4}\Biggl],\nn\\
\Sigma_1^{\rm FO}(\rho_c\ge 1/3) =\, & 2
\,,\nn
\end{align}
from which is straightforward to compute the cumulative non-singular $\Sigma_1(\rho_c)=\int_0^{\rho_c}\df \rho\,f_1(\rho)$.

The non-singular HJM distribution can be obtained numerically at $\ord{\alpha_s^2}$ from the Fortran program {\sc event2}~\cite{Catani:1996jh, Catani:1996vz}, and at $\ord{\alpha_s^3}$ from {\sc eerad3}~\cite{GehrmannDeRidder:2009dp,Gehrmann-DeRidder:2014hxk,Aveleira:2025svg}, {\sc Mercutio}~\cite{Weinzierl:2008iv,Weinzierl:2009ms}, or {\sc\mbox{CoLoRFulNNLO}}~\cite{DelDuca:2016ily}.
At $\ord{\alpha_s^3}$ we use the latter since it has significantly smaller statistical uncertainties than {\sc eerad3} or {\sc Mercutio}.

\begin{figure*}[t!]
\subfigure[]{\includegraphics[width=0.477\textwidth]{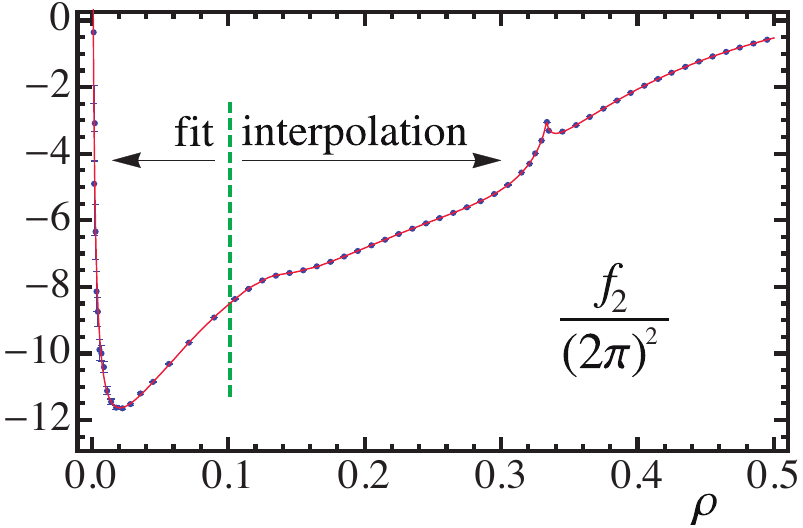}
\label{fig:NS-LO}}
\subfigure[]{\includegraphics[width=0.475\textwidth]{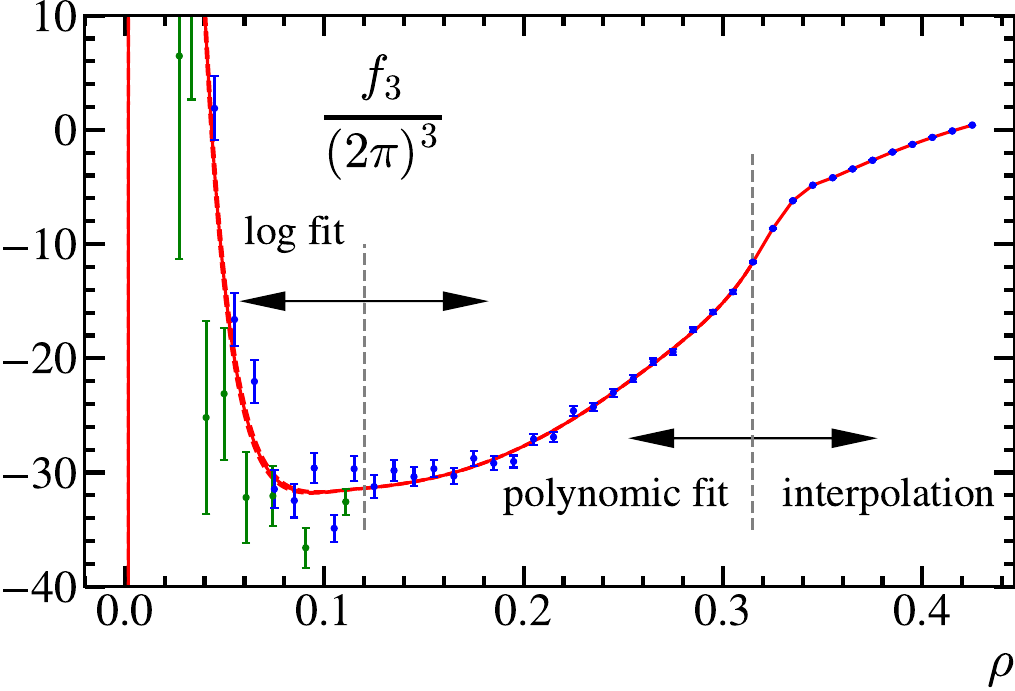}
\label{fig:NS-NLO}}
\vspace{-0.2cm}
\caption{Non-singular distributions for heavy jet mass at $\mathcal{O}(\alpha_s^2)$ (NLO) in panel (a) and $\mathcal{O}(\alpha_s^3)$ (NNLO) in panel (b). At NLO we use log binning and a fit function for $\rho < 0.1$ (left of green dashed line), and linear binning with an interpolation for $\rho>0.1$ (right of the green dashed line). While we show all {\sc event2} data used for the fit (in (a)), only a representative set of the points used in our interpolation function is displayed. The fit error function is too small to be visible in this plot. At NNLO (b) results from {\sc CoLoRFulNNLO} are shown. We use linear binning everywhere, but a fit function for $\rho<0.28$ (left of green dashed line), where errors are still visible, and an interpolation for $\rho>0.28$ (right of green dashed line), where the errors are almost negligible. The fit error function at the one-sigma level is shown with red dashed lines. The heavy jet mass non-singular distribution at $\mathcal{O}(\alpha_s)$ coincides with thrust, is known analytically, and can be seen in Fig.~3 of Ref.~\cite{Abbate:2010xh}.
\label{fig:NS-plots}}
\end{figure*}

At $\ord{\alpha_s^2}$ we use {\sc event2} results with log binning for $\rho<0.1$ and linear binning (with bin size $\Delta\rho = 0.001$) for $0.1<\rho<0.420204$. We used runs with a total of $3 \times 10^{11}$ events and an infrared cutoff $y_0=10^{-8}$. In the regions of linear binning, the statistical uncertainties are quite low and we can use a numerical interpolation for $f_2(\rho)$. This interpolation uses $153$ nodes which result from combining the finer binning described above. For $\rho<0.1$ we use the ansatz $f^{\rm fit}_2(\rho)=\sum_{i=0}^3 a_i \ln^i \rho + a_4\,\rho \ln^3\rho$ and fit the coefficients from {\sc event2} output, including the constraint that the total fixed-order cross section gives the known $\ord{\alpha_s^2}$ coefficient. The final result for the two-loop non-singular cross section for $\rho<0.1$ then has the form
\begin{equation}
f_2(\rho<0.1) = f^{\rm fit}_2(\rho)\,+\,\epsilon_2\, \delta f_2^{\rm fit}(\rho) \,.
\end{equation}
Here $f^{\rm fit}_2(\rho)$ gives the best-fit result and $\delta f_2^{\rm fit}(\rho)$ gives the $1$-$\sigma$ error function from the fit. The variable $\epsilon_2$ is varied between $-1$ and $1$ during our theory scans in order to account for the error in the non-singular function. In Fig.~\ref{fig:NS-LO} the {\sc event2} data is shown as dots and the best-fit non-singular function as a solid blue line. The statistical uncertainties are invisible on the scale of this plot.

We follow a very similar procedure to determine the $\ord{\alpha_s^3}$ non-singular cross section $f_3(\rho)$. The results are clustered in linear bins of size $0.005$. The {\sc CoLoRFulNNLO}
numerical results have larger uncertainties than the {\sc event2} two-loop ones. Therefore we employ a fit function for $0 <\rho \leq \rho_0=0.12$ with the functional form
\begin{equation}
f_3^{\rm log}(\rho) = \sum_{i=0}^5 \ell_i \ln^i\biggl(\frac{\rho}{\rho_0}\biggr)+\ell_6\, \rho\ln^5\biggl(\frac{\rho}{\rho_0}\biggr)\,,
\end{equation}
smoothly joined at $\rho=\rho_0$ with another polynomic fit function $f_3^{\rm pol}(\rho) = \sum_{i=0}^{15} p_i (\rho- \rho_0)^{i}$ used in the range $\rho_0 \leq \rho \leq 0.315$,
and use an interpolation from that point onwards.
The coefficients of both fit functions depend on the
result for the three-loop partonic dihemisphere soft function, which is not completely known. Using the polar expansion up to terms of order $L^4$, the linear combination contributing to the coefficient multiplying $\delta(\rho)$ reads: $s_3 - 3.2899\,s_{32} +19.4818\,s_{34}$ with $s_3$ given in Eq.~\eqref{eq:s3}. Due to the sizable uncertainties in the {\sc CoLoRFulNNLO}
results, it is not possible to fit for this combination, so $s_{32}$ and $s_{34}$
are left as variables that are varied in our theory scans in an uncorrelated way. All in all, the final result for the three-loop non-singular cross section function can once again be written in the form
\begin{equation}
f_3(\rho<0.27) = f_3^{\rm fit}(\rho)+\epsilon_3\, \delta f_3^{\rm fit}(\rho) \,,
\end{equation}
where $f_3^{\rm fit}(\rho)=\theta(\rho_0-\rho)f_4^{\rm log}(\rho) + \theta(\rho - \rho_0) f_3^{\rm pol}(\rho)$ is the best-fit function and $\delta f_3^{\rm fit}(\rho)$ gives the $1$-$\sigma$ error function for the fit, rescaled such that the $\chi^2_{\rm min}/{\rm d.o.f.}$ is exactly one. This is a conservative approach and we find it justified because the {\sc CoLoRFulNNLO} bins for small $\rho$ might suffer from cutoff effects. Exactly as we did for the $\ord{\alpha_s^2}$ distribution, $\epsilon_3$ is continuously varied in the final error analysis. In Fig.~\ref{fig:NS-NLO}, we plot the {\sc CoLoRFulNNLO}
data as dots, the best-fit function $f_3$ as a solid line, and the non-singular results with $\epsilon_3=\pm\,1$ as dashed lines almost on top of the central curve.

We conclude this section noting that
the non-singular distribution described here is valid in the dijet region. However, for $\rho \gtrsim 0.2$ the distribution approaches the trijet region where the cross section must be matched to a resummed treatment to account for the large Sudakov shoulder logarithms. A matching procedure valid for the dijet, tail and shoulder region was described recently in Ref.~\cite{Bhattacharya:2023qet}.

\subsection{Non-singular dihemisphere jet mass}

The non-singular DHM distribution is computed using
\begin{align} \label{eq:nssubtdhm}
\frac{\df \hat{\sigma}_{\rm ns}}{\df s_1\df s_2}(Q,\mu_{\rm ns}) \,=\,
\frac{\df \hat{\sigma}_{\text{full}}^{\text{FO}}}{\df s_1\df s_2}(Q,\mu_{\rm ns}) \,-\,
\frac{\df \hat{\sigma}_{s}^{\text{FO}}}{\df s_1\df s_2}(Q,\mu_{\rm ns})\,.
\end{align}
As we can see from \eq{HJM-hadronic} we also actually need the non-singular DHM distribution to determine the hadron-level non-singular HJM distribution. This is because we can determine which of the two hemispheres is heavier at the observable level only after hadronization effects are included.

Up to $\mathcal{O}(\alpha_s)$ it is trivial to write down the double differential hemisphere mass distribution, since there are at most three (massless) partons, one of them in the light hemisphere and the others in the heavy one. The former contributes as a Dirac delta function, yielding:
\begin{equation}
\label{eq:oneloopnsdouble}
\frac{\df^2 \hat\sigma^{(0+1)}}{\df s_1\df s_2} = \frac{1}{2Q^2}
\Biggl[\delta(s_1)\, \frac{\df \hat\sigma^{(0+1)}}{\df\rho}\bigg|_{\rho=s_2/Q^2}+
\delta(s_2)\, \frac{\df \hat\sigma^{(0+1)}}{\df\rho}\bigg|_{\rho=s_1/Q^2}\Biggr]\,,
\end{equation}
where (single differential) HJM distribution appears explicitly.

Starting at $\mathcal{O}(\alpha_s^2)$, the non-singular dihemisphere mass differential cross sections develops a richer structure
and becomes a non-trivial function of $s_1$ and $s_2$. On the one hand, one can have events with a single parton in one of the hemispheres along with two or more partons in the other, contributing to terms of the form $F_{\rm ns}(s_1,s_2)=[\,\delta(s_{1})f_{\rm ns}(s_{2})+\delta(s_{2})f_{\rm ns}(s_{1})\,]/2$ with $f_{\rm ns}(s)$ non-singular as $s\to 0$, which makes $F_{\rm ns}$ single-singular. (Such partonic configurations also contribute to the double-singular structure $F_{\rm s}(s_1,s_2)=[\,\delta(s_{1})f_{\rm s}(s_{2})+\delta(s_{2})f_{\rm s}(s_{1})\,]/2$ with $f_{\rm s}(s)$ singular for $s\to 0$, which is fully accounted for by the singular distribution.) On the other hand, events with more than one parton in each hemisphere contribute to terms $G_{\rm ns}(s_1,s_2)=G_{\rm ns}(s_2,s_1)$ without Dirac delta functions, which contain single-singular and non-singular contributions:
\begin{equation}\label{eq:GDJM}
G_{\rm ns}(s_1,s_2) = \frac{1}{2}\sum_{i=0}c_i[\,\mathcal{L}^i(s_1) h^i_{\rm ns}(s_2) + \mathcal{L}^i(s_2) h^i_{\rm ns}(s_1)\,] + g_{\rm ns}(s_1,s_2)\,,
\end{equation}
where $\mathcal{L}^i(s)=[\,Q^2\ln^i(s/Q^2)/s\,]_+/Q^2$ for $i\geq 0$ represent the single-singular contribution, and the function $g_{\rm ns}(s_1,s_2)=g_{\rm ns}(s_2,s_1)$ is non-singular in both $s_1$ and $s_2$.
While $G_{\rm ns}(s_1,s_2)$ is in principle accessible
from Monte-Carlo programs using two-dimensional binning,
it is much harder to obtain the $f_{\rm ns}(s)$ terms. Furthermore, $F_{\rm ns}(s_1,s_2)$ does not contribute to the differential non-singular DHM partonic cross section for $s_{1,2}>0$.

While $F_{\rm ns}(s_1,s_2)$ yields exponentially suppressed contributions in the DHM tail region when convolved with the two-dimensional shape function, $G_{\rm ns}(s_1,s_2)$ renders contributions that are not.
On the other hand, the contribution of $F_{\rm ns}$ to the doubly cumulative DHM distribution setting both $s_1^c=s_2^c=Q^2\rho$ (which is the partonic non-singular cumulative heavy jet mass distribution) gives \mbox{$f^\Sigma_{\rm ns}(Q^2 \rho)=\int_0^{Q^2\rho} \df s f_{\rm ns}(s)$}, implying for the corresponding differential distribution the term $Q^2 f_{\rm ns}(Q^2\rho)$.
Due to the numerical complexity, unfortunately, we were not able to reliably isolate these different contributions in the double differential non-singular distribution at $\mathcal{O}(\alpha_s^2)$ and $\mathcal{O}(\alpha_s^3)$.

However, as we show in the next section, for predictions of the HJM tail region a description of the dominant ${\cal O}(\Lambda_{\rm QCD})$ non-singular power correction is possible, since a cross section in the form of Eq.~(\ref{eq:oneloopnsdouble}) is fully compatible with the OPE based on the shape function also at ${\cal O}(\alpha_s^{2})$ and beyond.
Non-perturbative corrections to the non-singular distribution that arise beyond this approximation, such as those from 3-jet configurations, or alternate forms for the non-singular power corrections, are explored elsewhere~\cite{Benitez:2024nav}.

\section{Power Corrections for DHM and HJM} \label{sec:power}
In this section we will discuss hadronization in heavy jet mass from the point of view of SCET. We will first derive an OPE formula that works in the tail of the distribution, which allows for two-dimensional fits to data if the fit-region is limited to the part of the spectrum
for which the dijet approximation applies. Finally we will explain for DHM and HJM how to switch to a scheme in which the leading renormalon of the soft function is removed.

\subsection{Operator Product Expansion}\label{sec:OPE}
We start this section by reviewing the derivation of the OPE for the simpler case of the sum of jet
masses, referred to as the SHM.
We start from Eq.~\eqref{eq:thrust-hadronic}. The OPE relies on
the fact that the shape function effectively has only support for values of $k\sim \lqcd \ll Q\,\taus$, and is exponentially suppressed for $k\gg \lqcd$. Therefore
if $Q\,\taus\gg \lqcd $ one can expand in powers of $k$:
\begin{align}\label{eq:thrust-OPE}
\frac{\df \sigma}{\df \taus} \,=& \int_0^{Q\taus} \df k\,\, \frac{\df\hat \sigma}{\df \taus}
\biggl(\!\taus - \frac{k}{Q}\biggr) F_\tau(k)
=\sum_{i=0}^\infty\frac{(-1)^i}{Q^i}\frac{\df ^i}{\df \tau_s^i}
\biggl[\frac{\df \hat{\sigma}}{\df \taus}(\taus)\biggr]\int_0^{\infty} \df k\,k^i\, F_\tau(k)\\\nonumber
\,=& \sum_{i=0}^\infty(-1)^i\frac{\df^i}{\df \tau_s^i}
\biggl[\frac{\df \hat{\sigma}}{\df \taus}(\taus)\biggr]\frac{\Omega^\tau_i}{Q^i}\,.
\end{align}
In the second equality we extended the integration range to infinity due to the exponential suppression for the shape function $F_\tau$ for $k\gg \lqcd$. The $\Omega^\tau_i$ moments are directly related to the DHM shape function moments defined in Eq.~(\ref{eq:Omega-ij}):
\begin{equation}\label{eq:OmegaThrust}
\Omega^\tau_i=\sum_{j=0}^i\binom{i}{j}\Omega_{j-i,i}\,,\qquad
\Omega_{0,0}=1\,,\qquad \binom{i}{j} = \frac{i!}{j!(i-j)!}\,.
\end{equation}
In particular one has $\Omega_1^\tau=\Omega_{1,0}+\Omega_{0,1}=2\Omega_{1,0}$. The expression for the OPE in Eq.~\eqref{eq:thrust-OPE} can be applied to the cumulative $\taus$ distribution simply by substituting $\df\hat\sigma/\df\taus\to\widehat\Sigma^{\taus}(\taus)$.

In order to ease
comparison to the HJM OPE, it is instructive (although more complicated) to rederive Eq.~\eqref{eq:thrust-OPE}
using the DHM in Eq.~\eqref{eq:double-tot} and the projection onto the SHM given in Eq.~\eqref{eq:cumulative-projection}. For the DHM distribution the OPE is
\begin{align}\label{eq:s1s2-OPE}
\frac{\df^2\sigma}{\df s_1\df s_2} \,=&\! \int \!\df k_1\,\df k_2\,
\frac{\df^2\hat\sigma}{\df s_1\df s_2}(s_1-Qk_1,s_2-Qk_2)\,F(k_1,k_2)\\\nonumber
\,=& \sum_{i=0}^\infty\frac{(-Q)^i}{i!}\sum_{j=0}^i\binom{i}{j}
\frac{\partial^i }{\partial^{i-j}s_1\partial^j s_2}\biggl[\frac{\df^2\hat\sigma}{\df s_1\df s_2}\biggr]
\Omega_{j,i-j}\,.
\end{align}
An analogous OPE can be found for $\Sigma(s_1,s_2)$ by simply substituting
$\df^2\hat\sigma/(\df s_1\df s_2)$ by $\widehat\Sigma(s_1,s_2)$ everywhere. We will also need to express the $i$-th
derivative of the partonic SHM distribution in terms of the double differential hemisphere one.
\begin{equation}\label{eq:der-Sigma-tau}
\frac{\df^i}{\df \tau_s^i}\biggl[\frac{\df \hat{\sigma}}{\df \taus}(\taus)\biggr] =
Q^{2(i+1)}\!\!\int_0^{Q^2\taus}\!\!\df s_1\frac{\partial^i}{\partial s_2^i}
\biggl[\frac{\df^2\hat\sigma}{\df s_1\df s_2}\biggr]_{s_2\,=\,Q^2\taus\,-\,s_1}\,.
\end{equation}
To project Eq.~\eqref{eq:s1s2-OPE} onto the sum of hemisphere masses it is useful to
note that an integral identity holds since $f(s_1,s_2)$ vanishes for $s_1=0$ or $s_2=0$:
\begin{equation}\label{eq:IBP}
0 =\! \int_0^{Q^2\taus}\!\df s_1\, \frac{\df }{\df s_1}f(s_1,Q^2\taus-s_1) = \! \int_0^{Q^2\taus}\!\!\df s_1
\biggl[ \biggl(\frac{\partial }{\partial s_1}-\frac{\partial }{\partial s_2}\biggr)f(s_1,s_2)\biggr]_{s_2\,=\,Q^2\taus\,-\,s_1}\,,
\end{equation}
where $f(s_1,s_2)$ has support for positive $s_i$. Therefore one can replace $\partial/ \partial s_1\to
\partial/ \partial s_2$ under the SHM projection operation. With that in hand we find
\begin{align}\label{eq:thrust-OPE-2}
\frac{\df \sigma}{\df \taus} \,=&\sum_{i=0}^\infty\frac{(-Q)^{i+2}}{i!}\sum_{j=0}^i\binom{i}{j}\Omega_{j,i-j}\!\!
\int_0^{Q^2\taus}\!\df s_1
\frac{\partial^i }{\partial^{i-j}s_1\partial^j s_2}\biggl[\frac{\df^2\hat\sigma}{\df s_1\df s_2}\biggr]_{s_2\,=\,Q^2\taus\,-\,s_1}\\
\,=&\sum_{i=0}^\infty\frac{(-1)^i}{i!}\frac{\df^i}{\df \tau_s^i}\biggl[\frac{\df \hat{\sigma}}{\df \taus}(\taus)\biggr]\!
\sum_{j=0}^i\binom{i}{j}\frac{\Omega_{j,i-j}}{Q^i}\,,\nonumber
\end{align}
which reduces to Eq.~\eqref{eq:thrust-OPE} when the relation~\eqref{eq:OmegaThrust} is used. To get to the second line of \eq{thrust-OPE-2} we have used the identity in Eq.~\eqref{eq:IBP} to turn all derivatives into $\partial/\partial s_2$, which allows to pull the partial derivatives out of the sum in $j$ and hence to factor the SHM power corrections. Unfortunately, a similar manipulation is not possible for heavy jet mass.

The OPE for the HJM distribution is easiest derived for the cumulative distribution. Let us start again by expressing the $i$-th derivative of $\widehat\Sigma^\rho(\rho)$ in terms of derivatives of $\widehat\Sigma(s_1,s_2)$, simply applying the chain rule:
\begin{equation}\label{eq:der-Sigma-rho}
\frac{\df^i}{\df \rho^i}\widehat\Sigma^\rho(\rho) = Q^{2i} \sum_{j=0}^{i}\binom{i}{j}
\frac{\partial^i \widehat\Sigma(s_1,s_2)}{\partial^{i-j}s_1\partial^j s_2}\Bigg|_{s_1=s_2=Q^2\rho}
\!\!\equiv \sum_{j=0}^{i} \widehat\Sigma^{j,i-j}(Q,\rho)\,,
\end{equation}
where the $\widehat\Sigma^{i,j}(Q,\rho)$ functions are dimensionless and depend only on $Q$ and $\rho$. Unlike
for the sum of hemisphere masses, they depend on both $i$ and $j$ since there is no identity that holds
under the HJM projection integral in \eq{HJM-hadronic},
given that the integrand does not vanish in the upper integration limit.
To compute the $i$-th derivative of the differential heavy jet mass distribution one just replaces
$i\to i\,+\,1$ in Eq.~\eqref{eq:der-Sigma-rho}. Next we project the cumulative version of
Eq.~\eqref{eq:s1s2-OPE} into HJM under the assumption $Q\rho\gg \lqcd$, finding
\begin{equation}\label{eq:HJM-OPE}
\Sigma^\rho(\rho) = \sum_{i=0}^\infty\frac{(-1)^i}{i!}\sum_{j=0}^i\widehat\Sigma^{j,i-j}(Q,\rho)\frac{\Omega_{j,i-j}}{Q^i}\,.
\end{equation}
The OPE for the differential distribution can be obtained trivially substituting
$\sum_{j=0}^i\widehat\Sigma^{j,i-j}$ by $\sum_{j=0}^{i+1}\widehat\Sigma^{j,i+1-j}$.
Note, if
$\Omega_{j,i-j}$ only depended on the index $i$ (that is, if $\Omega_{i,j}$ only depended on $i+j$), then one could trivially do the $j$ sum using Eq.~(\ref{eq:der-Sigma-rho}) to arrive at an OPE with a form analogous to the SHM distribution in Eq.~\eqref{eq:thrust-OPE}.
Unfortunately, there is no physical reason why this identity should hold for the $\Omega_{i,j}$ moments
defined in Eq.~(\ref{eq:Omega-ij}), which implies that the OPE for HJM is in general more complicated.
However, when accounting only for the leading ${\cal O}(\Lambda_{\rm QCD})$ power correction ($i = 1$), the situation remains simple and we indeed obtain an expression analogous to that of SHM:
\begin{equation}\label{eq:OPE-truncated}
\Sigma^\rho(\rho) \,= \widehat\Sigma^\rho(\rho) -
\frac{\Omega_{1,0}}{Q}\frac{\df\widehat\Sigma^\rho(\rho)}{\df \rho}+\ldots\,,
\end{equation}
with a corresponding expression for the differential HJM distribution.
Fortunately, truncating the OPE after the first linear power correction is sufficient in the tail region so that the HJM distribution can be accurately described by $\alpha_s$ and the non-perturbative ${\cal O}(\Lambda_{\rm QCD})$ matrix element $\Omega_{1,0}$.

Experimental measurements convert energies and angles to four vectors of particles and thus to observables. Since these measurements are done on massive hadrons, the exact assumptions about the mass of the particles effects what observable is measured.
If such hadron-mass effects are ignored (which is not correct),
then the leading power corrections for thrust $\tau$ and HJM are related by $\Omega_{1,0}\equiv \Omega_1^\rho = \Omega_1^\tau/2$. However,
$\Omega_1^\rho$ and $\Omega_1^\tau$ are not related since they are in different universality classes, as was shown in Ref.~\cite{Mateu:2012nk}, so the factor of $1/2$ in this relation has an ${\cal O}(1)$ uncertainty. In practice, Monte Carlo simulations predict that
non-zero hadron masses bring $\Omega_1^\rho$ and $\Omega_1^\tau$ much closer.
The $\Omega_1^\rho$ and $\Omega_1^\tau$ correspond to different weighted integrals over an underlying non-perturbative function $\Omega_1(r)$ that fully describes hadron-mass effects. $\Omega_1^\rho$ and $\Omega_1^\tau$ are the leading terms in a basis parametrizing all possible hadron mass effects, and are therefore useful for describing the parameters in other dijet event shape universality classes.
On the other hand, the sum of hemisphere masses $\tau_s$ and HJM belong to the same universality class and therefore the relation between their leading power corrections remains $2\Omega_1^\rho = \Omega_1^\taus$ even once hadron-mass effects are properly accounted for.

To have a universal formula which is applicable to peak, tail and multi-tail, a valid approach is to convolve the partonic doubly differential DJM cross section with the two-dimensional shape function as shown in Eq.~\eqref{eq:HJM-hadronic}. This can be done exactly for the singular terms up to N$^3$LL$^\prime$ and the non-singular up to $\mathcal{O}(\alpha_s)$, as the corresponding DJM partonic contributions are known exactly. However, as already stated in Sec.~\ref{sec:nonsing}, at order $\mathcal{O}(\alpha_s^2)$ and beyond we do not have the exact double differential non-singular distribution. Interestingly, due to Eq.~(\ref{eq:OPE-truncated}), in the tail region where the inclusion of the ${\cal O}(\Lambda_{\rm QCD})$ power correction $\Omega_{1,0}$ is sufficient, we can use the following simplified ansatz for the double differential non-singular distribution,
\begin{equation}
\label{eq:NSansatz}
\frac{\df^2 \hat\sigma_{\rm ns}}{\df s_1\df s_2} \simeq \frac{1}{2Q^2}
\Biggl[\delta(s_1)\, \frac{\df \hat\sigma_{\rm ns}}{\df\rho}\bigg|_{\rho=s_2/Q^2}+
\delta(s_2)\, \frac{\df \hat\sigma_{\rm ns}}{\df\rho}\bigg|_{\rho=s_1/Q^2}\Biggr]\,.
\end{equation}
This approach is exact at tree-level and ${\cal O}(\alpha_s)$, but also correctly reproduces the leading order OPE in the tail region. Therefore, it yields uncertainties of $\mathcal{O}[\alpha_s^2 \Lambda_{\rm QCD}^2/(Q^2\rho^2)]$ which can be expected to be negligible for tail-region applications.
Using Eq.(\ref{eq:NSansatz}), the hadronization of the non-singular HJM distribution takes the form of a one-dimensional convolution:
\begin{align}\label{eq:ns-hadronic}
\frac{\df\sigma_{\rm ns}}{\df \rho} &= \int_0^{Q\rho}\!\df k\,
\frac{\df\hat\sigma_{\rm ns}}{\df \rho}\biggl(\!\rho-\frac{k}{Q}\biggr)
\Bigl[F^\Xi(Q\rho,k) + F^\Xi(k,Q\rho)\Bigr]\,,
\end{align}
where $F^\Xi(s_1^c,s_2)$ is defined in Eq.~\eqref{eq:Shape-cumulative}. For $Q\rho \gg \lqcd$ the term $F^\Xi(k,Q\rho)$ is exponentially suppressed and can be dropped. Taking the $Q\rho\to \infty$ limit in the first $F^\Xi(Q\rho,k)$ term, and expanding the partonic HJM distribution in small $k$ to linear order yields the leading $\Omega_{1,0}$ power correction as shown in Eq.~\eqref{eq:OPE-truncated}.
For the cumulative, analogous manipulations yield
\begin{align}
\Sigma^{\rm ns}_\rho(\rho)&= \int_0^{Q\rho}\!\df k\,
\frac{\df\hat\sigma_{\rm ns}}{\df \rho}\biggl(\!\rho-\frac{k}{Q}\biggr)
F^\Sigma(Q\rho,k)\,,
\end{align}
which also reproduces the leading power OPE term since in the tail $F^\Sigma(Q\rho,k)\simeq F^\Sigma(\infty,k)$.

\subsection{Power Corrections for Moments of the HJM Distribution}\label{sec:OPEmom}
In Sec.~\ref{sec:OPE} we derived an OPE for the tail of the HJM distribution in terms of the
non-perturbative shape function moments $\Omega_{i,j}$ defined in Eq.~(\ref{eq:Omega-ij}).
It turns out that moments of the HJM distribution
involve additional non-perturbative moments, not present for the tail distribution.
This occurs because in the tail of the HJM the hadronization corrections do not change the decision about which hemisphere is heavier, whereas in the peak of the distribution they can directly modify the outcome.\footnote{We discuss this interesting issue further at the end of Sec.~\ref{sec:shape}.}

To quantify this unusual feature, which to the best of our knowledge is not present in other familiar event shapes, let us first review the case of SHM (or thrust), where the OPE for the moments involves the same shape function moments as the OPE for distribution given in Eq.~(\ref{eq:thrust-OPE}).
From \eq{thrust-hadronic} we can derive the following OPE expansion for the moments:
\begin{align}\label{eq:tmomOPE}
M_n^\tau = & \!\int_0^{\tau_{\rm max}}\!\!\df\tau\,\tau^n\,\frac{\df\sigma}{\df\tau} =
\!\int_0^{\tau_{\rm max}}\!\!\df\tau\,\tau^n\!\!\int_0^{Q\tau}\!\!\df k\,
\frac{\df\hat\sigma}{\df\tau}\biggl(\!\tau-\frac{k}{Q}\biggr)F_\tau(k)\\\nn
=&\int_0^{Q\,\tau_{\rm max}}\!\!\df k\,F_\tau(k)\!\int_{k/Q}^{\tau_{\rm max}}\!\!\df\tau\,\tau^n
\frac{\df\hat\sigma}{\df\tau}\biggl(\!\tau-\frac{k}{Q}\biggr) \\\nn
=&
\int_0^{Q\,\tau_{\rm max}}\!\!\df k\,F_\tau(k)\!\int_0^{\tau_{\rm max}-k/Q}\!\!\df\tau
\biggl(\!\tau+\frac{k}{Q}\biggr)^{\!\!n}
\frac{\df\hat\sigma}{\df\tau}(\tau)\\\nn
\simeq&\sum_{i=0}^n\binom{n}{i}\!\!\int\!\df k\,F_\tau(k)\,\biggl(\frac{k}{Q}\biggr)^{\!\!n-i}\!\!\!
\int_0^{\tau_{\rm max}}\!\!\df\tau\,\tau^i\,\frac{\df\hat\sigma}{\df\tau} =
\sum_{i=0}^n\binom{n}{i}\frac{\Omega_{n-i}^\tau}{Q}\widehat M_i^\tau\,,
\end{align}
with ${\widehat M}_i^\tau$ being the $i$-th moment of the partonic thrust distribution. Here $\tau_{\rm max} = 0.5$
is the endpoint of the thrust distribution. To get the second line we have reversed the order of
integration, next we have redefined the inner integration variable $\tau\to\tau + k/Q$. In the last line
we have used that $k\gg\lqcd$ is exponentially suppressed by the shape function to set the inner integration
limit to $\tau_{\rm max}$, and that $Q\tau_{\rm max}\gg \lqcd$ to extend the outer integration limit
to infinity. The shape function moments $\Omega_{k}^\tau$ are the same that also appear in Eq.~(\ref{eq:thrust-OPE}).

For HJM the situation is not as simple, and perturbative and non-perturbative effects do not factor
in the same simple way when taking moments. Starting from \eq{HJM-hadronic} we get:
\begin{align}\label{eq:HJMmomOPE}
M_n^\rho = &\, 2Q^2\!\!\int_0^{\rho_{\rm max}}\!\!\df\rho\,\rho^n\,\frac{\df\sigma}{\df\rho}=
2\!\int_0^{\rho_{\rm max}}\!\!\df\rho\,\rho^n\!\!\int_0^{Q\rho}\!\!\df k_1\!\int_0^{Q\rho}\!\!\df k_2\,
\widehat\Xi(Q^2\rho-Qk_1,Q^2\rho-Qk_2)\,F(k_1,k_2)\nonumber\\\nonumber
= &\, 2Q^{1-n}\!\!\!\int_0^{Q\rho_{\rm max}}\!\df k_1\!\int_0^{Q\rho_{\rm max} - k_1}\!\!\df\ell\,(\ell+k_1)^n\!\!
\int_0^{Q\rho\,+\,k_1}\!\!\df k_2\,\widehat\Xi[\,Q\ell,Q(\ell+k_1-k_2)\,]F(k_1,k_2)\\\nonumber
= &\,2Q^{1-n}\!\!\! \int_0^{Q\rho_{\rm max}}\!\df k_1
\Biggl[\int_0^{k_1}\df k_2\!\int_0^{Q\rho_{\rm max} - k_1}\!\!\df\ell\,(\ell+k_1)^n
\,\widehat\Xi[\,Q\ell,Q(\ell+k_1-k_2)\,]\\\nn
&+\int_{k_1}^{Q\,\rho_{\rm max}-k_1}\!\!\df k_2\!
\int_0^{Q\rho_{\rm max} - k_2}\!\!\df \ell\,(\ell+k_2)^n\,
\widehat\Xi[\,Q(\ell+k_2-k_1),Q\ell\,]\Biggr]F(k_1,k_2)\\\nn
\simeq&\,2Q^{1-n}\sum_{i=0}^n\binom{n}{i}\!\! \int_0^\infty \!\!\df k_1 \!\!
\int_0^\infty \!\!\!\df k_2\,\theta(k_1-k_2)\,k_1^{n-i}\!\!\!
\int_0^{Q\rho_{\rm max}}\!\!\df\ell\:\ell^i\,
\Bigl(\widehat\Xi[\,Q\ell,Q(\ell+k_1-k_2)\,] \\
&+ \widehat\Xi[\,Q(\ell+k_1-k_2),Q\ell\,]\Bigr)F(k_1,k_2)\,,
\end{align}
with $\rho_{\rm max} \approx 0.477$ being the endpoint of the perturbative HJM distribution, see \app{maxrho}.
The endpoint again satisfies $Q\rho_{\rm max}\gg \lqcd$.
Thus we see that there is no OPE for the HJM moments analogous to \eq{tmomOPE}.

The steps in the derivation of \eq{HJMmomOPE} are analogous to those in \eq{tmomOPE}. To get to the second line we have
rescaled $\rho\to \ell/Q$ and switched the order of integration between $k_1$ and $\ell$, followed by a shift
$\ell\to\ell+k_1$. In the next step, when switching the order of integration between $\ell$ and $k_2$, one is
left with two integration regions, one for $k_1>k_2$ and another one for $k_2>k_1$, the former corresponding
to a rectangle and the latter to a triangle on the $(k_2-\ell)$ plane. This triangle can be mapped into a
rectangle shifting $\ell\to\ell+k_1-k_2$. To get the last equality, the second region can be mapped into
the first with the following three operations: renaming $k_1\leftrightarrow k_2$, switching the integration
order between $k_1$ and $k_2$ and using $F(k_1,k_2) = F(k_2,k_1)$. We extend the $k_i$ integrations to the
range $(0,\infty)$ inserting a Heaviside theta function $\theta(k_2-k_1)$ and using once more that the shape function
exponentially suppresses large values of $k_i$. Finally, we use this same argument to set the upper integration
limit of the $\ell$ integration to $Q\rho_{\rm max}$. Therefore, all approximations take place only in the
last equality and do not imply a separation of short- and long-distance physics with a simple product form. However, one can still
draw some general conclusions by having a closer look at the result.
To see that the non-perturbative dependence in \eq{HJMmomOPE} differs from the non-perturbative parameters appearing in the thrust case in \eq{tmomOPE}, we can consider what would happen if an expansion in small $k_1\sim k_2\ll \ell$ could be performed.\footnote{The reason that this expansion is not fully justified is because the region of small $\ell\sim k_1\sim k_2$ is equally important.}
Upon Taylor expanding the $\hat\Xi$ functions, we see that the dependence on the non-perturbative
momenta $k_i$ is through the combination $\theta(k_1-k_2)(k_1-k_2)^j k_1^iF(k_1,k_2)$ only.
Under this assumption
the hadronization effects will come in the form of shape function moments
$\Omega_{i,j}$ and $\Upsilon_{i,j}$.

As an illustration for how this relates to the physical picture, let us consider as a simple case a cross section with only non-perturbative radiation, where $\frac{\df^2 \hat\sigma}{\df s_1 \df s_2} = \delta(s_1)\delta(s_2)$, to show how the non-perturbative parameters appear in the moments. Then, still without resummation, $\widehat\Xi(s_1,s_2) = \theta(s_1)\delta(s_2)$, and taking $n>0$ we have
\begin{equation}
\big( M_n^{\rho} \big)^{{\alpha_s^0} + \rm nonpert}
=\frac{2\widehat\Upsilon_n}{Q^n}
\,,\qquad
\big( M_n^{\tau} \big)^{{\alpha_s^0} + \rm nonpert}
=\frac{\Omega^\tau_n}{Q^n}\,.
\end{equation}
Here for $i>0$ we defined
\begin{align}\label{eq:barUpsilon}
\widehat\Upsilon_{i} \,=\!\int_0^\infty\!\df k_1\,\df k_2\,
\theta( k_1- k_2)\, k_1^i\, F( k_1, k_2) = \frac{1}{2}\Biggl[
\Omega_{i,0} + \sum_{j=0}^{i-1}\Upsilon_{j,i-1-j}\Biggr]\,.
\end{align}
The lowest order power correction occurs for $i=1$, such that $\widehat\Upsilon_1$ provides an important non-perturbative parameter for HJM in the peak region, which differs from the $\Omega_1$ which appears for thrust. Interestingly, while $\Upsilon_{0,0}$ is renormalon-free, $\widehat\Upsilon_{1}$ is not and inherits a renormalon through its dependence on $\Omega_{1,0}$. Since the shape function is positive definite the values of the $\widehat\Upsilon_{i}$ power corrections are bounded
from above and below:
\begin{align}
\frac12 \Omega_{i,0} < \widehat\Upsilon_i < \Omega_{i,0} \,,
\end{align}
where the lower bound can be seen from the $\Omega_{i,0}$ term on the RHS of \eq{barUpsilon}, and the upper one by setting the theta-function in \eq{barUpsilon} to unity.
While for thrust the non-perturbative parameters in the moments and tail OPE are the same, for HJM they are different even under this simple approximation. With this approximation, hadronization alone is responsible for determining which mass is heavier for $M_n^\rho$ and thus depends on $\widehat\Upsilon_n$, whereas for thrust it is the sum of the hemisphere masses that is relevant, leading to $\Omega_n^\tau$. This disfavors hadronization models, such as those in Refs.~\cite{Webber:1994cp,Akhoury:1995sp}, in which a single non-perturbative parameter describes universally all $1/Q$ corrections.

An exception to this is the $0$-th moment, which corresponds to the total hadronic cross section. If one integrates over all possible $\rho$ values the result should be accurately described in perturbation theory alone. This might not appear obvious from \eq{HJMmomOPE}, but it is easiest seen from \eq{HJMCumF}
setting $\rho\to\rho_{\rm max}$ and approximating $Q\rho_{\rm max} - k_i\to Q\rho_{\rm max}$. Using that the shape function is normalized, one finds that the $0$'th moment is the total cross section without any dependence on the shape function.

\subsection{Power Corrections in the R-gap scheme}\label{sec:Rgap}

The hadron-level singular distribution for the double hemisphere mass can be obtained from Eq.~\eqref{eq:hemidist}
substituting the partonic soft function $\widehat S$ by the full soft function $S$, which is the
convolution of $\widehat S$ and the shape function $F$ in two dimensions\,\cite{Hoang:2007vb}
\begin{equation}\label{eq:softShape}
S(\ell_1,\ell_2) =\! \int\! \df k_1\, \df k_2\,\widehat S(\ell_1-k_1,\ell_2-k_2)F(k_1,k_2,\Delta)\,.
\end{equation}
Here, the partonic soft function $\widehat S$ is defined in fixed-order perturbation theory in the $\msbar$
scheme. $\Delta$ is a parameter of the shape function that represents an offset from zero momentum. This
offset can compensate for the $u=1/2$ renormalon contained in the partonic soft function. The shape
function $F(k_1,k_2, \Delta)$ is therefore defined to satisfy the relation
\begin{equation} \label{eq:softShapev2}
F(k_1,k_2, \Delta) = F(k_1 - \Delta,k_2 - \Delta).
\end{equation}
The term $\Delta$ stands for a parameter in the shape function that governs the hadronic gap, i.e.\ the soft momentum below which the shape function vanishes. Depending on the analytic functional form adopted for the shape function, also other parameters of the shape function may in principle contribute to that gap. Our parametrization of the shape function discussed in Sec.~\ref{sec:shape} satisfies the condition $F(k_1 \le 0,k_2) = F(k_1,k_2 \le 0) = 0$. Introducing the term $\Delta$ in this explicit way as a shift, the ${\cal O}(\Lambda_{\rm QCD})$ ($u=1/2$) renormalon in the partonic soft function is fully contained in $\Delta$ and can be removed through a perturbative redefinition~\cite{Hoang:2007vb}. Retaining the term $\Delta$ in its original form, which we call the $\msbar$ scheme for the shape function, leaves the renormalon unsubtracted.
Taking into account $\Delta$ the computation of the $\msbar$ parameters
$\overline\Omega_{i,j}$ and $\overline\Upsilon_{n,m}$ defined in Eqs.~\eqref{eq:Omega-ij} and
\eqref{eq:Upsilon-ij} becomes:
\begin{align}\label{eq:gapmomentrelation}
\overline\Omega_{i,j} = &\sum_{n=0}^i\binom{i}{n}\sum_{\ell=0}^j\binom{j}{\ell}\Delta^{i+j-n-\ell}
\!\!\int\!\df k_1\,\df k_2\,
k_1^n\,k_2^\ell\,F(k_1,k_2)\,,\\
\overline\Upsilon_{i,j} = &\sum_{n=0}^i\binom{i}{n}\sum_{\ell=0}^j\binom{j}{\ell}\Delta^{i+j-n-\ell}
\!\!\int\!\df k_1\,\df k_2\,
\theta(k_1-k_2)\,(k_1-k_2)\,k_1^n\,k_2^\ell\,F(k_1,k_2)\,.\nonumber
\end{align}
Since the shape function is normalized to one upon integration of its two arguments, one finds that
$\overline\Omega_{1,0}$ has a linear dependence on $\Delta$,
\begin{align}\label{eq:o10case}
\overline\Omega_{1,0} = &
\Delta +
\!\!\int\!\!\df k_1\,\df k_2\,
k_1 \,F(k_1,k_2)\,.
\end{align}

In the tail region, where $s_{1,2}/Q \gg \Lambda_{\text{QCD}}$, Eq.~\eqref{eq:softShape} can be expanded
to give
\begin{align}\label{eq:soft-OPE}
S(\ell_1,\ell_2,\mu) = \widehat S(\ell_1,\ell_2) - \Biggl[\frac{\df \widehat S(\ell_1,\ell_2)}{\df \ell_1} +
\frac{\df \widehat S(\ell_1,\ell_2)}{\df \ell_2} \Biggr]
\overline{\Omega}_{1,0}
+ {\cal O}\!\Biggl(\! \frac{\alpha_s \lqcd }{(\ell_{1,2})^3},\frac{\lqcd^2}{\ell_{1,2}^4}\!\Biggr) ,
\end{align}
where $ \overline{\Omega}_{1,0}\equiv \overline\Omega_1^\rho$ is the leading non-perturbative power correction in
the $\msbar$ scheme. For thrust and HJM it causes a shift of the partonic distribution in the tail region known also for other event-shapes. Projecting Eq.~\eqref{eq:soft-OPE} onto the thrust soft function leads to
\begin{align}\label{eq:thrust-soft-OPE}
S_\tau(\ell,\mu) = \widehat S_\tau(\ell) - 2\,\frac{\df \widehat S_\tau(\ell)}{\df \ell}\,
\overline{\Omega}_{1,0}
+ {\cal O}\!\Biggl(\!\frac{\alpha_s \lqcd }{\ell^2},\frac{\lqcd ^2}{\ell^3}\!\Biggr) ,
\end{align}
To remove the leading $\mathcal{O}(\lqcd )$ renormalon ambiguity of the soft function we redefine
\begin{align} \label{eq:deltasplitting}
\Delta & = \bar{\Delta}(R,\mu) + \delta(R,\mu)\,.
\end{align}
The term $\delta(R,\mu)$, which is a perturbative series in $\alpha_s(\mu)$, is then shifted into the partonic soft function $\widehat S$ where it eliminates the $\mathcal{O}(\lqcd )$ renormalon. This series is designed such that it exactly eliminates the leading
asymptotic behavior as the partonic soft function.
The definite form adopted for the series $\delta(R,\mu)$ defines a particular scheme for the renormalon subtraction.
The term $\bar{\Delta}(R,\mu)$, which remains in the shape function, then defines that shape function in that particular scheme, and its size at some reference values for the scales $R$ and $\mu$ is non-perturbative. The evolution in $R$ and $\mu$ is governed by perturbative RG evolution that depends on the scheme choice for $\delta(R,\mu)$ given that $\Delta$ is scheme-invariant.

The scale $R$ is the momentum scale associated with the subtraction of the linear infrared sensitivity and $\mu$ stands to a potential additional logarithmic anomalous dimension. The scheme for $\delta(R,\mu)$ we adopt here is the same as the R-gap scheme~\cite{Hoang:2008fs} employed already for the thrust analysis in Ref.~\cite{Abbate:2010xh} and reexamined recently also in Ref.~\cite{Benitez:2024nav},
\begin{equation} \label{eq:delta-scheme}
\!\!\!\!\delta(R,\mu) = \frac{1}{2}R \, e^{\gamma_E} \frac{\df}{\df L_{y}}
\bigl[\, \ln \widetilde{S}(L_{y},L_{y},\alpha_s(\mu)) \,\bigr]_{y\, =\, (R\, e^{\gamma_E})^{-1}}\,,
\end{equation}
where $\widetilde{S}$ is defined in Eq.~\eqref{eq:lapdef}.
There are other possible definitions of the subtraction series that would also remove the renormalon, see~\cite{Bachu:2020nqn,Dehnadi:2023msm,Bell:2023dqs}.
Note that the non-global part of the 2-dimensional partonic soft function, which is renormalon-free~\cite{Hoang:2008fs}, does not contribute to
$\delta$, since it becomes a constant when setting $y_1 = y_2$. We write the R-gap subtraction series in the form
\begin{align} \label{eq:deltaseries}
\delta(R,\mu) =
R \,e^{\gamma_E} \!\sum_{i=1}^\infty \alpha_s^i(\mu)\, \delta^i(R,\mu)\,.
\end{align}
At order N$^3$LL$^\prime$ we need the first three coefficients. They have been calculated in
Ref.~\cite{Hoang:2008fs} and for five active flavors they take the following numerical values:
\begin{align} \label{eq:d123}
\delta^1(R,\mu_s) &= -\,0.848826\, L_R \,, \\
\delta^2(R,\mu_s) &= -\,0.156279 - 0.46663\, L_R - 0.517864\, L_R^2 \,, \nonumber \\
\delta^3(R,\mu_s) &= -\,0.552986\, - 0.622467\, L_R -\,0.777219\, L_R^2 - 0.421261\, L_R^3 \,,\nonumber
\end{align}
where $L_R = \ln(\mu/R)$. This yields the relation between $\bar\Omega_{1,0}$ in the $\msbar$ and
$\Omega_{1,0}(R,\mu)$ in the R-gap scheme,
\begin{equation} \label{eq:omegaschemechange}
\Omega_{1,0}(R,\mu) = \overline{\Omega}_{1,0} - \delta (R,\mu)\,,
\end{equation}
which is free of the $\mathcal{O}(\lqcd)$ renormalon. As shall be discussed in Sec.~\ref{sec:OPE}, the leading-power OPE depends only on $\Omega_{1,0}$, which is, along with the strong coupling, the only relevant parameter in the tail region. Interestingly, these subtractions do not affect the leading peak parameter $\Upsilon_{0,0}$, which is the same both in the R-gap and $\msbar$ schemes for the shape function, but does affect $\widehat\Upsilon_{1}$, whose R-gap scheme is $\widehat\Upsilon_{1}(R,\mu) = \overline{\Upsilon}_{1} - \delta (R,\mu)/2$, as can be seen from Eq.~\eqref{eq:barUpsilon}.
We therefore conclude that this parameter is free from the $u=1/2$ renormalon. From
Eq.~\eqref{eq:gapmomentrelation} we can then derive the relation
\begin{equation}
\label{eq:eq:omegaschemeRgap}
\Omega_{1,0}(R,\mu) =\! \int \!\df k_1\,\df k_2 \,k_1F(k_1,k_2) + \bar\Delta(R,\mu) \,,
\end{equation}
and the soft function factorization in Eq.~\eqref{eq:softShape} is rewritten as
\begin{align} \label{eq:soft-nonperturbative-subtract}
\!\!S(\ell_1,\ell_2,\mu) =\! &\int \!\df k_1\df k_2 \,e^{\delta(R,\mu)\bigl( \frac{\partial}{\partial \ell_1}+ \frac{\partial}{\partial \ell_2}\bigr)}
\widehat S(\ell_1-k_1,\ell_2-k_2,\mu) \\\nonumber
&\times F\bigl[k_1 - \bar{\Delta}(R,\mu),k_2 - \bar{\Delta}(R,\mu)\bigr]\,.
\end{align}
The logarithms in the $\delta^i$ coefficients in Eqs.~(\ref{eq:d123}) can become large if $\mu$ and $R$ are widely separated. This imposes that $R \sim \mu$, which requires that $R$ becomes a function of $\rho$ in an analogous way to $\mu$. On the other hand, we also must consider $\overline{\Omega}_{1,0} \sim \Lambda_\text{QCD}$, which makes the choice $R \simeq 1$\,GeV desirable. This is automatically satisfied in the peak region, where $\mu_s\sim\lqcd $, but not in the tail and far tail, where one has $\mu_s\gg\lqcd $. To remedy this we (i)~employ $R \sim \mu$ in $\delta(R,\mu)$ which avoids large logs within the renormalon subtracted partonic soft function, and (ii)~use the $\mu$- and $R$-evolution equations to relate $\bar\Delta(R\sim \mu,\mu)$ to a reference gap parameter $\bar\Delta(R_\Delta,\mu_\Delta)$ with $R_\Delta \sim \mu_\Delta \sim {\cal O}(1\,{\rm GeV})$
such that $\Omega_{1,0}(R_\Delta,\mu_\Delta)\sim\lqcd $~\cite{Hoang:2008yj,Hoang:2009yr,Hoang:2008fs,Hoang:2017suc}. For our numerics we choose $R_\Delta = \mu_\Delta=2\,$GeV. This $\bar \Delta(R,\mu)$ is an additional parameter of the non-perturbative shape function. The $\mu$- and $R$-evolution equations are
\begin{align} \label{eq:RandmuRGE}
R \frac{\df}{\df R} \bar{\Delta} (R,R) &= -R \sum_{n=0}^\infty \gamma_n^R \biggl[ \frac{\alpha_s(R)}{4 \pi} \biggr]^{n+1} , \\
\mu \frac{\df}{\df\mu} \bar{\Delta} (R,\mu) &= 2 R\, e^{\gamma_E} \sum_{n=0}^\infty \Gamma_n^\text{cusp} \biggl[ \frac{\alpha_s(\mu)}{4 \pi} \biggr]^{n+1} ,\nonumber
\end{align}
where for five flavors the $\Gamma_n^\text{cusp}$ are given in Eq.~(26) of Ref.~\cite{Abbate:2010xh} and the
$\gamma^R$ coefficients read
\begin{align} \label{eq:gammaR}
\gamma_0^R &= 0, \; \gamma_1^R = -\,43.954260\,,\;\gamma_2^R = -\,606.523329\,.
\end{align}
The solution to Eqs.~\eqref{eq:RandmuRGE} can be computed analytically and is given, at N$^{(k+1)}$LL in the
event-shape large-log counting, by
\begin{align} \label{eq:DeltaRevolution}
\bar{\Delta}(R,\mu) \,=\, &\, \bar{\Delta}(R_\Delta,\mu_\Delta) + R\, e^{\gamma_E}
\omega(\Gamma^\text{cusp},\mu,R)
+ R_\Delta e^{\gamma_E} \omega(\Gamma^\text{cusp},R_\Delta,\mu_\Delta) \\
&+ \Lambda_{\text{QCD}}^{(k)} \,\sum_{j=0}^{k} (-1)^j S_j\, e^{i \pi \hat{b}_1}
\bigr[\,\Gamma ( -\,\hat{b}_1 -j, t_1) - \Gamma(-\,\hat{b}_1 -j,t_0) \,\bigl]\nonumber\\[0.1cm]
\,\equiv\, &\,\bar{\Delta}(R_\Delta,\mu_\Delta) + \Delta^{\rm diff}(R_\Delta,R, \mu_\Delta,\mu)\,,\nonumber
\end{align}
where $\omega$ has been defined in the third line of Eq.~\eqref{eq:UH}, $\hat{b}_1 = \beta_1/(2\beta_0^2)$
with $\beta_i$ the strong coupling constant anomalous dimension coefficients, and the $S_j$ can be
obtained from $\gamma_i^R$, as shown in Eq.~(A.15) of Ref.~\cite{Hoang:2017suc}. Moreover,
$t_0 = -\,2\pi/[\,\beta_0\alpha_s(R_\Delta)\,]$ and $t_1 = -\,2\pi/[\,\beta_0\alpha_s(R)\,]$, and $\Lambda_{\text{QCD}}^{(k)}$
is the value of $\Lambda_{\text{QCD}}$ at $k$'th loop order in the $\beta$-function expansion~\cite{Hoang:2009yr}. Since
there are no double logs in the evolution of the gap parameter, its running starts at NLL in the
event-shape log counting, achieving N$^{k+1}$LL precision if $\gamma_i$ are included up to $i=k$.
Furthermore, since $\gamma_0^R = 0$, which is specific for the R-gap scheme, see also~\cite{Bachu:2020nqn,Dehnadi:2023msm}, R-evolution only contributes
at N$^2$LL order and beyond. This is summarized in Table~\ref{tab:ordercounting}.
For an automated and exact solution of generic R-evolution equations using a semi-analytic algorithm, see Ref.~\cite{Lepenik:2019jjk}.
In our
theoretical cross section predictions with power corrections and R-gap subtractions we specify a numerical input value for the gap parameter \mbox{$\bar{\Delta} (R_\Delta = \mu_\Delta = 2 \,{\rm GeV})$}. We then use
Eq.~\eqref{eq:DeltaRevolution} and evolve it to the scales $R(\rho)$ and $\mu_s(\rho)$ as given in
Sec.~\ref{sec:profiles}.
Note that via Eq.~(\ref{eq:eq:omegaschemeRgap}) $\Omega_{1,0}(R,\mu)$ follows the same $R$- and $\mu$-evolution equations as $\bar{\Delta}(R,\mu)$.

We also apply the R-gap scheme to the non-singular part of the cross section by using the convolution
\begin{align} \label{eq:nonsingular-shape}
\frac{\df^2\sigma_{\rm ns}}{\df s_1\df s_2} = &\! \int\!\!\df k_1\,\df k_2\,
\frac{\df^2\hat\sigma_{\rm ns}}{\df s_1\df s_2}(s_1 - Q\,k_1, s_2 - Q\,k_2)\\\nonumber
&\times e^{\delta(R,\mu_s)\bigl( \frac{\partial}{\partial k_1}+ \frac{\partial}{\partial k_2}\bigr)}\!
F\bigl[k_1 - \bar{\Delta}(R,\mu_s), k_2 - \bar{\Delta}(R,\mu_s)\bigr].
\end{align}
To project Eq.~\eqref{eq:nonsingular-shape} into HJM we use the ansatz in Eq.~\eqref{eq:NSansatz}, yielding
the following modification to Eq.~\eqref{eq:ns-hadronic}:
\begin{align}
\frac{\df\sigma_{\rm ns}}{\df \rho}= &\! \int_0^{Q\rho^\prime}\!\!\df k\,
\frac{\df\hat\sigma_{\rm ns}}{\df \rho}\biggl(\!\rho^\prime-\frac{k}{Q}\biggr)
e^{\delta(R,\mu_s)\bigl[ \frac{\partial}{\partial k}+ \frac{\partial}{\partial (Q\rho^\prime)}\bigr]}\!
\bigl[F^\Xi(Q\rho^\prime,k) + F^\Xi(k,Q\rho^\prime)\bigr],\\\nonumber
Q\rho^\prime=&\,\, Q\rho - \bar{\Delta}(R,\mu_s)\,,
\end{align}
and analogously for $\Sigma^\rho_{\rm ns}(\rho)$.

\section{Two-Dimensional Shape Functions}\label{sec:shape}

In this section we generalize the setup for a one-dimensional shape function expanded in basis functions employed in Ref.~\cite{Ligeti:2008ac} into a form suitable for
factorization theorems involving the two-dimensional non-perturbative shape function $F$ such as the DHM distribution in Eq.~(\ref{eq:double-tot}).
$F$ has mass dimension $-2$, is normalized with positive support
\begin{align} \label{eq:normF}
\int_0^\infty\!\! \df k_1 \df k_2 \: F(k_1,k_2) &= 1 \,,
\end{align}
and we can assume that it is positive definite $F(k_1,k_2)\ge 0$.

We want to provide a general basis for describing any function $F$ whose dominant support occurs
for the region $k_1,k_2 \lesssim$ few $\Lambda_{\rm QCD}$, vanishes for $k_{1,2}=0$ and is exponentially suppressed for
larger values. Any such two-dimensional function can be expanded in a basis of products of one-dimensional functions as
\begin{align} \label{eq:shapebasis}
\sqrt{ F(k_1,k_2)}
& = \frac{1}{\sqrt{\lambda_1\lambda_2}} \sum_{n=0}^\infty \sum_{m=0}^\infty
c_{n,m} \,
f_n\!\biggl(\frac{k_1}{\lambda_1}\biggr)
f_m\!\biggl(\frac{k_2}{\lambda_2}\biggr)
\,,
\end{align}
where $f_j(z)$ form a fixed, complete and orthogonal set of functions on $[0,\infty)$,
\begin{align}
\int_0^\infty\! \df z\: f_n(z) f_{n^\prime}(z) =\delta_{nn^\prime}\,,
\end{align}
and $\lambda_1$ and $\lambda_2$ are mass dimension one parameters that are both $\sim \Lambda_{\rm QCD}$.
The choice of $\lambda_1$ and $\lambda_2$ is part of the basis definition. \eq{shapebasis} with given
$\lambda_{1,2}$ and $c_{n,m}$ fully specify any such function $F(k_1,k_2)$.

If the shape function is symmetric $F(k_1,k_2)=F(k_2,k_1)$ then we must take $\lambda_1=\lambda_2=\lambda$
and we have $c_{n,m}=c_{m,n}$. This is the relevant case for our study of the DHM and HJM distributions.
However other examples of doubly differential distributions, such as for two different angularities, would not be symmetric.

For the $f_n(z)$ functions we use the basis of Ref.~\cite{Ligeti:2008ac},
\begin{align} \label{eq:basisfunctions}
f_n(z) &= 8 \sqrt{\frac{2z^3(2n+1)}{3}}\, e^{-2z} P_n [\,g(z)\,]\,, \\
g(z)&=\frac{2}{3} \bigl[3 - e^{-4z} (3 + 12 z + 24z^2 + 32 z^3) \bigr] - 1\,,\nonumber
\end{align}
where $P_n(x)$ are the Legendre polynomials. The exponential fall off for large-$k$ ensures that any moment
\begin{align}
\Omega_{i,j} = \int_0^\infty\!\! \df k_1 \df k_2\: k_1^i\, k_2^j \: F(k_1,k_2) \sim
\Lambda_{\rm QCD}^{i+j}\,,
\end{align}
exists. This also ensures
\begin{align}
\int_0^\infty\!\! \df k_1 \df k_2\: k_1^i\, k_2^j \:
h(k_1,k_2) \: F(k_1,k_2) \sim
\Lambda_{\rm QCD}^{i+j}\,,
\end{align}
where $h(k_1,k_2)$ is any dimensionless bounded function that is piecewise continuous on any finite interval in $k_1$
and $k_2$.

In practice one must truncate the infinite sums in \eq{shapebasis} and use a finite set of basis functions,
\begin{align} \label{eq:shapebasisT}
F(k_1,k_2)
& = \frac{1}{\lambda_1\lambda_2} \Biggl[\,
\sum_{n=0}^{N_1} \sum_{m=0}^{N_2}
c_{n,m} \,
f_n\!\biggl(\frac{k_1}{\lambda_1}\biggr)
f_m\!\biggl(\frac{k_2}{\lambda_2}\biggr) \!
\Biggr]^2
\,.
\end{align}
Here the normalization condition in \eq{normF} implies the constraint
\begin{align} \label{eq:normFN}
\sum_{n=0}^{N_1} \sum_{m=0}^{N_2}
\bigl( c_{n,m} \bigr)^2 =1
\,.
\end{align}
The simplest truncation takes $N_1=N_2=N$ which keeps only the coefficients within a square in the $(n,m)$ grid. We refer
to this as the square truncation. In this case after imposing the normalization constraint there are $N(N+3)/2$ independent
coefficients $c_{n,m}$ when $F$ is symmetric, and $N(N+2)$ independent $c_{n,m}$ coefficients for a non-symmetric $F$.
Another example of truncation is to take $N_1=N$ and $N_2 = N-n$ which keeps only the coefficients within a right-triangle
in the $(n,m)$ grid. We refer to this as the triangular truncation. In this case after imposing the normalization
constraint there are $[N(N+3)+2\,\text{Floor}(N/2)]/4$ independent coefficients $c_{n,m}$ when $F$ is symmetric, and
$N(N+3)/2$ independent $c_{n,m}$ coefficients for a non-symmetric $F$.
Choosing either one of these truncation schemes
depends on the region of the spectrum one is interested in. For instance the tail of a double differential distribution can
be expanded in an OPE in terms of the $\Omega_{i,j}$ coefficients only, such that at order $n$ in the
expansion only $\Omega_{i,j}$ with $i + j = n$ contribute. Therefore there are as many $\Omega_{i,j}$ as
$c_{i,j}$ if the series is triangular-truncated.
If we are interested in the peak and the tail,
also $\Upsilon_{i,j}$ play a role. For example, when looking into the $n$-th moment of the HJM
distribution $\Upsilon_{i,j}$ with $i + j + 1 = n$ also play a role, such that again there are as many
$c_{i,j}$ as $\Omega_{i,j}$ and $\Upsilon_{i,j}$ combined if the series is square-truncated.

With either truncation the parameter $\lambda_1$ is strongly correlated with $c_{0,1}$ and $\lambda_2$ is
strongly correlated with $c_{1,0}$. Since a proper choice for $\lambda_1$ and $\lambda_2$ is necessary to
obtain a basis with good convergence properties, it can be convenient to include $\lambda_1$ and
$\lambda_2$ as fit parameters and set $c_{0,1}=c_{1,0}=0$.

In the tail region, where only the first moment of the shape function $\Omega_{1,0}$ matters, one can simply
take $c_{0,0}=1$ and zero for the rest of $c$'s. In this case, following from Eq.~(\ref{eq:eq:omegaschemeRgap}), there is a one-to-one correspondence between
$\lambda$ and $\Omega_1$:
\begin{align} \label{eq:omega-from-lambda}
\Omega_{1,0}(R_\Delta,\mu_\Delta) = \lambda + \bar{\Delta}(R_\Delta,\mu_\Delta)\,.
\end{align}

\begin{figure*}[t!]
\subfigure[]{\includegraphics[width=0.475\textwidth]{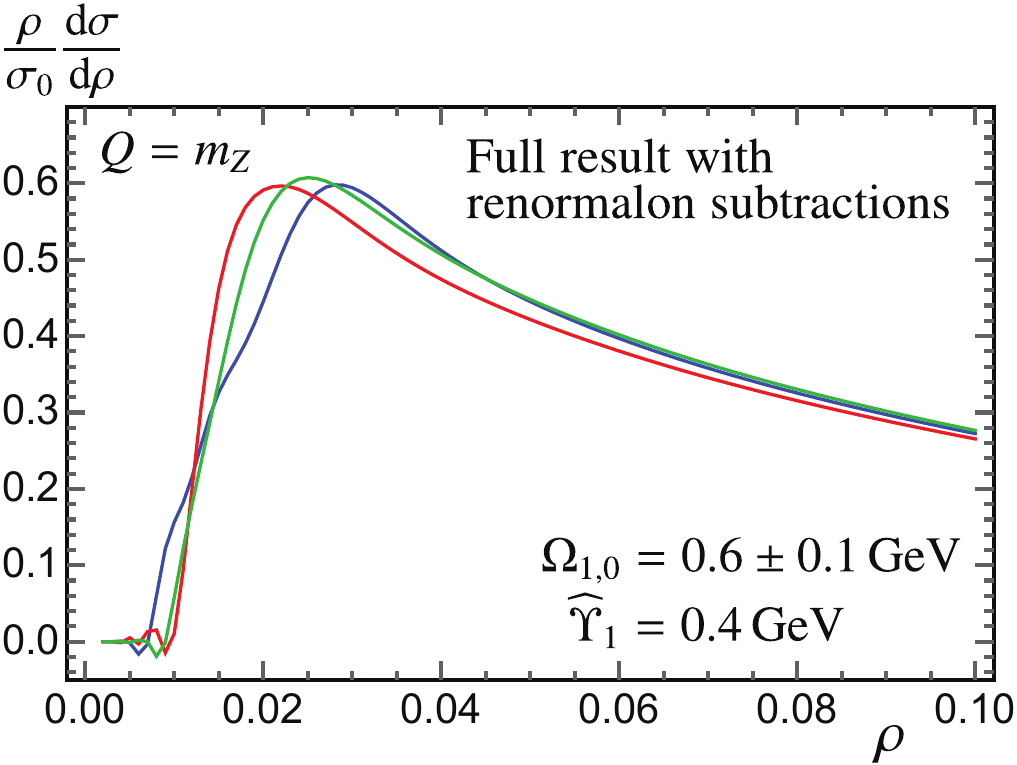}
\label{fig:Peak-Omega}}
\subfigure[]{\includegraphics[width=0.475\textwidth]{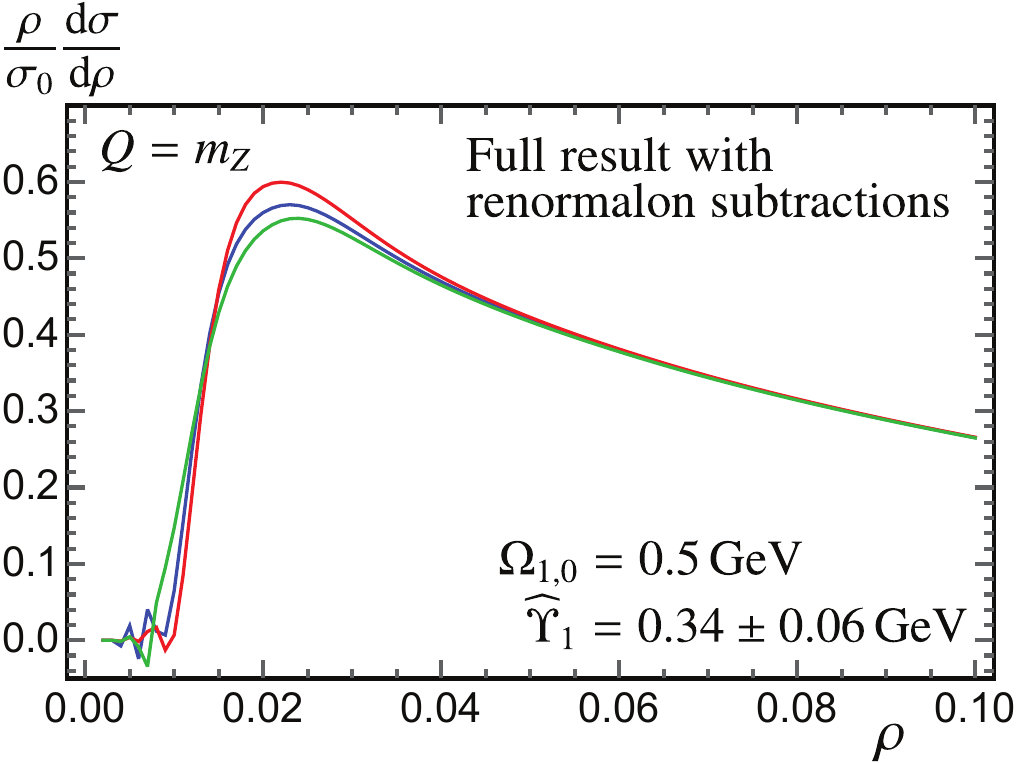}
\label{fig:Peak-Upsilon}}
\vspace{-0.2cm}
\caption{\label{fig:peak-plots}
Effects on the peak of the HJM distribution when $\Omega_{1,0}$ is varied holding $\widehat\Upsilon_{1}$
fixed (left panel) and when $\widehat\Upsilon_{1}$ is varied holding $\Omega_{1,0}$ fixed. The non-fixed parameter
is varied by $18\%$ upwards and downwards from a reference value.} %
\end{figure*}

With the parametrization of the two-dimensional shape function being set up, let us discuss the particular
impact of $\widehat\Upsilon_1$ in the peak region of the HJM distribution, which we have already mentioned in Sec.~\ref{sec:OPEmom}, by a concrete analysis.
In Fig.~\ref{fig:peak-plots} the effect of $\Omega_{1,0}$ and $\widehat\Upsilon_1$, the leading non-perturbative
parameters in the peak, is shown varying one while keeping the other fixed, always within the bound shown after
Eq.~\eqref{eq:barUpsilon}. This can be achieved using a model with certain square truncation,
varying the $c_{i,j}$ and adjusting $\lambda$ such that the parameter which we want fixed does not change its
value. If terms only up to $N = 1,\,2$, and $3$ are kept, we can vary either $\Omega_1$ or
$\widehat{\Upsilon}_1$ by $23\%,\,36.2\%$, and $43.6\%$ respectively, if the other one is fixed. To show
the effect of each parameter it is enough to use the truncated series with $N = 2$.
In Fig.~\ref{fig:Peak-Omega} we vary $\Omega_1$ keeping $\widehat\Upsilon_1$ fixed, and observe that the
effect of this parameter is mostly shifting the peak position right or left. In Fig.~\ref{fig:Peak-Upsilon}
we do the opposite variation, keeping $\Omega_1$ fixed while varying $\widehat{\Upsilon}_1$, and the effect
is mostly moving the maximum of the distribution up or down. If the plot is extended towards the tail, all curves
merge, which is consistent with our previous observations that $\widehat{\Upsilon}_n$ is irrelevant for the OPE in the tail region, but important for the moments. For a complete description of the peak both parameters are
relevant. We have also explored the effect of keeping both $\Omega_1$ and $\widehat{\Upsilon}_1$ fixed while
varying the model. For this exercise it is sufficient to truncate the model with $N = 1$. There is a
continuous set of such models with both parameters fixed, that can be found by finding a constraint between
$c_{1,1},\,c_{1,0}$ and $\lambda$ in the form of a finite curve on the three-dimensional space defined by
these three parameters. Scanning over the remaining degree of freedom one simply changes higher mass-dimension
non-perturbative parameters while keeping the two having mass-dimension one fixed. We find that all models
having the same $\Omega_1$ and $\widehat{\Upsilon}_1$ look nearly identical both in the peak and tail regions.
Hence we conclude that
these two parameters are the most relevant non-perturbative parameters in the whole distribution.

Note that one can project Eq.~\eqref{eq:shapebasis} into the thrust shape function using
the first equation in the last line
of Eq.~\eqref{eq:thrust-hadronic}. In this way it may seem that the $c_n$ coefficients of
Eq.~(55) in Ref.~\cite{Abbate:2010xh} can be easily related with the $c_{n,m}$ coefficients of
Eq.~\eqref{eq:shapebasis}. This is in practice not the case as the parametrization for thrust raises as
$k^3$ while the thrust projection of Eq.~\eqref{eq:shapebasis} starts off as $k^7$. One then finds that each
$c$ coefficient of the HJM parametrization depends on an infinite number of the thrust basis coefficients.
Moreover, the reconstructed thrust shape function turns out to oscillate around the projected
parametrization, making this approach unpractical. An easy way of having a two-dimensional shape function
that once projected onto thrust raises like $k^3$ for small $k$ is simply using for the basis a set of functions
such that $0$-th element behaves like a constant at the origin:
\begin{equation} \label{eq:basisfunctions2}
\bar f_n(z) = \sqrt{4(2n+1)}\, e^{-2z} P_n
(1 - 2e^{-4z})\,.
\end{equation}
Since we are not attempting an accurate peak description in this article, we will make use of Eqs.~\eqref{eq:shapebasis}.

\section{Profile Functions for heavy jet mass} \label{sec:profiles}
In this section we discuss constructing profile functions to smoothly transition between the peak and tail regions. As already mentioned, in this article we do not focus on the transition to the
trijet region, where heavy jet mass is affected by large Sudakov shoulder logarithms. These logarithms should also be resummed to have a trustworthy prediction for $\rho \gtrsim 0.2$, and we note that the matching between the dijet, tail and trijet region was discussed in~\cite{Bhattacharya:2023qet}.
In this analysis we treat this trijet region at fixed-order. Therefore we only consider three regions: the ``peak region'', where the singular cross section dominates and non-perturbative effects are described by a convolution over a shape function , the ``far-tail'' region where the full fixed order result is used, and the ``tail'' region, where the singular cross section is still dominating, but where the non-perturbative corrections can be described in terms of an OPE and the non-singular corrections slowly turn on. Profile functions allow us to smoothly transition between these regions.

The factorized expression for the singular partonic differential cross section
${\df \hat{\sigma}_{\rm s}}/{\df \rho}$ obtained from Eq.~(\ref{eq:HJM-dist}) depends on three renormalization scales:
hard $\muh$, jet $\muj$, and soft $\mus$. Each of these probes different physics. On top of that, the
non-singular correction contains a different scale, which will be referred to as the non-singular scale
$\mu_{\rm ns}$. To avoid large logarithms appearing in the corrections to the hard coefficient $H$, and
the jet $J$ and soft $S$ functions, the corresponding scales must satisfy the following
theoretical constraints in the three distinct regions of the spectrum:
\begin{align}
\label{eq:profile123}
& \text{1) peak:} & &
\muh\sim Q \,,\ \
\muj\sim \sqrt{\lqcd Q}\,,\
& \mus &\gtrsim \lqcd \,, \\
& \text{2) tail:} & &
\muh\sim Q \,,\ \
\muj\sim Q\sqrt{\rho} \,,\
&\mus &\sim Q\rho \,,\nonumber\\
& \text{3) far-tail:} & &
\muh=\muj = \mus \sim Q \,. \nonumber
\end{align}
To have renormalization scales that satisfy these requirements in all three regions in a continuous, smooth way,
we employ $\rho$-dependent profile functions. Note that the scaling relations in \eq{profile123} must be satisfied to obtain correct results, and that the freedom in the profile functions primarily occurs in how these conditions are smoothly connected to one another in the two transition regions.

The transition regions
can be identified by an inspection of the differential distribution~\cite{Abbate:2010xh,Hoang:2014wka,Benitez:2024nav}, but
there is some freedom in where exactly these regions start and end. In particular, we choose the transition interval between peak and tail using the observations made already in
thrust~\cite{Abbate:2010xh} and C-parameter~\cite{Hoang:2014wka} that the transition between these
regions occurs near the maximum of the peak. For thrust and HJM
this happens approximately at the same value of the event shape since, as we have already explained in Sec.~\ref{sec:OPE}, hadron masses effects make the leading power correction for this two variables to have very similar size.

The parametrization for the profile functions we use is as follows. The hard scale reads
\begin{equation}
\muh = \,e_H\,Q\,,
\end{equation}
where $e_H$ is a free parameter which we vary between $1/4$ to $1$ in our theory error analysis and with $e_H=1/2$ being the default. This default choice turns out to be advantageous when the Sudakov shoulder region is accounted for as well, see Ref.~\cite{Benitez:2025vsp}. In Tab.~\ref{tab:theoryerr} we have collected the default values and the interval of variation of all parameters used for the profile functions.

The profile function for the soft scale has the form
\begin{equation} \label{eq:muSprofile}
\!\mu_s(\rho) = \left\{ \begin{tabular}{p{.5\columnwidth} l}
$\mu_s(0)$ & $0 \le \rho < t_0$ \\
$\zeta(\mu_s(0),0,0,r_s\mu_H,t_0,t_1,\rho)$ & $t_0 \le \rho < t_1$ \\
$r_s \mu_H \rho$ & $t_1 \le \rho < t_2$ \\
$\zeta(0,r_s\mu_H,\mu_H,0,t_2,t_s,\rho)$ & $t_2 \le \rho < t_s$ \\
$\mu_H$ & $t_s \le \rho < 1$
\end{tabular}
\right.\!.
\end{equation}
The 1st, 3rd, and 5th lines satisfy the theoretical requirements in Eq.~\eqref{eq:profile123}, and the
2nd and fourth lines parametrize the
transitions in between. Here $\mu_s(0)$ parametrizes the intercept of the soft scale at $\rho=0$.
Since the location of the peak and the begin of the tail region scale inversely with $Q$,
the transition interval $[t_0,t_1]$ between these regions is parametrized by
\begin{equation}
n_0 \equiv t_0 \,Q/(1\,{\rm GeV})\,,\qquad
n_1 \equiv t_1 \,Q/(1\,{\rm GeV}) \,,
\end{equation}
In the tail region $t_1< \rho <t_2$, the parameter $r_s$ sets the slope of the soft scale.
In Ref.~\cite{Hoang:2014wka} it has been argued that for thrust and C-parameter it is convenient to set the default
a factor of two larger than the canonical choice ($1$ for thrust or HJM, $1/6$ for C-parameter). We therefore adopt $r_s=2$ as the default and vary around that value. The patch between $t_2$ and $t_s$ transitions
between the tail and far-tail regions, where all renormalization scales merge.
The function $\zeta(a_1,b_1,a_2,b_2,t_1,t_2,\,t)$ with $t_1 < t_2$, is used for the transitions between the peak, tail and far-tail regions. It smoothly connects two straight
lines of the form $\ell_1(t) = a_1 \,+\, b_1\,t$ for $t < t_1$
and $\ell_2(t) = a_2 \,+\, b_2\,t$ for $t > t_2$ at the meeting points $t_1$ and $t_2$, such that also the first derivative is continuous. We refer to Ref.~\cite{Hoang:2014wka} for its definition.

The parametrization of the soft scale in Eq.~\eqref{eq:muSprofile} was also used in Refs.~\cite{Stewart:2014nna,Hoang:2014wka,Butenschoen:2016lpz,Benitez:2024nav,Benitez:2025vsp}, but it differs from that of
Ref.~\cite{Abbate:2010xh}.
(In the latter reference the entire tail region was treated as a transition between the
peak and the far-tail, with the slope set by parameters that did not capture the adequate scaling behavior shown in the second line of Eq.~(\ref{eq:profile123}). Using the improved soft scale parametrization leads to thrust results that are stable against fit range variations, see Ref.~\cite{Benitez:2024nav}.)

Through the canonical scaling relations in Eq.~(\ref{eq:profile123}) the jet scale is tied to the soft scale. To allow for variations around that canonical scaling a $\rho$-dependent trumpet factor is used
which yields the parametrization
\begin{equation} \label{eq:muJprofile}
\!\!\!\!\!\mu_J(\rho) = \left\{\!\! \begin{array}{lr}\,
\bigl[\,1 + e_J (\rho-t_s)^2\,\bigr] \sqrt{ \mu_H \mu_s (\rho)} & ~~~~\rho \le t_s\\
\,\,\mu_H & ~~~~\rho > t_s
\end{array}
\right.\!.
\end{equation}
The trumpet and its first derivative vanishes at $\rho = t_s$ so that the transition to the far-tail
region is smooth. In our theory scans the parameter $e_J$ is varied between $\pm 1$.

For the renormalon subtraction scale $R$ the choice $R=\mus(\rho)$ is adequate in the tail region, as it ensures that the removal of the ${\cal O}(\Lambda_{\rm QCD})$ infrared renormalon from the soft function does not generate large logarithms. In the peak region, as discussed in~\cite{Abbate:2010xh}, it is convenient to modify this relation within the R-gap subtraction scheme
to avoid having a vanishing subtraction ${\cal O}(\alpha_s)$ (see the first line of Eq.~\eqref{eq:d123}). This yields the form
\begin{equation} \label{eq:muRprofile}
\!\!\!\!\!R(\rho) = \left\{\!\! \begin{array}{ll}\,\,
R(0) & ~~~~0 \le \rho < t_0 \\
\,\,\zeta(R_0,0,0,r_s\mu_H,t_0,t_1,\rho) & ~~~~t_0 \le \rho < t_1 \\
\,\,\mu_s(\rho) &~~~~ t_1 \le \rho \le 1
\end{array}
\right.\!\!.
\end{equation}
There is only one free parameter in this profile function, $R(0)$, which simply picks a constant value for $R$ for $\rho<t_0$. The connection function $\zeta$ again ensures a smooth transition between peak and tail.
The value of $R(0)$ (which can be interpreted as the scale below which infrared sensitive perturbative corrections are being subtracted), affects the very definition of the
non-perturbative shape function $F$.
As already mentioned above, in the R-gap subtraction scheme it is important to keep both parameters slightly different from each other. We adopt
[\,$\mu_s(0)-R(0)= 0.4\, {\rm GeV}$\,] to achieve stability in the peak (see Ref.~\cite{Abbate:2010xh}), and neither of these two parameters is varied to avoid changing the meaning of $F$.

Our EFT treatment does not account for large logs in the non-singular contributions, besides those summed up
by the RGE of the strong coupling. To estimate the effect of this missing resummation we vary the scale
$\mu_{\rm ns}$ as follows:
\begin{equation} \label{eq:muNSprofile}
\mu_{\rm ns}(\rho) = \mu_H - \frac{n_s}{2} \big[ \mu_H - \mu_J(\rho) \big],
\end{equation}
where $n_s$ is varied continuously between $-1$ and $1$, and $n_s$ larger (smaller) than zero probes non-singular scales larger (smaller) than the hard scale. We note that in Ref.~\cite{Hoang:2014wka} only three discrete choices were considered, namely $n_s=1,0,-1$.

\begin{figure*}[t!]
\subfigure[]{\includegraphics[width=0.49\textwidth]{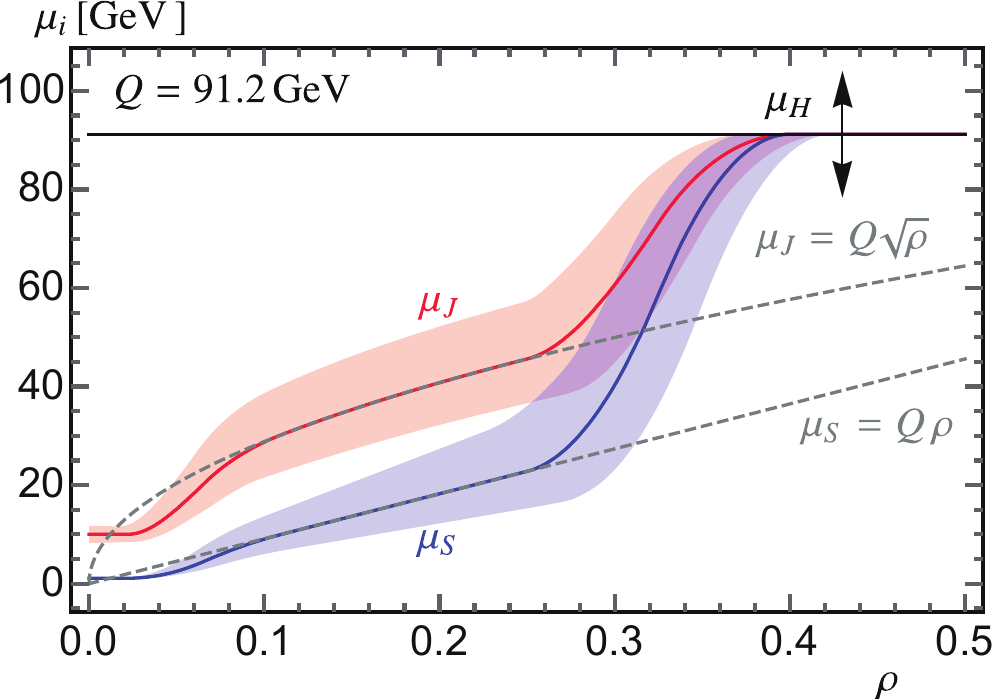}
\label{fig:profiles-default}}
\subfigure[]{\includegraphics[width=0.485\textwidth]{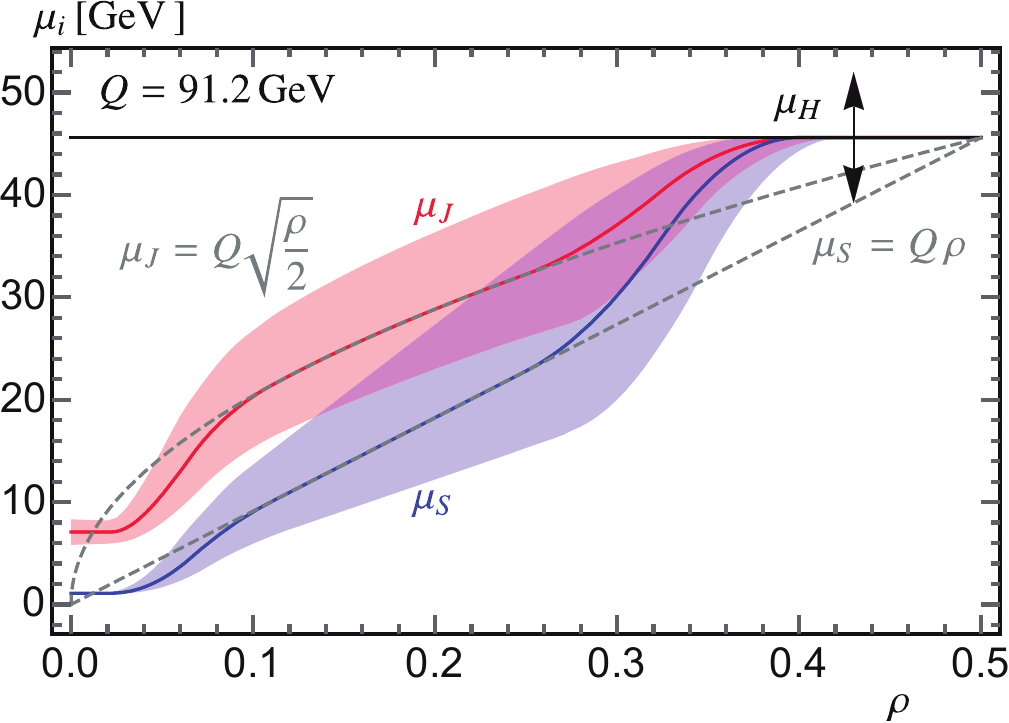}
\label{fig:profiles-Canonical}}
\vspace{-0.2cm}
\caption{Profile functions for the HJM distribution. Solid lines are the central results for the renormalization scales $\mu_H$ (black), $\mu_J(\rho)$ (red), and $\mu_s(\rho)$ (blue) at $Q=m_Z$. The vertical $\updownarrow$ arrow and colored bands indicate the regions covered by variation of profile parameters.
$R(\rho)$ and $\mu_s(\rho)$ are indistinguishable in this figure, being different
only at small $\rho$. Above $\rho=t_s\simeq 0.4$ all the scales merge, $\mu_H=\mu_J=\mu_s=R$.
Left panel (a) shows the standard choice $e_H = r_s = 1$, while right panel (b) has our preferred
values $e_H = 0.5$ and $r_s=2$. The dashed gray lines show the canonical profile functions in the
tail region.}
\label{fig:profiles}
\end{figure*}
\begin{table}[tbh!]
\centering
\begin{tabular}{ccc}
parameter\ & \ default value\ & \ range of values \ \\
\hline
$\mu_s(0)$ & $1.1$\,GeV & -\\
$R(0)$ & $0.7$\,GeV & -\\
$n_0$ & $2$ & $1.5$ to $2.5$ \\
$n_1$ & $10$ & $8.5$ to $11.5$\\
$t_2$ & $0.25$ & $0.225$ to $0.275$\\
$t_s$ & $0.4$ & $0.375$ to $0.425$\\
$r_s$ & $2$ & $1.33$ to $3$\\
$e_J$ & $0$ & $-\,1$ to $1$\\
$e_H$ & $0.5$ & $0.25$ to $1.0$\\
$n_s$ & $0$ & $-\,1$ to $1$\\
\hline
$s_{32}$ & $0$ & $-\,250$ to $+250$ \\
$s_{34}$ & $0$ & $-\,7$ to $7$ \\
\hline
$\epsilon_{2}$ & $0$ & $-\,1$ to $1$ \\
$\epsilon_{3}$ & $0$ & $-\,1$ to $1$ \\
\end{tabular}
\caption{Heavy jet mass perturbative parameters used to estimate the truncation uncertainty. The second column
shows default values, while the range used for our random scan appears on the third column.}
\label{tab:theoryerr}
\end{table}

In Fig.~\ref{fig:profiles} the behavior and variations (colored bands) of the profile functions and our choice of the default parameter setting (solid lines) are illustrated for $Q=m_Z$. The left panel shows the profile functions using $r_s=1$ and $e_H = 1$ as the default (solid lines) for the soft slope and hard scale parameters.
The gray dashed lines show the canonical scale choices $\mu_s = Q\,\rho$ and $\mu_J=Q\sqrt{\rho}$. We
can see that these canonical scales do not meet in the upper endpoint region. For our preferred default with $r_s=2$ and $e_H = 0.5$, shown in the right panel, the canonical lines meet at the endpoint. For this choice of the default parameters the soft profile function remains unchanged, while the jet scale profile function is reduced by a factor of $1/\sqrt{2}$.
In addition this choice also fits together smoothly with the addition of shoulder resummation in the region near $\rho=1/3$, see Refs.~\cite{Bhattacharya:2023qet, Benitez:2025vsp}.
Physically, the lower hard scale can be understood by the fact that the
far-tail of the HJM distribution
is not populated by isotropically distributed particles, as is the case of thrust, but contains, for example, configurations
with one jet recoiling against a bunch of particles laying on a conic surface.

In total we have introduced ten free parameters,
eight of which we vary to account for renormalization scale uncertainties. The default central values for other profile parameters are listed in Table~\ref{tab:theoryerr}. In the two lower sections of the table we also collected the parameters $\epsilon_1$ and $\epsilon_2$ that parametrize the non-singular statistical errors and the two unknown non-singular three loop constants $s_{23}$ and $s_{34}$ (see Sec.~\ref{sec:nonsing}). We note that apart from the default choice $e_H = 0.5$ and the two constants $s_{32}$, $s_{34}$ that specify the non-global structure of the three-loop soft function,
the central values and variations that we use here are essentially identical to those used for thrust in Ref.~\cite{Benitez:2024nav}.\footnote{In Ref.~\cite{Benitez:2024nav} $r_s$ and $e_J$ where varied in the slightly different ranges $[1.5,2.5]$ and $[-1.5,1.5]$, respectively, which would yield similar final results.}

These choices for the defaults and the variations of the dijet profile function parameters,
along with profile functions for the shoulder resummation~\cite{Bhattacharya:2023qet}, have been employed for the $\alpha_s$ fits from the HJM distribution carried out recently in Ref.~\cite{Benitez:2025vsp}.

\section{Results}\label{sec:results}

In this section we analyze our theoretical results, including the size of various components of the cross section, order-by-order convergence of the resummed perturbation theory, and correlation between $\alpha_s(m_Z)$ and $\Omega_1$.

The ingredients required for predicting the differential cross section at various (resummed) orders are shown
in Table~\ref{tab:ordercounting}. This includes the order for the cusp and non-cusp anomalous
dimensions for the factorization factors and functions $H$, $J_\tau$, and $\widehat S$; their perturbative order; the beta function $\beta(\alpha_s)$;
and the order for the non-singular corrections discussed in Sec.~\ref{sec:nonsing}. It also includes the R-anomalous dimension $\gamma_R$ and the renormalon subtraction $\delta(R)$ discussed in Sec.~\ref{sec:Rgap}. For our
analyses we use only primed orders, and we set the renormalization scales to the profile functions at the
distribution level. As discussed in Ref.~\cite{Almeida:2014uva}, in distribution scale setting, the unprimed orders
miss single logarithms in the usual logarithmic counting. The primed-order distribution scale-setting
properly resums the desired series of logarithms for $\rho$, and was also used in
Refs.~\cite{Ligeti:2008ac,Abbate:2010xh,Abbate:2012jh,Hoang:2014wka,Hoang:2015hka} to make predictions for
$B\to s\,\gamma$, thrust and C-parameter.

\begin{table}[t!]\centering
\begin{tabular}{c|ccccccc}
& cusp & non-cusp & matching, non-singular & $\beta[\alpha_s]$ & non-singular & $\gamma_\Delta^{\mu,R}$ & $\delta$ \\
\hline
LL & $1$ & - & tree & $1$ & - & - & - \\
NLL & $2$ & $1$ & tree & $2$ & - & $1$ & - \\
N$^2$LL & $3$ & $2$ & $1$ & $3$ & $1$ & $2$ & $1$ \\
N$^3$LL & $4$ & $3$ & $2$ & $4$ & $2$ & $3$ & $2$ \\
\hline
NLL$^\prime$ & $2$ & $1$ & $1$ & $2$ & $1$ & $1$ & $1$ \\
N$^2$LL$^\prime$ & $3$ & $2$ & $2$ & $3$ & $2$ & $2$ & $2$ \\
N$^3$LL$^\prime$ & $4$ & $3$ & $3$ & $4$ & $3$ & $3$ & $3$ \\
\end{tabular}
\caption{Loop corrections required for specified orders.}\label{tab:ordercounting}
\end{table}

\subsection{Components of the Cross Section and Transition Regions}
\label{sec:fartail}

\begin{figure*}[t]
\subfigure[]
{\includegraphics[width=0.48\textwidth]{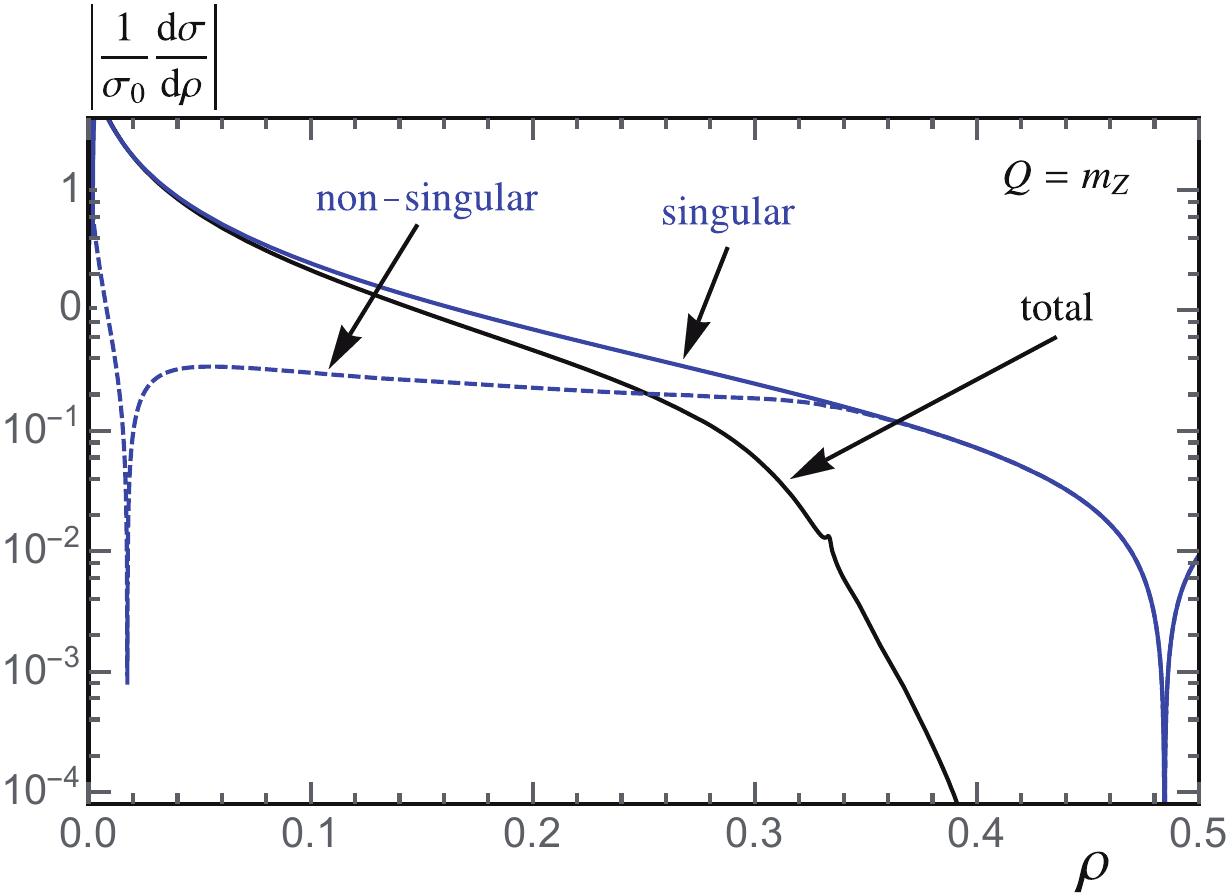}
\label{fig:FO-components}}\subfigure[]
{\includegraphics[width=0.48\textwidth]{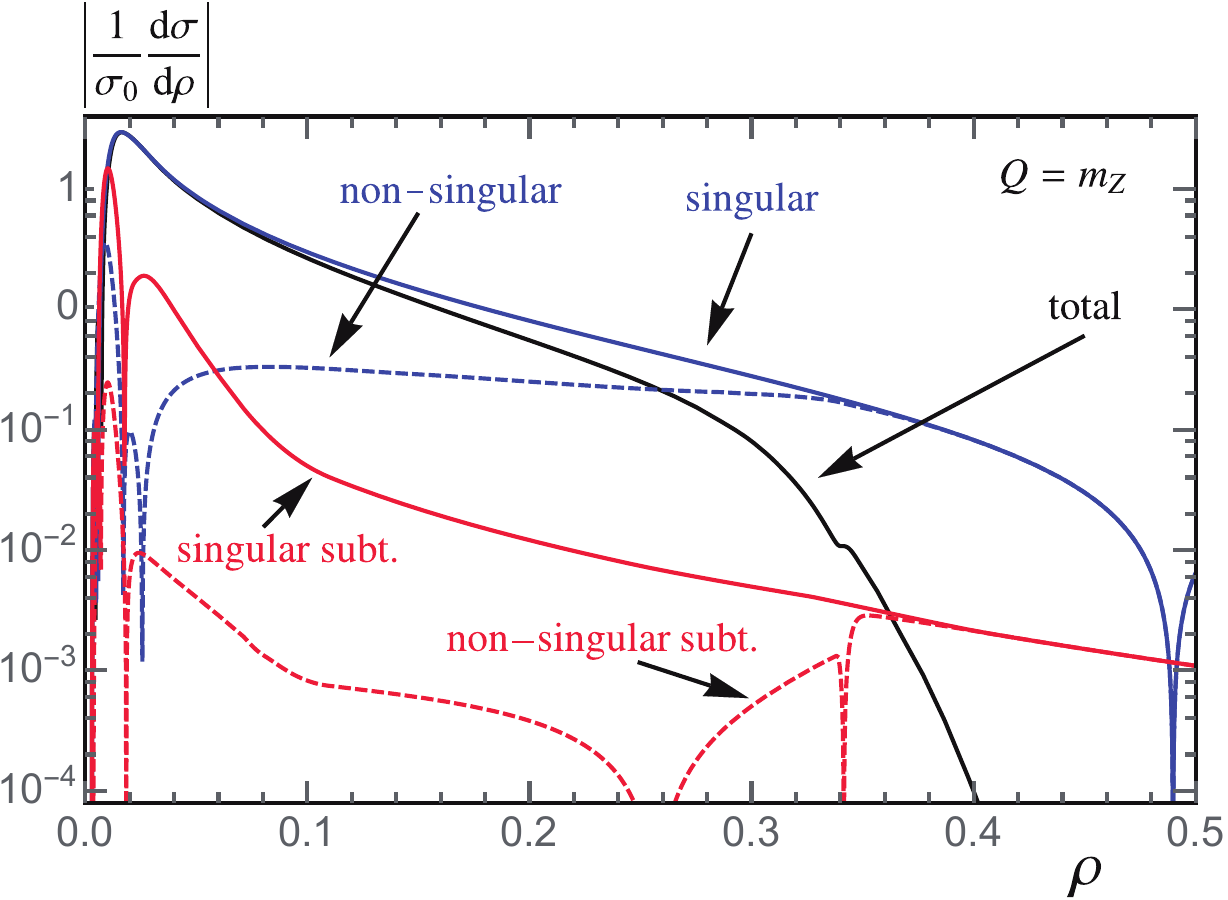}
\label{fig:Resum-components}}
\caption{Components of the HJM distribution for $Q=m_Z$ taking $\alpha_s(m_Z)=0.1181$ at fixed order (left) and with resummation (right).
The left plot (a) shows singular (solid blue) and non-singular (dashed blue) components of the fixed-order results at ${\cal O}(\alpha_s^3)$.
In the right plot,
the solid blue is resummed distribution at N$^3$LL$^\prime$ including the shape function but not matched to fixed order and not including the renormalon subtraction. The solid red is the effect the renormalon subtraction on the resummed distribution. The dashed blue is the non-singular distribution used to match to fixed order (also including the shape function) and the dashed red shows how the renormalon subtraction affects the non-singular. The solid black line is the final prediction, given by a sum of the other four distributions.
For this plot we use $\Omega_1=600\,\rm MeV$.}
\label{fig:components}
\end{figure*}
In Fig.~\ref{fig:components}, the resummed matched distribution is broken down into the different components. The left panel shows the singular and non-singular contributions at fixed-order at $\mathcal{O}(\alpha_s^3)$. The right panel shows the singular and non-singular for the fully resummed distribution including the shape function as well as the effect of the renormalon subtraction.
The distribution is N$^3$LL$^\prime+\,{\cal O}(\alpha_s^3)\,+\,\Omega_1(R,\mu)$ for $Q=m_Z$.
The components of
the cross section are: the singular (solid blue) and non-singular terms (dashed blue), and the contributions from terms
that involve the gap-subtraction and its $R$-revolution, for both singular (solid red) and non-singular subtractions
(dashed red), which we plot independently. The sum of these four components yields the total distribution
(solid black line). The hierarchy between singular and non-singular in the tail region is very similar to the case of
thrust and C-parameter, see Figs.~7, 12 and 4 in Refs.~\cite{Abbate:2010xh,Benitez:2024nav} and \cite{Hoang:2014wka}, respectively. However, the contribution from the non-singular terms for HJM (and thrust) are larger than they are for the C-parameter
in the region $\sim [0.15,0.33]$, both at fixed order and with resummation.
The subtractions are a small part of the cross section in the tail region but are essential at the level of precision
obtained in our prediction. Near $\rho\sim 0.33$ singular and non-singular (as well as their respective subtractions)
become almost equal in size but with opposite sign, rendering the total distribution very small. This becomes clear from
the figure where each individual singular and non-singular curve is much larger than the sum.
The shoulder region shows up near $\rho=0.33$ and is visible in the total distribution curve as a little hump.
Achieving complete theoretical control in this region requires Sudakov shoulder resummation which is understood~\cite{Catani:1998sf,Bhattacharya:2022dtm,Bhattacharya:2023qet,Benitez:2025vsp}, but not the focus of the current analysis.

\subsection{Perturbative Uncertainties from Theory Scans}
\label{sec:pertuncertainty}
\begin{figure*}[t!]
\subfigure[]
{\includegraphics[width=0.48\textwidth]{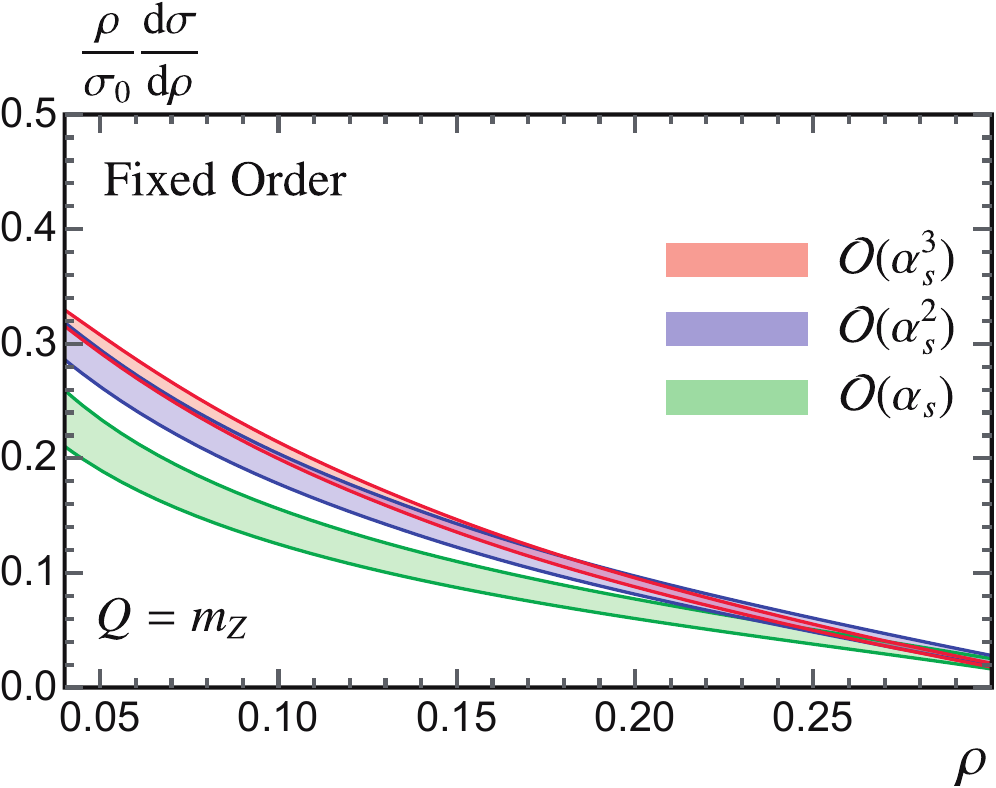}
\label{fig:tailbandFO}}
\subfigure[]{\includegraphics[width=0.48\textwidth]{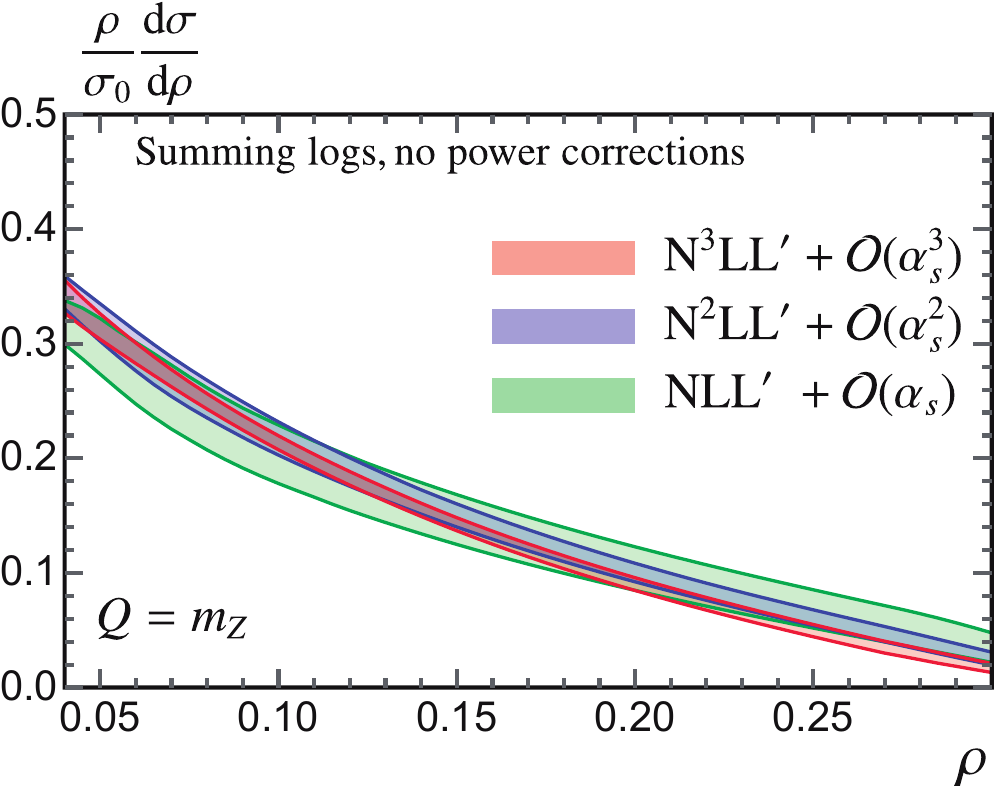}
\label{fig:tailbandnoF}}
\subfigure[]{\includegraphics[width=0.48\textwidth]{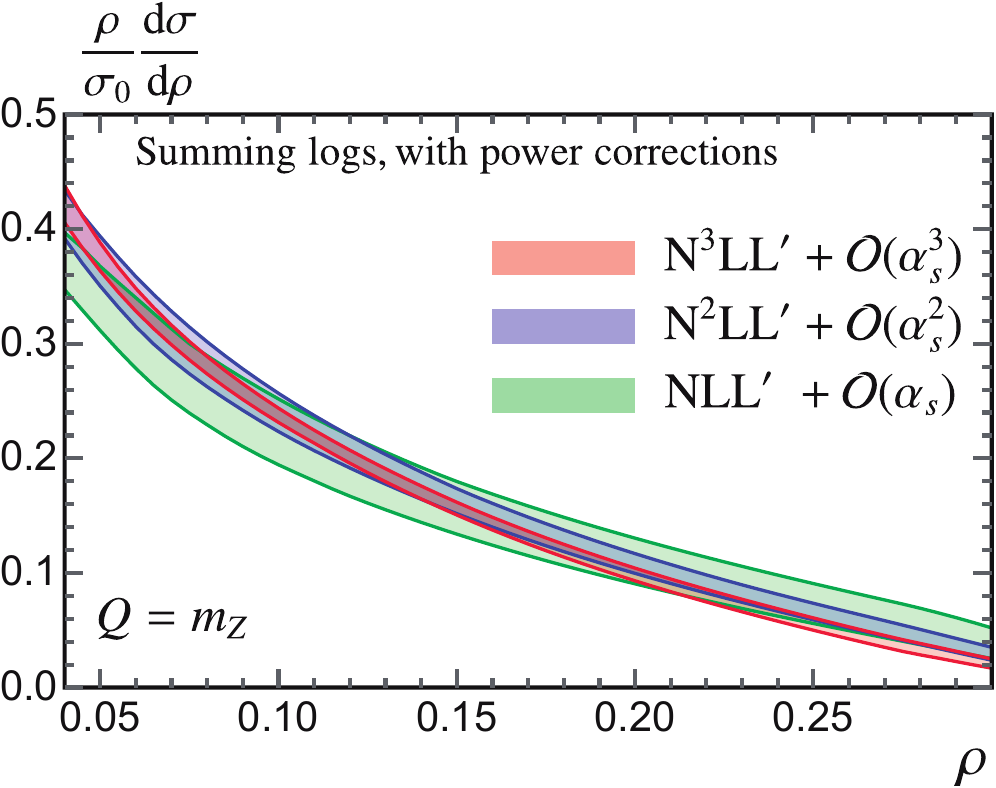}
\label{fig:tailbandnogap}}
\subfigure[]{\includegraphics[width=0.48\textwidth]{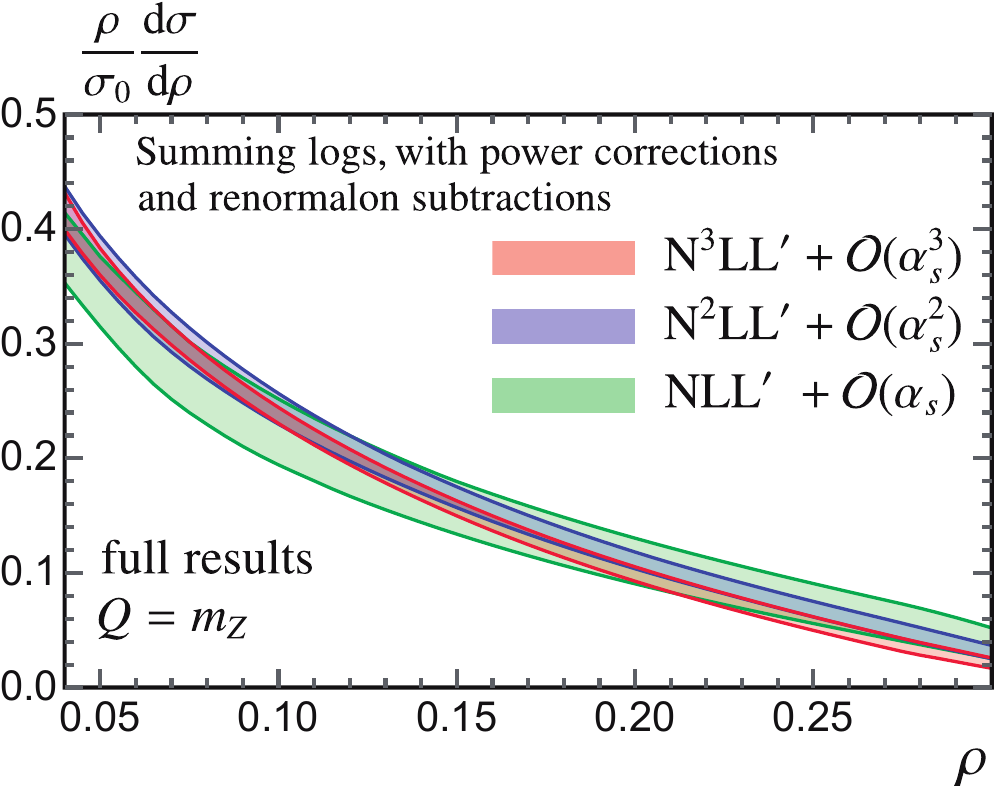}
\label{fig:tailbandwithgap}}
\caption{Theory scan for cross section uncertainties in the heavy jet mass distribution. The panels are
(a)~fixed order, (b)~resummation with no hadronization corrections, (c)~resummation with a shape function
using the $\msbar$ scheme for the leading power correction, without renormalon subtractions, and
(d)~resummation with a non-perturbative function using the R-gap scheme for $\Omega_1^\rho$, with
renormalon subtractions.}
\label{fig:4-plots}%
\end{figure*}

In Fig.~\ref{fig:4-plots} we show the results for the HJM differential cross sections in the tail region at $Q=m_Z$, for
various levels of theoretical refinement. All plots have been produced with \mbox{$\alpha_s(m_Z) = 0.1181$}. When non-perturbative corrections are included, given
that we show only the tail of the distribution, we can adopt the simplest two-dimensional shape
function, truncated with $N=0$ and with $\lambda = 0.5\,$GeV. When renormalon subtractions are included
in the R-gap scheme, the reference value for the gap parameter is
\mbox{$\bar\Delta(R_0=\mu_0=2\,{\rm GeV})=0.05\,$GeV}. With this setup one has
$\Omega_{1,0}=\lambda + \bar\Delta(R_0,\mu_0)$ and
\mbox{$\widehat\Upsilon_1=0.6367\lambda $
$+ \bar\Delta(R_0,\mu_0)/2$}.

In Fig.~\ref{fig:tailbandFO} we show the simplest setup, namely partonic fixed-order ${\cal O}(\alpha_s^k)$
QCD predictions, obtained setting all renormalization scales equal to $e_H\,Q$
(no resummation or power corrections included), for $k = 1, 2, 3$. The displayed uncertainty bands are
determined by the variation of the $e_H$ parameter only.
In the dijet region, we can see that there
is not good overlap between the $\mathcal{O}(\alpha_s)$ (green), $\mathcal{O}(\alpha_s^2)$ (blue), and
$\mathcal{O}(\alpha_s^3)$ (red) results, particularly for $\rho<0.15$, even though the uncertainty bands become smaller at higher orders. In this region, the fixed-order uncertainty bands do not reflect the true uncertainties since they do not account for the resummation of Sudakov shoulder logarithms, necessary to restore a good perturbative behavior. Once the perturbative resummation is included, as shown in Fig.~\ref{fig:tailbandnoF}, all curves nicely overlap in the dijet region.
We show the resummed results with N$^k$LL$^\prime+{\cal O}(\alpha_s^k)$
accuracy for $k = 1$ (green), $k = 2$ (blue), and $k = 3$ (red). The uncertainties for
$0.05<\rho<0.25$ are roughly constant for $k = 1$ ($10.5\%$) and $k = 2$ ($6.1\%$), while for $k=3$ they
vary between $5\%$ to $2.5\%$ in the dijet region.

For $\rho > 0.2$ the uncertainties actually overlap better for the fixed order results in Fig.~\ref{fig:tailbandFO} than those obtained from including dijet resummation in Fig.~\ref{fig:tailbandnoF}. This region can not be treated accurately without also considering shoulder resummation, as done in Refs.~\cite{Bhattacharya:2022dtm,Bhattacharya:2023qet}, which yields convergent and overlapping uncertainty bands.
In panel (c) the additional impact of the convolution with the non-perturbative shape function is shown. We see that it predominantly leads to an overall shift of the distribution without affecting much the order-by-order behavior.
Panel (d) includes soft renormalon subtractions in the R-gap scheme,
which might impact the perturbative accuracy. It is therefore instructive to compare our full setup, order by
order, with the results that use the $\msbar$ scheme for the hadronization corrections. At NLL$^\prime$ both
schemes have a similar accuracy, but including R-gap subtractions makes the relative uncertainty flatter and
slightly larger towards the peak. At N$^2$LL$^\prime$ uncertainties are $1.5\%$ smaller in average if renormalon
subtractions are performed, achieving $6\%$ precision at $\rho \sim 0.18$. At the highest order both schemes
have almost identical uncertainty bands, being the R-gap $<1\%$ more imprecise on average.
In other words, while the renormalon subtraction is important up to N$^2$LL$^\prime$, beyond that order it is not so critical for a precision prediction of the HJM distribution in the tail.

\subsection[Degeneracy between $\alpha_s$ and $\Omega_{1,0}$]
{Degeneracy between $\boldsymbol{\alpha_s}$ and $\boldsymbol{\Omega_{1,0}}$}
\label{sec:degeneracy}
\begin{figure*}[t!]
\begin{center}\!\!\!\!
\subfigure[]
{\includegraphics[width=0.315\textwidth]{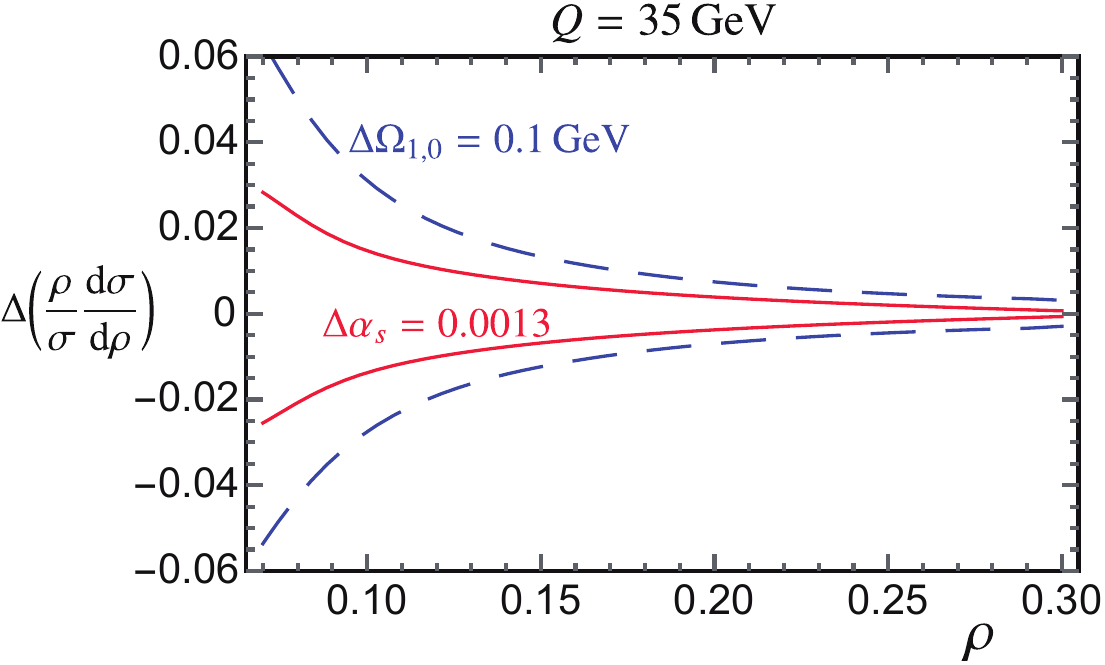}
\label{fig:35}}\!\!\!\!
\subfigure[]{\includegraphics[width=0.335\textwidth]{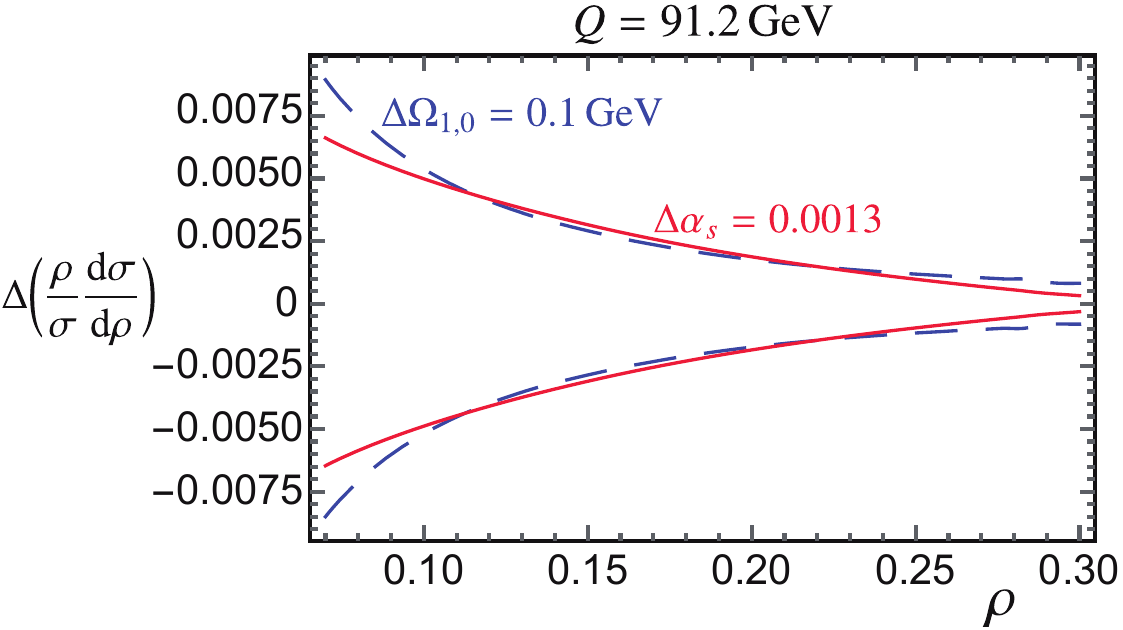}
\label{fig:91}}\!\!\!\!
\subfigure[]{\includegraphics[width=0.315\textwidth]{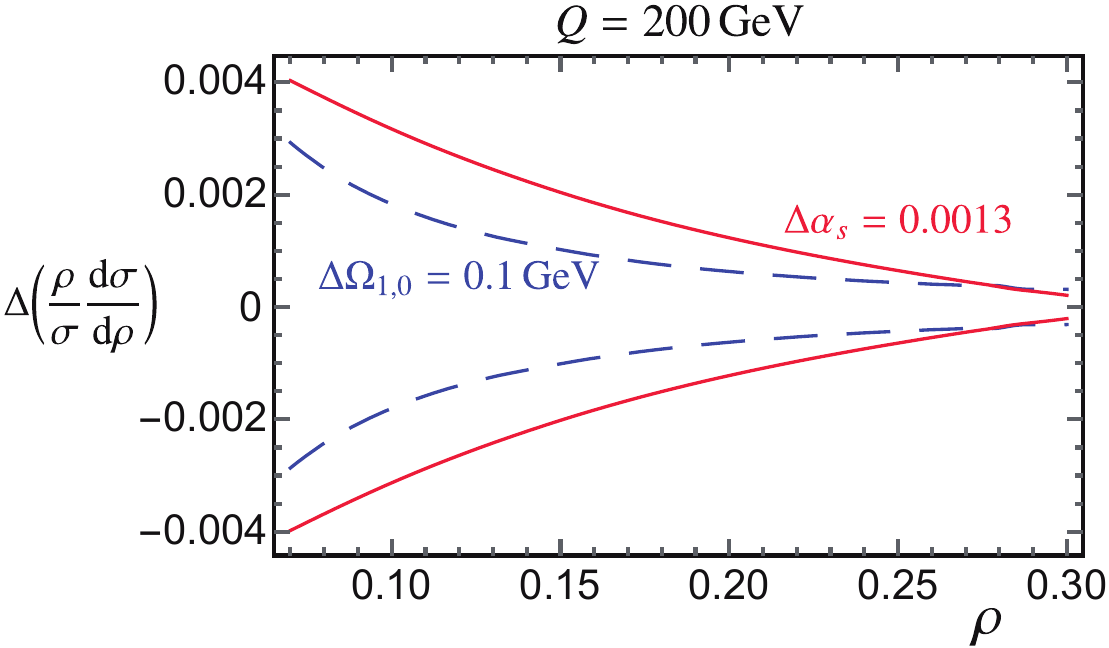}
\label{fig:200}}
\caption{\label{fig:degeneracy} Difference between the default normalized cross sections when a one fit parameter
is varied at a time: $\alpha_s(m_Z)$ by $\pm\, 0.0013$ (solid red) and $\Omega_{1,0}$ by $\pm \,0.1$\,GeV
(dashed blue). The three panels correspond to different center of mass energies: (left)~$Q=35$\,GeV,
(center)~$Q=91.2$\,GeV, (right)~$Q=200$\,GeV. }
\end{center}
\end{figure*}

Our results open the possibility of extracting the strong coupling constant with high precision fitting
experimental data in the tail
region, where the OPE is valid and can be truncated at leading power.
Much as it happened for thrust and C-parameter tail distributions, there is a strong degeneracy between the
two theoretical parameters relevant in this region, $\alpha_s$ and $\Omega_{1,0}$. This degeneracy depends on
the center of mass energy, as shown in Fig.~\ref{fig:degeneracy}, and therefore can be broken by including disparate $Q$ values into fits.

\section{Conclusions} \label{sec:conclusion}
Providing a precise theoretical description of the heavy jet mass distribution (HJM) requires resolving a number of novel complications not found in the simpler thrust observable.
These complications arise largely because the factorization theorem for HJM in the dijet region involves the two hemispheres in a way which does not admit the reduction to single-scale functions. Consequently, both the perturbative treatment of the soft function
and the non-perturbative treatment must be handled with particular care.

On the perturbative side, we have extended the accuracy of previous work~\cite{Catani:1991bd,Chien:2010kc} on the DHM and HJM distributions by providing an analytic formula for the resummation of logarithms in the two-loop dihemisphere soft function~\cite{Kelley:2011ng,Hornig:2011iu,Monni:2011gb}. We showed that the exact resummation, involving a challenging
integral in Fourier space, can in practice be replaced by a rapidly convergent expansion which involves only analytic expressions in terms of
a truncated series in powers of $\ln(y_1/y_2)$,
where $y_i$ are the Fourier-transform variables of the hemisphere masses. This `polar expansion'
can be carried out to reach arbitrary precision, and we have tested that if six terms are included the cross
section is already accurate to a one in $10^{7}$. The polar expansion allows one to sum up large logs in an
efficient way when a convolution with a two-dimensional shape function is also included. This approach only sums global logs, so for the double differential DHM distribution, it can only be applied when both hemisphere masses are of the same order. However, for the HJM distribution it is
an efficient way
to reach N${}^3$LL$^\prime$ accuracy. The method is general and can be applied to higher resummation order once the corresponding three-loop dihemisphere soft function is known.

On the non-perturbative side, we studied the OPE and provided an operator-based definition of the power corrections. We showed that even if a priori the entire dihemisphere non-singular cross section is needed when the partonic result is convolved with a shape
function, a simple ansatz allows to effectively include non-perturbative corrections in a way which correctly
reproduces the leading-power tail OPE and smoothly joins with the singular cross section in the peak region, where the
non-singular is strongly suppressed.
We have generalized the parametrization of a one-dimensional shape function in terms of a basis of orthogonal
polynomials to the case of a two-dimensional shape function as needed for DHM and HJM. Each element of our basis is the product of
two one-dimensional functions with different arguments.
Our parametrization also covers the more general
situation of non-symmetric shape functions (e.g.\ for cross sections differential in two angularities), but for DHM or HJM, only the symmetric formulation is needed.
We have discussed two ways how the infinite sum of functions can be truncated, one of them suited for a
tail analysis, the other more useful for a peak plus tail description.

On a practical side, we found that at leading power the OPE for HJM has the same form as the thrust OPE, so that the ${\cal O}(\Lambda_{\rm QCD})$ power corrections of these two variables are described by two analogous operator matrix elements $\Omega_1^\tau$ and $\Omega_1^\rho$.
As with thrust, the leading power correction for HJM can be defined in a scheme that subtracts the $u=1/2$ renormalon of the partonic soft function. Since thrust and HJM are in different universality classes due to hadron-mass effects, one should expect $\Omega_1^\tau$ and $\Omega_1^\rho$ to differ by an $\mathcal{O}(1)$ factor (Monte Carlo studies suggest that $\Omega_1^\tau\approx \Omega_1^\rho$, in contrast to the na\"\i ve expectation that $\Omega_1^\tau\approx 2\Omega_1^\rho$). When fitting for $\alpha_s$ using the tail of the distribution, where $\rho \gtrsim { \Lambda_{\rm QCD}}/{Q}$, a single non-perturbative parameter $\Omega_{1}^{\rho}$ should be sufficient. Beyond ${\cal O}(\Lambda_{\rm QCD})$ the OPE becomes more complicated than for thrust and C-parameter, involving partial derivatives of the partonic DHM cross section and the shape-function tail parameters $\Omega_{i,j}$. A full-fledged OPE does not exist for moments of the HJM distribution, but an inspection on the simplifications that occur when integrating over the full distribution with powers of $\rho$ reveals that these depend on a new kind of hadronic parameters which we call $\widehat\Upsilon_{i}$. These new power corrections have little effect in the tail, but can modify the peak position and the peak height of the distribution. Because of the complications of the HJM power corrections, trying to rigorously fit the HJM data in the peak region is expected to be challenging.

The framework established here provides a coherent, field-theory-based toolkit for high-precision studies of heavy-jet-mass observables and paves the way for a global fit of the strong coupling $\alpha_s$ using heavy-jet-mass data on par with recent thrust and C-parameter extractions. The extension to the far-tail region, where Sudakov-shoulder logarithms become important, can be achieved by matching this resummed dijet analysis with existing trijet factorization theorems. Such developments enable next-generation extractions of QCD parameters from event-shape measurements and deepen our insight into non-perturbative dynamics in high-energy collisions. A global $\alpha_s$ fit to HJM experimental data incorporating the advancements described here is
presented in a separate paper~\cite{Benitez:2025vsp}.

\acknowledgments
This work was supported in part by FWF Austrian Science Fund under the Project No. P32383-N27, the U.S.\ Department of Energy Office of Nuclear Physics under contract DE-SC0011090, the U.S.\ Department of Energy Office of High-energy Physics under contract DE-SC0013607, the Spanish MICIU/AEI/10.13039/501100011033 grant No.\ PID2022-141910NB-I00, the JCyL grant SA091P24 under program EDU/841/2024, the EU STRONG-2020 project under Program No.\ H2020-INFRAIA-2018-1, Grant Agreement No.\ 824093 and the Research in Teams project ``Improving the Precision of Predictions for Event-Shape Distributions" of the Erwin Schr\"odinger International Institute for Mathematics and Physics. IS was also supported by the Simons Foundation through the Investigator grant 327942, the Delta ITP consortium, a program of the Netherlands Organisation for Scientific Research (NWO) that is funded by the Dutch Ministry of Education, Culture and Science (OCW), and also thanks Nikhef and DESY for hospitality during the completion of this work.
We thank the University of Vienna for hospitality while parts of this work were completed. We are grateful to the Erwin-Schr\"odinger International Institute for Mathematics and Physics for partial support during the Thematic Programme “Quantum Field Theory at the Frontiers of the Strong Interactions”, July 31 - September 1, 2023.
We thank Thomas Gehrmann for providing us with numerical results from the program {\sc eerad3}, and Zolt\'an Tr\'ocs\'anyi for providing the numerical tables from the {\sc CoLoRFulNNLO} program.

\appendix

\section{Maximum value of HJM}
\label{app:maxrho}
\begin{figure}[t]
\centering
\reflectbox{\includegraphics[width=0.3\linewidth]{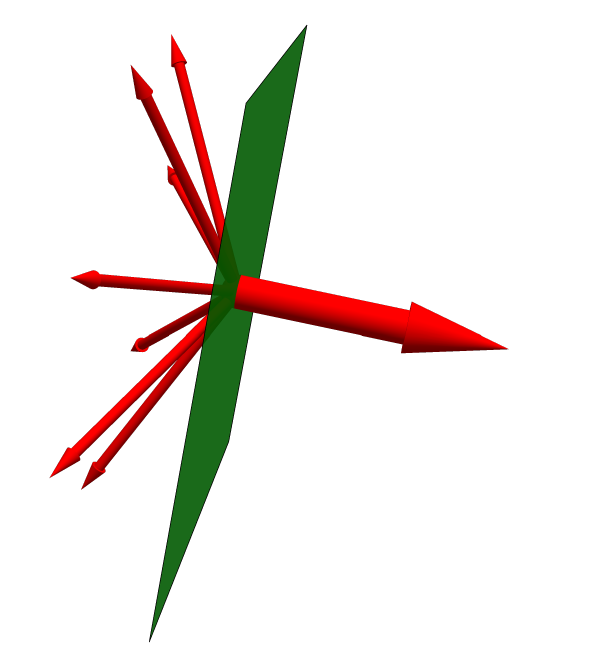}}
\hspace{10pt}
\includegraphics[width=0.5\linewidth]{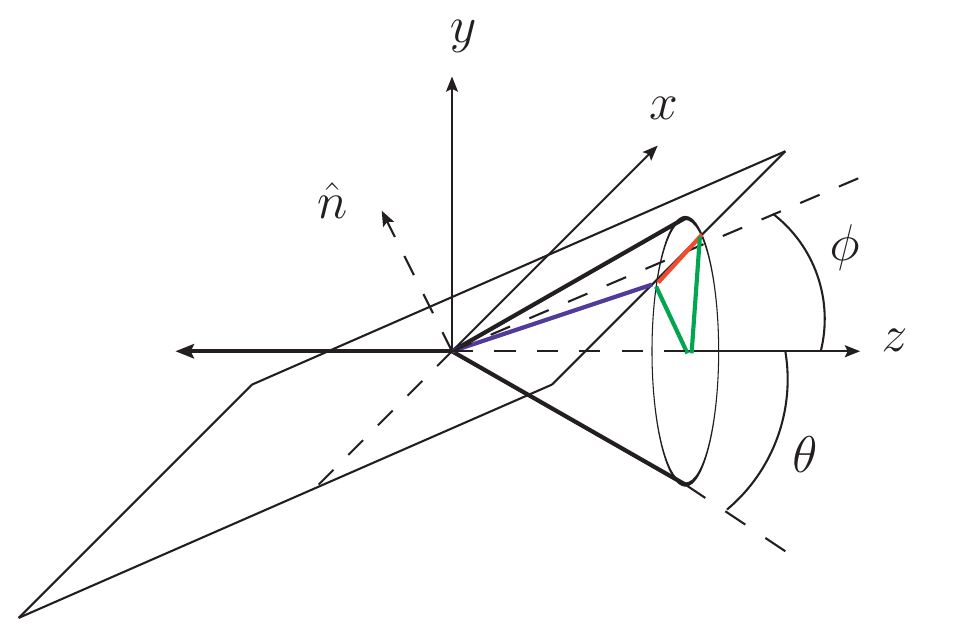}
\caption{The maximum possible value of $\rho$ for $n$ massless partons comes from configurations with $n-1$ partons in the heavy hemisphere equally distributed on the surface of a cone recoiling against a single parton in the light hemisphere, as shown on the left. If the cone angle is too large, the thrust axis jumps to a configuration like that on the right. The maximum value of $\rho$ for any collection of massless partons is $\rho_\text{max} \approx 0.477$.}
\label{fig:maxrho}
\end{figure}

The maximum value of $\rho$ is obtained from the configuration of $n$ massless particles shown in Fig. \ref{fig:maxrho}. These maximum values $\rho_{\text{max}}^n$ for $n$ massless partons are
\begin{align}
\rho_{\text{max}}^3 &= \frac{1}{3} \approx 0.3333\,,\\
\rho_{\text{max}}^4 &= \frac{1}{5} \left(7-2 \sqrt{6}\right) \approx 0.4202\,,\nonumber\\
\rho_{\text{max}}^5 &= \frac{1}{3} \left(4-\sqrt{7}\right) \approx 0.4514\,.\nonumber
\end{align}
The asymptotic limit as $n\to\infty$ is given by
\begin{equation}
\rho_{\rm max}^\infty = \tan^2 \biggl[ \frac{1}{2} \arctan \biggl( \sqrt{\pi^2 \cos^2 (\phi) + \tan^2(\phi)}\, \biggr)\! \biggr]\,,
\end{equation}
where $\phi$ satisfies:
\begin{equation}
\pi \sin (\phi) = \arcsin \Biggl[ \frac{\tan (\phi)}{\sqrt{\pi^2 \cos^2 (\phi)+ \tan^2 (\phi)}} \Biggr] + \frac{\pi}{2}\,.
\end{equation}
The numerical solution to this equation is $\phi = 0.6298$ leading to $\rho_\text{max}^{\infty} = 0.4770$.

If massive particles are allowed, a single massless particle can recoil against a particle with mass $Q$, both nearly at rest, giving $\rho_{\text{max}} = 1$.

\bibliography{thrust3}

\providecommand{\href}[2]{#2}\begingroup\raggedright\begin{thebibliography}{100}

\bibitem{GehrmannDeRidder:2007bj}
A.~{Gehrmann-De Ridder}, T.~Gehrmann, E.~W.~N. Glover and G.~Heinrich,
  \emph{{Second-order QCD corrections to the thrust distribution}},
  \href{http://dx.doi.org/10.1103/PhysRevLett.99.132002}{\emph{Phys. Rev.
  Lett.} {\bfseries 99} (2007) 132002},
  [\href{https://arxiv.org/abs/0707.1285}{{\ttfamily 0707.1285}}].

\bibitem{GehrmannDeRidder:2007hr}
A.~{Gehrmann-De Ridder}, T.~Gehrmann, E.~W.~N. Glover and G.~Heinrich,
  \emph{{NNLO corrections to event shapes in $e^+e^-$ annihilation}},
  \href{http://dx.doi.org/10.1088/1126-6708}{\emph{JHEP} {\bfseries 12} (2007)
  094}, [\href{https://arxiv.org/abs/0711.4711}{{\ttfamily 0711.4711}}].

\bibitem{Weinzierl:2008iv}
S.~Weinzierl, \emph{{NNLO corrections to 3-jet observables in electron-positron
  annihilation}},
  \href{http://dx.doi.org/10.1103/PhysRevLett.101.162001}{\emph{Phys. Rev.
  Lett.} {\bfseries 101} (2008) 162001},
  [\href{https://arxiv.org/abs/0807.3241}{{\ttfamily 0807.3241}}].

\bibitem{Gehrmann-DeRidder:2014hxk}
A.~Gehrmann-De~Ridder, T.~Gehrmann, E.~W.~N. Glover and G.~Heinrich,
  \emph{{EERAD3: Event shapes and jet rates in electron-positron annihilation
  at order $\alpha_s^3$}},
  \href{http://dx.doi.org/10.1016/j.cpc.2014.07.024}{\emph{Comput. Phys.
  Commun.} {\bfseries 185} (2014) 3331},
  [\href{https://arxiv.org/abs/1402.4140}{{\ttfamily 1402.4140}}].

\bibitem{Weinzierl:2009ms}
S.~Weinzierl, \emph{{Event shapes and jet rates in electron-positron
  annihilation at NNLO}},
  \href{http://dx.doi.org/10.1088/1126-6708}{\emph{JHEP} {\bfseries 06} (2009)
  041}, [\href{https://arxiv.org/abs/0904.1077}{{\ttfamily 0904.1077}}].

\bibitem{DelDuca:2016ily}
V.~Del~Duca, C.~Duhr, A.~Kardos, G.~Somogyi, Z.~Sz{\H o}r, Z.~Tr{\'o}cs{\'a}nyi
  et~al., \emph{{Jet production in the CoLoRFulNNLO method: event shapes in
  electron-positron collisions}},
  \href{http://dx.doi.org/10.1103/PhysRevD.94.074019}{\emph{Phys. Rev.}
  {\bfseries D94} (2016) 074019},
  [\href{https://arxiv.org/abs/1606.03453}{{\ttfamily 1606.03453}}].

\bibitem{Ellis:1980wv}
R.~K. Ellis, D.~A. Ross and A.~E. Terrano, \emph{{The Perturbative Calculation
  of Jet Structure in $e^+ e^-$ Annihilation}},
  \href{http://dx.doi.org/10.1016/0550-3213(81)90165-6}{\emph{Nucl. Phys.}
  {\bfseries B178} (1981) 421}.

\bibitem{Catani:1996jh}
S.~Catani and M.~H. Seymour, \emph{{The Dipole Formalism for the Calculation of
  QCD Jet Cross Sections at Next-to-Leading Order}},
  \href{http://dx.doi.org/10.1016/0370-2693(96)00425-X}{\emph{Phys. Lett.}
  {\bfseries B378} (1996) 287--301},
  [\href{https://arxiv.org/abs/hep-ph/9602277}{{\ttfamily hep-ph/9602277}}].

\bibitem{Catani:1996vz}
S.~Catani and M.~H. Seymour, \emph{{A general algorithm for calculating jet
  cross sections in NLO QCD}},
  \href{http://dx.doi.org/10.1016/S0550-3213(96)00589-5}{\emph{Nucl. Phys.}
  {\bfseries B485} (1997) 291--419},
  [\href{https://arxiv.org/abs/hep-ph/9605323}{{\ttfamily hep-ph/9605323}}].

\bibitem{Dixon:2018qgp}
L.~J. Dixon, M.-X. Luo, V.~Shtabovenko, T.-Z. Yang and H.~X. Zhu,
  \emph{{Analytical Computation of Energy-Energy Correlation at Next-to-Leading
  Order in QCD}},
  \href{http://dx.doi.org/10.1103/PhysRevLett.120.102001}{\emph{Phys. Rev.
  Lett.} {\bfseries 120} (2018) 102001},
  [\href{https://arxiv.org/abs/1801.03219}{{\ttfamily 1801.03219}}].

\bibitem{Fleming:2007qr}
S.~Fleming, A.~H. Hoang, S.~Mantry and I.~W. Stewart, \emph{{Jets from massive
  unstable particles: Top-mass determination}},
  \href{http://dx.doi.org/10.1103/PhysRevD.77.074010}{\emph{Phys. Rev.}
  {\bfseries D77} (2008) 074010},
  [\href{https://arxiv.org/abs/hep-ph/0703207}{{\ttfamily hep-ph/0703207}}].

\bibitem{Schwartz:2007ib}
M.~D. Schwartz, \emph{{Resummation and NLO Matching of Event Shapes with
  Effective Field Theory}},
  \href{http://dx.doi.org/10.1103/PhysRevD.77.014026}{\emph{Phys. Rev.}
  {\bfseries D77} (2008) 014026},
  [\href{https://arxiv.org/abs/0709.2709}{{\ttfamily 0709.2709}}].

\bibitem{Bauer:2008dt}
C.~W. Bauer, S.~P. Fleming, C.~Lee and G.~F. Sterman, \emph{{Factorization of
  $e^+e^-$ Event Shape Distributions with Hadronic Final States in Soft
  Collinear Effective Theory}},
  \href{http://dx.doi.org/10.1103/PhysRevD.78.034027}{\emph{Phys. Rev.}
  {\bfseries D78} (2008) 034027},
  [\href{https://arxiv.org/abs/0801.4569}{{\ttfamily 0801.4569}}].

\bibitem{Bauer:2000ew}
C.~W. Bauer, S.~Fleming and M.~E. Luke, \emph{{Summing Sudakov logarithms in $B
  \to X_s \gamma$ in effective field theory}},
  \href{http://dx.doi.org/10.1103/PhysRevD.63.014006}{\emph{Phys. Rev.}
  {\bfseries D63} (2000) 014006},
  [\href{https://arxiv.org/abs/hep-ph/0005275}{{\ttfamily hep-ph/0005275}}].

\bibitem{Bauer:2000yr}
C.~W. Bauer, S.~Fleming, D.~Pirjol and I.~W. Stewart, \emph{{An Effective field
  theory for collinear and soft gluons: Heavy to light decays}},
  \href{http://dx.doi.org/10.1103/PhysRevD.63.114020}{\emph{Phys. Rev.}
  {\bfseries D63} (2001) 114020},
  [\href{https://arxiv.org/abs/hep-ph/0011336}{{\ttfamily hep-ph/0011336}}].

\bibitem{Bauer:2001ct}
C.~W. Bauer and I.~W. Stewart, \emph{{Invariant operators in collinear
  effective theory}},
  \href{http://dx.doi.org/10.1016/S0370-2693(01)00902-9}{\emph{Phys. Lett.}
  {\bfseries B516} (2001) 134--142},
  [\href{https://arxiv.org/abs/hep-ph/0107001}{{\ttfamily hep-ph/0107001}}].

\bibitem{Bauer:2001yt}
C.~W. Bauer, D.~Pirjol and I.~W. Stewart, \emph{{Soft-Collinear Factorization
  in Effective Field Theory}},
  \href{http://dx.doi.org/10.1103/PhysRevD.65.054022}{\emph{Phys. Rev.}
  {\bfseries D65} (2002) 054022},
  [\href{https://arxiv.org/abs/hep-ph/0109045}{{\ttfamily hep-ph/0109045}}].

\bibitem{Bauer:2002nz}
C.~W. Bauer, S.~Fleming, D.~Pirjol, I.~Z. Rothstein and I.~W. Stewart,
  \emph{{Hard scattering factorization from effective field theory}},
  \href{http://dx.doi.org/10.1103/PhysRevD.66.014017}{\emph{Phys. Rev.}
  {\bfseries D66} (2002) 014017},
  [\href{https://arxiv.org/abs/hep-ph/0202088}{{\ttfamily hep-ph/0202088}}].

\bibitem{Becher:2008cf}
T.~Becher and M.~D. Schwartz, \emph{{A Precise determination of $\alpha_s$ from
  LEP thrust data using effective field theory}},
  \href{http://dx.doi.org/10.1088/1126-6708/2008/07/034}{\emph{JHEP} {\bfseries
  07} (2008) 034}, [\href{https://arxiv.org/abs/0803.0342}{{\ttfamily
  0803.0342}}].

\bibitem{Hoang:2014wka}
A.~H. Hoang, D.~W. Kolodrubetz, V.~Mateu and I.~W. Stewart,
  \emph{{$C$-parameter distribution at N$^3$LL$^\prime$ including power
  corrections}},
  \href{http://dx.doi.org/10.1103/PhysRevD.91.094017}{\emph{Phys. Rev.}
  {\bfseries D91} (2015) 094017},
  [\href{https://arxiv.org/abs/1411.6633}{{\ttfamily 1411.6633}}].

\bibitem{Moult:2018jzp}
I.~Moult and H.~X. Zhu, \emph{{Simplicity from Recoil: The Three-Loop Soft
  Function and Factorization for the Energy-Energy Correlation}},
  \href{http://dx.doi.org/10.1007/JHEP08(2018)160}{\emph{JHEP} {\bfseries 08}
  (2018) 160}, [\href{https://arxiv.org/abs/1801.02627}{{\ttfamily
  1801.02627}}].

\bibitem{Chien:2010kc}
Y.-T. Chien and M.~D. Schwartz, \emph{{Resummation of heavy jet mass and
  comparison to LEP data}},
  \href{http://dx.doi.org/10.1007/JHEP08(2010)058}{\emph{JHEP} {\bfseries 08}
  (2010) 058}, [\href{https://arxiv.org/abs/1005.1644}{{\ttfamily 1005.1644}}].

\bibitem{Hornig:2009vb}
A.~Hornig, C.~Lee and G.~Ovanesyan, \emph{{Effective Predictions of Event
  Shapes: Factorized, Resummed, and Gapped Angularity Distributions}},
  \href{http://dx.doi.org/10.1088/1126-6708}{\emph{JHEP} {\bfseries 05} (2009)
  122}, [\href{https://arxiv.org/abs/0901.3780}{{\ttfamily 0901.3780}}].

\bibitem{Bell:2018gce}
G.~Bell, A.~Hornig, C.~Lee and J.~Talbert, \emph{{$e^+ e^-$ angularity
  distributions at NNLL$^\prime$ accuracy}},
  \href{http://dx.doi.org/10.1007/JHEP01(2019)147}{\emph{JHEP} {\bfseries 01}
  (2019) 147}, [\href{https://arxiv.org/abs/1808.07867}{{\ttfamily
  1808.07867}}].

\bibitem{Becher:2012qc}
T.~Becher and G.~Bell, \emph{{NNLL Resummation for Jet Broadening}},
  \href{http://dx.doi.org/10.1007/JHEP11(2012)126}{\emph{JHEP} {\bfseries 1211}
  (2012) 126}, [\href{https://arxiv.org/abs/1210.0580}{{\ttfamily 1210.0580}}].

\bibitem{Becher:2011pf}
T.~Becher, G.~Bell and M.~Neubert, \emph{{Factorization and Resummation for Jet
  Broadening}},
  \href{http://dx.doi.org/10.1016/j.physletb.2011.09.005}{\emph{Phys. Lett.}
  {\bfseries B704} (2011) 276--283},
  [\href{https://arxiv.org/abs/1104.4108}{{\ttfamily 1104.4108}}].

\bibitem{Chiu:2011qc}
J.-y. Chiu, A.~Jain, D.~Neill and I.~Z. Rothstein, \emph{{The Rapidity
  Renormalization Group}},
  \href{http://dx.doi.org/10.1103/PhysRevLett.108.151601}{\emph{Phys. Rev.
  Lett.} {\bfseries 108} (2012) 151601},
  [\href{https://arxiv.org/abs/1104.0881}{{\ttfamily 1104.0881}}].

\bibitem{Chiu:2012ir}
J.-Y. Chiu, A.~Jain, D.~Neill and I.~Z. Rothstein, \emph{{A Formalism for the
  Systematic Treatment of Rapidity Logarithms in Quantum Field Theory}},
  \href{http://dx.doi.org/10.1007/JHEP05(2012)084}{\emph{JHEP} {\bfseries 1205}
  (2012) 084}, [\href{https://arxiv.org/abs/1202.0814}{{\ttfamily 1202.0814}}].

\bibitem{Collins:1981uk}
J.~C. Collins and D.~E. Soper, \emph{{Back-To-Back Jets in QCD}},
  \href{http://dx.doi.org/10.1016/0550-3213(81)90339-4}{\emph{Nucl. Phys.}
  {\bfseries B193} (1981) 381}.

\bibitem{Korchemsky:1998ev}
G.~P. Korchemsky, \emph{{Shape functions and power corrections to the event
  shapes}},  \href{https://arxiv.org/abs/hep-ph/9806537}{{\ttfamily
  hep-ph/9806537}}.

\bibitem{Korchemsky:1999kt}
G.~P. Korchemsky and G.~Sterman, \emph{{Power corrections to event shapes and
  factorization}},
  \href{http://dx.doi.org/10.1016/S0550-3213(99)00308-9}{\emph{Nucl. Phys.}
  {\bfseries B555} (1999) 335--351},
  [\href{https://arxiv.org/abs/hep-ph/9902341}{{\ttfamily hep-ph/9902341}}].

\bibitem{Korchemsky:2000kp}
G.~P. Korchemsky and S.~Tafat, \emph{{On power corrections to the event shape
  distributions in QCD}},
  \href{http://dx.doi.org/10.1088/1126-6708/2000/10/010}{\emph{JHEP} {\bfseries
  10} (2000) 010}, [\href{https://arxiv.org/abs/hep-ph/0007005}{{\ttfamily
  hep-ph/0007005}}].

\bibitem{Berger:2003iw}
C.~F. Berger, T.~K{\'u}cs and G.~Sterman, \emph{{Event shape / energy flow
  correlations}},
  \href{http://dx.doi.org/10.1103/PhysRevD.68.014012}{\emph{Phys. Rev.}
  {\bfseries D68} (2003) 014012},
  [\href{https://arxiv.org/abs/hep-ph/0303051}{{\ttfamily hep-ph/0303051}}].

\bibitem{Catani:1992ua}
S.~Catani, L.~Trentadue, G.~Turnock and B.~R. Webber, \emph{{Resummation of
  large logarithms in $e^+\, e^-$ event shape distributions}},
  \href{http://dx.doi.org/10.1016/0550-3213(93)90271-P}{\emph{Nucl. Phys.}
  {\bfseries B407} (1993) 3--42}.

\bibitem{Banfi:2004yd}
A.~Banfi, G.~P. Salam and G.~Zanderighi, \emph{{Principles of general
  final-state resummation and automated implementation}},
  \href{http://dx.doi.org/10.1088/1126-6708/2005/03/073}{\emph{JHEP} {\bfseries
  03} (2005) 073}, [\href{https://arxiv.org/abs/hep-ph/0407286}{{\ttfamily
  hep-ph/0407286}}].

\bibitem{deFlorian:2004mp}
D.~de~Florian and M.~Grazzini, \emph{{The back-to-back region in $e^+ e^-$
  energy energy correlation}},
  \href{http://dx.doi.org/10.1016/j.nuclphysb.2004.10.051}{\emph{Nucl. Phys.}
  {\bfseries B704} (2005) 387--403},
  [\href{https://arxiv.org/abs/hep-ph/0407241}{{\ttfamily hep-ph/0407241}}].

\bibitem{Banfi:2014sua}
A.~Banfi, H.~McAslan, P.~F. Monni and G.~Zanderighi, \emph{{A general method
  for the resummation of event-shape distributions in $e^+ e^-$ annihilation}},
  \href{http://dx.doi.org/10.1007/JHEP05(2015)102}{\emph{JHEP} {\bfseries 05}
  (2015) 102}, [\href{https://arxiv.org/abs/1412.2126}{{\ttfamily 1412.2126}}].

\bibitem{Bell:2018vaa}
G.~Bell, R.~Rahn and J.~Talbert, \emph{{Two-loop anomalous dimensions of
  generic dijet soft functions}},
  \href{http://dx.doi.org/10.1016/j.nuclphysb.2018.09.026}{\emph{Nucl. Phys. B}
  {\bfseries 936} (2018) 520--541},
  [\href{https://arxiv.org/abs/1805.12414}{{\ttfamily 1805.12414}}].

\bibitem{Bell:2018oqa}
G.~Bell, R.~Rahn and J.~Talbert, \emph{{Generic dijet soft functions at
  two-loop order: correlated emissions}},
  \href{http://dx.doi.org/10.1007/JHEP07(2019)101}{\emph{JHEP} {\bfseries 07}
  (2019) 101}, [\href{https://arxiv.org/abs/1812.08690}{{\ttfamily
  1812.08690}}].

\bibitem{Bell:2020yzz}
G.~Bell, R.~Rahn and J.~Talbert, \emph{{Generic dijet soft functions at
  two-loop order: uncorrelated emissions}},
  \href{http://dx.doi.org/10.1007/JHEP09(2020)015}{\emph{JHEP} {\bfseries 09}
  (2020) 015}, [\href{https://arxiv.org/abs/2004.08396}{{\ttfamily
  2004.08396}}].

\bibitem{Bell:2021dpb}
G.~Bell, K.~Brune, G.~Das and M.~Wald, \emph{{Automation of Beam and Jet
  functions at NNLO}},
  \href{http://dx.doi.org/10.21468/SciPostPhysProc.7.021}{\emph{SciPost Phys.
  Proc.} {\bfseries 7} (2022) 021},
  [\href{https://arxiv.org/abs/2110.04804}{{\ttfamily 2110.04804}}].

\bibitem{Bhattacharya:2022dtm}
A.~Bhattacharya, M.~D. Schwartz and X.~Zhang, \emph{{Sudakov shoulder
  resummation for thrust and heavy jet mass}},
  \href{http://dx.doi.org/10.1103/PhysRevD.106.074011}{\emph{Phys. Rev. D}
  {\bfseries 106} (2022) 074011},
  [\href{https://arxiv.org/abs/2205.05702}{{\ttfamily 2205.05702}}].

\bibitem{Bhattacharya:2023qet}
A.~Bhattacharya, J.~K.~L. Michel, M.~D. Schwartz, I.~W. Stewart and X.~Zhang,
  \emph{{NNLL resummation of Sudakov shoulder logarithms in the heavy jet mass
  distribution}}, \href{http://dx.doi.org/10.1007/JHEP11(2023)080}{\emph{JHEP}
  {\bfseries 11} (2023) 080},
  [\href{https://arxiv.org/abs/2306.08033}{{\ttfamily 2306.08033}}].

\bibitem{Catani:1997xc}
S.~Catani and B.~Webber, \emph{{Infrared safe but infinite: Soft gluon
  divergences inside the physical region}}, {\emph{JHEP} {\bfseries 9710}
  (1997) 005}, [\href{https://arxiv.org/abs/hep-ph/9710333}{{\ttfamily
  hep-ph/9710333}}].

\bibitem{Luisoni:2020efy}
G.~Luisoni, P.~F. Monni and G.~P. Salam, \emph{{$C$-parameter hadronisation in
  the symmetric 3-jet limit and impact on $\alpha_s$ fits}},
  \href{http://dx.doi.org/10.1140/epjc/s10052-021-08941-z}{\emph{Eur. Phys. J.
  C} {\bfseries 81} (2021) 158},
  [\href{https://arxiv.org/abs/2012.00622}{{\ttfamily 2012.00622}}].

\bibitem{Caola:2021kzt}
F.~Caola, S.~Ferrario~Ravasio, G.~Limatola, K.~Melnikov and P.~Nason, \emph{{On
  linear power corrections in certain collider observables}},
  \href{http://dx.doi.org/10.1007/JHEP01(2022)093}{\emph{JHEP} {\bfseries 01}
  (2022) 093}, [\href{https://arxiv.org/abs/2108.08897}{{\ttfamily
  2108.08897}}].

\bibitem{Caola:2022vea}
F.~Caola, S.~Ferrario~Ravasio, G.~Limatola, K.~Melnikov, P.~Nason and M.~A.
  Ozcelik, \emph{{Linear power corrections to e$^{+}$e$^{-}$ shape variables in
  the three-jet region}},
  \href{http://dx.doi.org/10.1007/JHEP12(2022)062}{\emph{JHEP} {\bfseries 12}
  (2022) 062}, [\href{https://arxiv.org/abs/2204.02247}{{\ttfamily
  2204.02247}}].

\bibitem{Hagiwara:2010cd}
K.~Hagiwara and G.~Kirilin, \emph{{Angular distribution of thrust axis with
  power-suppressed contribution in $e^+e^-$ annihilation}},
  \href{http://dx.doi.org/10.1007/JHEP10(2010)093}{\emph{JHEP} {\bfseries 10}
  (2010) 093}, [\href{https://arxiv.org/abs/1006.5330}{{\ttfamily 1006.5330}}].

\bibitem{Mateu:2013gya}
V.~Mateu and G.~Rodrigo, \emph{{Oriented Event Shapes at N$^{3}$LL$\, +\,
  O(\alpha_s^{2})$}},
  \href{http://dx.doi.org/10.1007/JHEP11(2013)030}{\emph{JHEP} {\bfseries 1311}
  (2013) 030}, [\href{https://arxiv.org/abs/1307.3513}{{\ttfamily 1307.3513}}].

\bibitem{Bris:2022cdr}
A.~Bris, N.~G. Gracia and V.~Mateu, \emph{{NLO oriented event-shape
  distributions for massive quarks}},
  \href{http://dx.doi.org/10.1007/JHEP02(2023)247}{\emph{JHEP} {\bfseries 02}
  (2023) 247}, [\href{https://arxiv.org/abs/2211.10239}{{\ttfamily
  2211.10239}}].

\bibitem{Fleming:2007xt}
S.~Fleming, A.~H. Hoang, S.~Mantry and I.~W. Stewart, \emph{{Top Jets in the
  Peak Region: Factorization Analysis with NLL Resummation}},
  \href{http://dx.doi.org/10.1103/PhysRevD.77.114003}{\emph{Phys. Rev.}
  {\bfseries D77} (2008) 114003},
  [\href{https://arxiv.org/abs/0711.2079}{{\ttfamily 0711.2079}}].

\bibitem{Lepenik:2019jjk}
C.~Lepenik and V.~Mateu, \emph{{NLO Massive Event-Shape Differential and
  Cumulative Distributions}},
  \href{http://dx.doi.org/10.1007/JHEP03(2020)024}{\emph{JHEP} {\bfseries 03}
  (2020) 024}, [\href{https://arxiv.org/abs/1912.08211}{{\ttfamily
  1912.08211}}].

\bibitem{Bachu:2020nqn}
B.~Bachu, A.~H. Hoang, V.~Mateu, A.~Pathak and I.~W. Stewart, \emph{{Boosted
  top quarks in the peak region with N$^3$LL resummation}},
  \href{http://dx.doi.org/10.1103/PhysRevD.104.014026}{\emph{Phys. Rev. D}
  {\bfseries 104} (2021) 014026},
  [\href{https://arxiv.org/abs/2012.12304}{{\ttfamily 2012.12304}}].

\bibitem{Bris:2020uyb}
A.~Bris, V.~Mateu and M.~Preisser, \emph{{Massive event-shape distributions at
  N$^2$LL}}, \href{http://dx.doi.org/10.1007/JHEP09(2020)132}{\emph{JHEP}
  {\bfseries 09} (2020) 132},
  [\href{https://arxiv.org/abs/2006.06383}{{\ttfamily 2006.06383}}].

\bibitem{Gritschacher:2013pha}
S.~Gritschacher, A.~H. Hoang, I.~Jemos and P.~Pietrulewicz, \emph{{Secondary
  Heavy Quark Production in Jets through Mass Modes}},
  \href{http://dx.doi.org/10.1103/PhysRevD.88.034021}{\emph{Phys. Rev.}
  {\bfseries D88} (2013) 034021},
  [\href{https://arxiv.org/abs/1302.4743}{{\ttfamily 1302.4743}}].

\bibitem{Pietrulewicz:2014qza}
P.~Pietrulewicz, S.~Gritschacher, A.~H. Hoang, I.~Jemos and V.~Mateu,
  \emph{{Variable Flavor Number Scheme for Final State Jets in Thrust}},
  \href{http://dx.doi.org/10.1103/PhysRevD.90.114001}{\emph{Phys.Rev.}
  {\bfseries D90} (2014) 114001},
  [\href{https://arxiv.org/abs/1405.4860}{{\ttfamily 1405.4860}}].

\bibitem{Bris:2024bcq}
A.~Bris and V.~Mateu, \emph{{Secondary massive quarks with the Mellin-Barnes
  expansion}}, \href{http://dx.doi.org/10.1007/JHEP05(2024)146}{\emph{JHEP}
  {\bfseries 05} (2024) 146},
  [\href{https://arxiv.org/abs/2402.09536}{{\ttfamily 2402.09536}}].

\bibitem{Hoang:2018zrp}
A.~H. Hoang, S.~Pl\"atzer and D.~Samitz, \emph{{On the Cutoff Dependence of the
  Quark Mass Parameter in Angular Ordered Parton Showers}},
  \href{http://dx.doi.org/10.1007/JHEP10(2018)200}{\emph{JHEP} {\bfseries 10}
  (2018) 200}, [\href{https://arxiv.org/abs/1807.06617}{{\ttfamily
  1807.06617}}].

\bibitem{Hoang:2024nqi}
A.~H. Hoang, O.~L. Jin, S.~Pl{\"a}tzer and D.~Samitz, \emph{{Matching
  hadronization and perturbative evolution: the cluster model in light of
  infrared shower cutoff dependence}},
  \href{http://dx.doi.org/10.1007/JHEP07(2025)005}{\emph{JHEP} {\bfseries 07}
  (2025) 005}, [\href{https://arxiv.org/abs/2404.09856}{{\ttfamily
  2404.09856}}].

\bibitem{Freedman:2013vya}
S.~M. Freedman, \emph{{Subleading Corrections To Thrust Using Effective Field
  Theory}},  \href{https://arxiv.org/abs/1303.1558}{{\ttfamily 1303.1558}}.

\bibitem{Freedman:2014uta}
S.~M. Freedman and R.~Goerke, \emph{{Renormalization of Subleading Dijet
  Operators in Soft-Collinear Effective Theory}},
  \href{http://dx.doi.org/10.1103/PhysRevD.90.114010}{\emph{Phys. Rev.}
  {\bfseries D90} (2014) 114010},
  [\href{https://arxiv.org/abs/1408.6240}{{\ttfamily 1408.6240}}].

\bibitem{Boughezal:2016zws}
R.~Boughezal, X.~Liu and F.~Petriello, \emph{{Power Corrections in the
  N-jettiness Subtraction Scheme}},
  \href{http://dx.doi.org/10.1007/JHEP03(2017)160}{\emph{JHEP} {\bfseries 03}
  (2017) 160}, [\href{https://arxiv.org/abs/1612.02911}{{\ttfamily
  1612.02911}}].

\bibitem{Moult:2016fqy}
I.~Moult, L.~Rothen, I.~W. Stewart, F.~J. Tackmann and H.~X. Zhu,
  \emph{{Subleading Power Corrections for N-Jettiness Subtractions}},
  \href{http://dx.doi.org/10.1103/PhysRevD.95.074023}{\emph{Phys. Rev.}
  {\bfseries D95} (2017) 074023},
  [\href{https://arxiv.org/abs/1612.00450}{{\ttfamily 1612.00450}}].

\bibitem{Feige:2017zci}
I.~Feige, D.~W. Kolodrubetz, I.~Moult and I.~W. Stewart, \emph{{A Complete
  Basis of Helicity Operators for Subleading Factorization}},
  \href{http://dx.doi.org/10.1007/JHEP11(2017)142}{\emph{JHEP} {\bfseries 11}
  (2017) 142}, [\href{https://arxiv.org/abs/1703.03411}{{\ttfamily
  1703.03411}}].

\bibitem{Moult:2017rpl}
I.~Moult, I.~W. Stewart and G.~Vita, \emph{{A subleading operator basis and
  matching for gg $\to$ H}},
  \href{http://dx.doi.org/10.1007/JHEP07(2017)067}{\emph{JHEP} {\bfseries 07}
  (2017) 067}, [\href{https://arxiv.org/abs/1703.03408}{{\ttfamily
  1703.03408}}].

\bibitem{Moult:2017jsg}
I.~Moult, L.~Rothen, I.~W. Stewart, F.~J. Tackmann and H.~X. Zhu,
  \emph{{N-jettiness subtractions for $gg\to H$ at subleading power}},
  \href{http://dx.doi.org/10.1103/PhysRevD.97.014013}{\emph{Phys. Rev.}
  {\bfseries D97} (2018) 014013},
  [\href{https://arxiv.org/abs/1710.03227}{{\ttfamily 1710.03227}}].

\bibitem{Goerke:2017lei}
R.~Goerke and M.~Inglis-Whalen, \emph{{Renormalization of Dijet Operators at
  Order $1/Q^2$ in Soft-Collinear Effective Theory}},
  \href{http://dx.doi.org/10.1007/JHEP05(2018)023}{\emph{JHEP} {\bfseries 05}
  (2018) 023}, [\href{https://arxiv.org/abs/1711.09147}{{\ttfamily
  1711.09147}}].

\bibitem{Beneke:2017ztn}
M.~Beneke, M.~Garny, R.~Szafron and J.~Wang, \emph{{Anomalous dimension of
  subleading-power N-jet operators}},
  \href{http://dx.doi.org/10.1007/JHEP03(2018)001}{\emph{JHEP} {\bfseries 03}
  (2018) 001}, [\href{https://arxiv.org/abs/1712.04416}{{\ttfamily
  1712.04416}}].

\bibitem{Moult:2018jjd}
I.~Moult, I.~W. Stewart, G.~Vita and H.~X. Zhu, \emph{{First Subleading Power
  Resummation for Event Shapes}},
  \href{http://dx.doi.org/10.1007/JHEP08(2018)013}{\emph{JHEP} {\bfseries 08}
  (2018) 013}, [\href{https://arxiv.org/abs/1804.04665}{{\ttfamily
  1804.04665}}].

\bibitem{Moult:2019uhz}
I.~Moult, I.~W. Stewart, G.~Vita and H.~X. Zhu, \emph{{The Soft Quark
  Sudakov}}, \href{http://dx.doi.org/10.1007/JHEP05(2020)089}{\emph{JHEP}
  {\bfseries 05} (2020) 089},
  [\href{https://arxiv.org/abs/1910.14038}{{\ttfamily 1910.14038}}].

\bibitem{Beneke:2022obx}
M.~Beneke, M.~Garny, S.~Jaskiewicz, J.~Strohm, R.~Szafron, L.~Vernazza et~al.,
  \emph{{Next-to-leading power endpoint factorization and resummation for
  off-diagonal \textquotedblleft{}gluon\textquotedblright{} thrust}},
  \href{http://dx.doi.org/10.1007/JHEP07(2022)144}{\emph{JHEP} {\bfseries 07}
  (2022) 144}, [\href{https://arxiv.org/abs/2205.04479}{{\ttfamily
  2205.04479}}].

\bibitem{Agarwal:2023fdk}
N.~Agarwal, M.~van Beekveld, E.~Laenen, S.~Mishra, A.~Mukhopadhyay and
  A.~Tripathi, \emph{{Next-to-leading power corrections to event-shape
  variables}},
  \href{http://dx.doi.org/10.1007/s12043-024-02743-0}{\emph{Pramana} {\bfseries
  98} (2024) 60}, [\href{https://arxiv.org/abs/2306.17601}{{\ttfamily
  2306.17601}}].

\bibitem{Hoang:2007vb}
A.~H. Hoang and I.~W. Stewart, \emph{{Designing Gapped Soft Functions for Jet
  Production}},
  \href{http://dx.doi.org/10.1016/j.physletb.2008.01.040}{\emph{Phys. Lett.}
  {\bfseries B660} (2008) 483--493},
  [\href{https://arxiv.org/abs/0709.3519}{{\ttfamily 0709.3519}}].

\bibitem{Ligeti:2008ac}
Z.~Ligeti, I.~W. Stewart and F.~J. Tackmann, \emph{{Treating the b quark
  distribution function with reliable uncertainties}},
  \href{http://dx.doi.org/10.1103/PhysRevD.78.114014}{\emph{Phys. Rev.}
  {\bfseries D78} (2008) 114014},
  [\href{https://arxiv.org/abs/0807.1926}{{\ttfamily 0807.1926}}].

\bibitem{Abbate:2010xh}
R.~Abbate, M.~Fickinger, A.~H. Hoang, V.~Mateu and I.~W. Stewart, \emph{{Thrust
  at N${}^3$LL with Power Corrections and a Precision Global Fit for
  $\alpha_s(m_Z)$}},
  \href{http://dx.doi.org/10.1103/PhysRevD.83.074021}{\emph{Phys. Rev.}
  {\bfseries D83} (2011) 074021},
  [\href{https://arxiv.org/abs/1006.3080}{{\ttfamily 1006.3080}}].

\bibitem{Agarwal:2020uxi}
N.~Agarwal, A.~Mukhopadhyay, S.~Pal and A.~Tripathi, \emph{{Power corrections
  to event shapes using eikonal dressed gluon exponentiation}},
  \href{http://dx.doi.org/10.1007/JHEP03(2021)155}{\emph{JHEP} {\bfseries 03}
  (2021) 155}, [\href{https://arxiv.org/abs/2012.06842}{{\ttfamily
  2012.06842}}].

\bibitem{Manohar:1994kq}
A.~V. Manohar and M.~B. Wise, \emph{{Power suppressed corrections to hadronic
  event shapes}},
  \href{http://dx.doi.org/10.1016/0370-2693(94)01504-6}{\emph{Phys. Lett.}
  {\bfseries B344} (1995) 407--412},
  [\href{https://arxiv.org/abs/hep-ph/9406392}{{\ttfamily hep-ph/9406392}}].

\bibitem{Webber:1994cp}
B.~R. Webber, \emph{{Estimation of power corrections to hadronic event
  shapes}}, \href{http://dx.doi.org/10.1016/0370-2693(94)91147-9}{\emph{Phys.
  Lett.} {\bfseries B339} (1994) 148--150},
  [\href{https://arxiv.org/abs/hep-ph/9408222}{{\ttfamily hep-ph/9408222}}].

\bibitem{Dokshitzer:1995zt}
Y.~L. Dokshitzer and B.~R. Webber, \emph{{Calculation of power corrections to
  hadronic event shapes}},
  \href{http://dx.doi.org/10.1016/0370-2693(95)00548-Y}{\emph{Phys. Lett.}
  {\bfseries B352} (1995) 451--455},
  [\href{https://arxiv.org/abs/hep-ph/9504219}{{\ttfamily hep-ph/9504219}}].

\bibitem{Akhoury:1995sp}
R.~Akhoury and V.~I. Zakharov, \emph{{On the universality of the leading, 1/Q
  power corrections in QCD}},
  \href{http://dx.doi.org/10.1016/0370-2693(95)00866-J}{\emph{Phys. Lett.}
  {\bfseries B357} (1995) 646--652},
  [\href{https://arxiv.org/abs/hep-ph/9504248}{{\ttfamily hep-ph/9504248}}].

\bibitem{Nason:1995hd}
P.~Nason and M.~H. Seymour, \emph{{Infrared renormalons and power suppressed
  effects in $e^+\, e^-$ jet events}},
  \href{http://dx.doi.org/10.1016/0550-3213(95)00461-Z}{\emph{Nucl. Phys.}
  {\bfseries B454} (1995) 291--312},
  [\href{https://arxiv.org/abs/hep-ph/9506317}{{\ttfamily hep-ph/9506317}}].

\bibitem{Korchemsky:1994is}
G.~P. Korchemsky and G.~Sterman, \emph{{Nonperturbative corrections in resummed
  cross-sections}},
  \href{http://dx.doi.org/10.1016/0550-3213(94)00006-Z}{\emph{Nucl. Phys.}
  {\bfseries B437} (1995) 415--432},
  [\href{https://arxiv.org/abs/hep-ph/9411211}{{\ttfamily hep-ph/9411211}}].

\bibitem{MovillaFernandez:2001ed}
P.~A. Movilla~Fernandez, S.~Bethke, O.~Biebel and S.~Kluth, \emph{{Tests of
  power corrections for event shapes in $e^+ e^-$ annihilation}},
  \href{http://dx.doi.org/10.1007/s100520100750}{\emph{Eur. Phys. J.}
  {\bfseries C22} (2001) 1--15},
  [\href{https://arxiv.org/abs/hep-ex/0105059}{{\ttfamily hep-ex/0105059}}].

\bibitem{Gardi:2001ny}
E.~Gardi and J.~Rathsman, \emph{{Renormalon resummation and exponentiation of
  soft and collinear gluon radiation in the thrust distribution}},
  \href{http://dx.doi.org/10.1016/S0550-3213(01)00284-X}{\emph{Nucl. Phys.}
  {\bfseries B609} (2001) 123--182},
  [\href{https://arxiv.org/abs/hep-ph/0103217}{{\ttfamily hep-ph/0103217}}].

\bibitem{Gehrmann:2009eh}
T.~Gehrmann, M.~Jaquier and G.~Luisoni, \emph{{Hadronization effects in event
  shape moments}},
  \href{http://dx.doi.org/10.1140/epjc/s10052-010-1288-4}{\emph{Eur. Phys. J.}
  {\bfseries C67} (2010) 57--72},
  [\href{https://arxiv.org/abs/0911.2422}{{\ttfamily 0911.2422}}].

\bibitem{Gehrmann:2012sc}
T.~Gehrmann, G.~Luisoni and P.~F. Monni, \emph{{Power corrections in the
  dispersive model for a determination of the strong coupling constant from the
  thrust distribution}},
  \href{http://dx.doi.org/10.1140/epjc/s10052-012-2265-x}{\emph{Eur. Phys. J.}
  {\bfseries C73} (2013) 2265},
  [\href{https://arxiv.org/abs/1210.6945}{{\ttfamily 1210.6945}}].

\bibitem{Abbate:2012jh}
R.~Abbate, M.~Fickinger, A.~H. Hoang, V.~Mateu and I.~W. Stewart,
  \emph{{Precision Thrust Cumulant Moments at N$^3$LL}},
  \href{http://dx.doi.org/10.1103/PhysRevD.86.094002}{\emph{Phys. Rev.}
  {\bfseries D86} (2012) 094002},
  [\href{https://arxiv.org/abs/1204.5746}{{\ttfamily 1204.5746}}].

\bibitem{Hoang:2015hka}
A.~H. Hoang, D.~W. Kolodrubetz, V.~Mateu and I.~W. Stewart, \emph{{Precise
  determination of $\alpha_s$ from the $C$-parameter distribution}},
  \href{http://dx.doi.org/10.1103/PhysRevD.91.094018}{\emph{Phys. Rev.}
  {\bfseries D91} (2015) 094018},
  [\href{https://arxiv.org/abs/1501.04111}{{\ttfamily 1501.04111}}].

\bibitem{Bell:2023dqs}
G.~Bell, C.~Lee, Y.~Makris, J.~Talbert and B.~Yan, \emph{{Effects of renormalon
  scheme and perturbative scale choices on determinations of the strong
  coupling from e+e- event shapes}},
  \href{http://dx.doi.org/10.1103/PhysRevD.109.094008}{\emph{Phys. Rev. D}
  {\bfseries 109} (2024) 094008},
  [\href{https://arxiv.org/abs/2311.03990}{{\ttfamily 2311.03990}}].

\bibitem{Nason:2023asn}
P.~Nason and G.~Zanderighi, \emph{{Fits of \ensuremath{\alpha}$_{s}$ using
  power corrections in the three-jet region}},
  \href{http://dx.doi.org/10.1007/JHEP06(2023)058}{\emph{JHEP} {\bfseries 06}
  (2023) 058}, [\href{https://arxiv.org/abs/2301.03607}{{\ttfamily
  2301.03607}}].

\bibitem{Salam:2001bd}
G.~P. Salam and D.~Wicke, \emph{{Hadron masses and power corrections to event
  shapes}}, {\emph{JHEP} {\bfseries 05} (2001) 061},
  [\href{https://arxiv.org/abs/hep-ph/0102343}{{\ttfamily hep-ph/0102343}}].

\bibitem{Mateu:2012nk}
V.~Mateu, I.~W. Stewart and J.~Thaler, \emph{{Power Corrections to Event Shapes
  with Mass-Dependent Operators}},
  \href{http://dx.doi.org/10.1103/PhysRevD.87.014025}{\emph{Phys. Rev.}
  {\bfseries D87} (2013) 014025},
  [\href{https://arxiv.org/abs/1209.3781}{{\ttfamily 1209.3781}}].

\bibitem{Farhi:1977sg}
E.~Farhi, \emph{{A QCD Test for Jets}},
  \href{http://dx.doi.org/10.1103/PhysRevLett.39.1587}{\emph{Phys. Rev. Lett.}
  {\bfseries 39} (1977) 1587--1588}.

\bibitem{Clavelli:1979md}
L.~Clavelli, \emph{{Jet Invariant Mass in Quantum Chromodynamics}},
  \href{http://dx.doi.org/10.1016/0370-2693(79)90789-5}{\emph{Phys. Lett.}
  {\bfseries B85} (1979) 111}.

\bibitem{Chandramohan:1980ry}
T.~Chandramohan and L.~Clavelli, \emph{{Consequences of Second Order QCD for
  Jet Structure in $e^+ e^-$ annihilation}},
  \href{http://dx.doi.org/10.1016/0550-3213(81)90224-8}{\emph{Nucl.Phys.}
  {\bfseries B184} (1981) 365}.

\bibitem{Clavelli:1981yh}
L.~Clavelli and D.~Wyler, \emph{{Kinematica Bounds on Jet Variables and the
  Heavy Jet Mass Distribution}}, {\emph{Phys.Lett.} {\bfseries B103} (1981)
  383}.

\bibitem{Dasgupta:2001sh}
M.~Dasgupta and G.~P. Salam, \emph{{Resummation of nonglobal QCD observables}},
  \href{http://dx.doi.org/10.1016/S0370-2693(01)00725-0}{\emph{Phys. Lett.}
  {\bfseries B512} (2001) 323--330},
  [\href{https://arxiv.org/abs/hep-ph/0104277}{{\ttfamily hep-ph/0104277}}].

\bibitem{Banfi:2002hw}
A.~Banfi, G.~Marchesini and G.~Smye, \emph{{Away from jet energy flow}},
  \href{http://dx.doi.org/10.1088/1126-6708/2002/08/006}{\emph{JHEP} {\bfseries
  08} (2002) 006}, [\href{https://arxiv.org/abs/hep-ph/0206076}{{\ttfamily
  hep-ph/0206076}}].

\bibitem{Khelifa-Kerfa:2015mma}
K.~Khelifa-Kerfa and Y.~Delenda, \emph{{Non-global logarithms at finite N$_{c}$
  beyond leading order}},
  \href{http://dx.doi.org/10.1007/JHEP03(2015)094}{\emph{JHEP} {\bfseries 03}
  (2015) 094}, [\href{https://arxiv.org/abs/1501.00475}{{\ttfamily
  1501.00475}}].

\bibitem{Larkoski:2015zka}
A.~J. Larkoski, I.~Moult and D.~Neill, \emph{{Non-Global Logarithms,
  Factorization, and the Soft Substructure of Jets}},
  \href{http://dx.doi.org/10.1007/JHEP09(2015)143}{\emph{JHEP} {\bfseries 09}
  (2015) 143}, [\href{https://arxiv.org/abs/1501.04596}{{\ttfamily
  1501.04596}}].

\bibitem{Becher:2015hka}
T.~Becher, M.~Neubert, L.~Rothen and D.~Y. Shao, \emph{{Effective Field Theory
  for Jet Processes}},
  \href{http://dx.doi.org/10.1103/PhysRevLett.116.192001}{\emph{Phys. Rev.
  Lett.} {\bfseries 116} (2016) 192001},
  [\href{https://arxiv.org/abs/1508.06645}{{\ttfamily 1508.06645}}].

\bibitem{Becher:2016mmh}
T.~Becher, M.~Neubert, L.~Rothen and D.~Y. Shao, \emph{{Factorization and
  Resummation for Jet Processes}},
  \href{http://dx.doi.org/10.1007/JHEP11(2016)019,
  10.1007/JHEP05(2017)154}{\emph{JHEP} {\bfseries 11} (2016) 019},
  [\href{https://arxiv.org/abs/1605.02737}{{\ttfamily 1605.02737}}].

\bibitem{Larkoski:2016zzc}
A.~J. Larkoski, I.~Moult and D.~Neill, \emph{{The Analytic Structure of
  Non-Global Logarithms: Convergence of the Dressed Gluon Expansion}},
  \href{http://dx.doi.org/10.1007/JHEP11(2016)089}{\emph{JHEP} {\bfseries 11}
  (2016) 089}, [\href{https://arxiv.org/abs/1609.04011}{{\ttfamily
  1609.04011}}].

\bibitem{Schwartz:2014wha}
M.~D. Schwartz and H.~X. Zhu, \emph{{Nonglobal logarithms at three loops, four
  loops, five loops, and beyond}},
  \href{http://dx.doi.org/10.1103/PhysRevD.90.065004}{\emph{Phys. Rev.}
  {\bfseries D90} (2014) 065004},
  [\href{https://arxiv.org/abs/1403.4949}{{\ttfamily 1403.4949}}].

\bibitem{Kelley:2011ng}
R.~Kelley, M.~D. Schwartz, R.~M. Schabinger and H.~X. Zhu, \emph{{The two-loop
  hemisphere soft function}},
  \href{http://dx.doi.org/10.1103/PhysRevD.84.045022}{\emph{Phys. Rev.}
  {\bfseries D84} (2011) 045022},
  [\href{https://arxiv.org/abs/1105.3676}{{\ttfamily 1105.3676}}].

\bibitem{Hornig:2011iu}
A.~Hornig, C.~Lee, I.~W. Stewart, J.~R. Walsh and S.~Zuberi, \emph{{Non-global
  Structure of the $\mathcal{O}(\alpha_s^2)$ Dijet Soft Function}},
  \href{http://dx.doi.org/10.1007/JHEP08(2011)054}{\emph{JHEP} {\bfseries 1108}
  (2011) 054}, [\href{https://arxiv.org/abs/1105.4628}{{\ttfamily 1105.4628}}].

\bibitem{Monni:2011gb}
P.~F. Monni, T.~Gehrmann and G.~Luisoni, \emph{{Two-Loop Soft Corrections and
  Resummation of the Thrust Distribution in the Dijet Region}},
  \href{http://dx.doi.org/10.1007/JHEP08(2011)010}{\emph{JHEP} {\bfseries 1108}
  (2011) 010}, [\href{https://arxiv.org/abs/1105.4560}{{\ttfamily 1105.4560}}].

\bibitem{Matsuura:1987wt}
T.~Matsuura and W.~L. van Neerven, \emph{{Second Order Logarithmic Corrections
  to the {Drell-Yan} Cross-section}},
  \href{http://dx.doi.org/10.1007/BF01624369}{\emph{Z. Phys.} {\bfseries C38}
  (1988) 623}.

\bibitem{Matsuura:1988sm}
T.~Matsuura, S.~C. van~der Marck and W.~L. van Neerven, \emph{{The Calculation
  of the Second Order Soft and Virtual Contributions to the Drell-Yan
  Cross-Section}},
  \href{http://dx.doi.org/10.1016/0550-3213(89)90620-2}{\emph{Nucl. Phys.}
  {\bfseries B319} (1989) 570}.

\bibitem{Gehrmann:2005pd}
T.~Gehrmann, T.~Huber and D.~Maitre, \emph{{Two-loop quark and gluon form
  factors in dimensional regularisation}},
  \href{http://dx.doi.org/10.1016/j.physletb.2005.07.019}{\emph{Phys. Lett.}
  {\bfseries B622} (2005) 295--302},
  [\href{https://arxiv.org/abs/hep-ph/0507061}{{\ttfamily hep-ph/0507061}}].

\bibitem{Moch:2005id}
S.~Moch, J.~A.~M. Vermaseren and A.~Vogt, \emph{{The Quark form-factor at
  higher orders}},
  \href{http://dx.doi.org/10.1088/1126-6708/2005/08/049}{\emph{JHEP} {\bfseries
  08} (2005) 049}, [\href{https://arxiv.org/abs/hep-ph/0507039}{{\ttfamily
  hep-ph/0507039}}].

\bibitem{Lee:2010cg}
R.~N. Lee, A.~V. Smirnov and V.~A. Smirnov, \emph{{Analytic Results for
  Massless Three-Loop Form Factors}},
  \href{http://dx.doi.org/10.1007/JHEP04(2010)020}{\emph{JHEP} {\bfseries 04}
  (2010) 020}, [\href{https://arxiv.org/abs/1001.2887}{{\ttfamily 1001.2887}}].

\bibitem{Baikov:2009bg}
P.~A. Baikov, K.~G. Chetyrkin, A.~V. Smirnov, V.~A. Smirnov and M.~Steinhauser,
  \emph{{Quark and gluon form factors to three loops}},
  \href{http://dx.doi.org/10.1103/PhysRevLett.102.212002}{\emph{Phys. Rev.
  Lett.} {\bfseries 102} (2009) 212002},
  [\href{https://arxiv.org/abs/0902.3519}{{\ttfamily 0902.3519}}].

\bibitem{Lunghi:2002ju}
E.~Lunghi, D.~Pirjol and D.~Wyler, \emph{{Factorization in leptonic radiative
  $B \to \gamma e \nu$ decays}},
  \href{http://dx.doi.org/10.1016/S0550-3213(02)01032-5}{\emph{Nucl. Phys.}
  {\bfseries B649} (2003) 349--364},
  [\href{https://arxiv.org/abs/hep-ph/0210091}{{\ttfamily hep-ph/0210091}}].

\bibitem{Bauer:2003pi}
C.~W. Bauer and A.~V. Manohar, \emph{{Shape function effects in $B \to X_s
  \gamma$ and $B \to X_u \ell \nu$ decays}},
  \href{http://dx.doi.org/10.1103/PhysRevD.70.034024}{\emph{Phys. Rev.}
  {\bfseries D70} (2004) 034024},
  [\href{https://arxiv.org/abs/hep-ph/0312109}{{\ttfamily hep-ph/0312109}}].

\bibitem{Becher:2006qw}
T.~Becher and M.~Neubert, \emph{{Toward a NNLO calculation of the ${\bar B} \to
  X_s \gamma$ decay rate with a cut on photon energy. II: Two-loop result for
  the jet function}},
  \href{http://dx.doi.org/10.1016/j.physletb.2006.04.046}{\emph{Phys. Lett.}
  {\bfseries B637} (2006) 251--259},
  [\href{https://arxiv.org/abs/hep-ph/0603140}{{\ttfamily hep-ph/0603140}}].

\bibitem{Bruser:2018rad}
R.~Br{\"u}ser, Z.~L. Liu and M.~Stahlhofen, \emph{{Three-Loop Quark Jet
  Function}},
  \href{http://dx.doi.org/10.1103/PhysRevLett.121.072003}{\emph{Phys. Rev.
  Lett.} {\bfseries 121} (2018) 072003},
  [\href{https://arxiv.org/abs/1804.09722}{{\ttfamily 1804.09722}}].

\bibitem{Banerjee:2018ozf}
P.~Banerjee, P.~K. Dhani and V.~Ravindran, \emph{{Gluon jet function at three
  loops in QCD}},
  \href{http://dx.doi.org/10.1103/PhysRevD.98.094016}{\emph{Phys. Rev.}
  {\bfseries D98} (2018) 094016},
  [\href{https://arxiv.org/abs/1805.02637}{{\ttfamily 1805.02637}}].

\bibitem{Gracia:2021nut}
N.~G. Gracia and V.~Mateu, \emph{{Toward massless and massive event shapes in
  the large-\ensuremath{\beta}$_{0}$ limit}},
  \href{http://dx.doi.org/10.1007/JHEP07(2021)229}{\emph{JHEP} {\bfseries 07}
  (2021) 229}, [\href{https://arxiv.org/abs/2104.13942}{{\ttfamily
  2104.13942}}].

\bibitem{vonManteuffel:2020vjv}
A.~von Manteuffel, E.~Panzer and R.~M. Schabinger, \emph{{Cusp and collinear
  anomalous dimensions in four-loop QCD from form factors}},
  \href{http://dx.doi.org/10.1103/PhysRevLett.124.162001}{\emph{Phys. Rev.
  Lett.} {\bfseries 124} (2020) 162001},
  [\href{https://arxiv.org/abs/2002.04617}{{\ttfamily 2002.04617}}].

\bibitem{Henn:2019swt}
J.~M. Henn, G.~P. Korchemsky and B.~Mistlberger, \emph{{The full four-loop cusp
  anomalous dimension in $\mathcal{N}=4$ super Yang-Mills and QCD}},
  \href{http://dx.doi.org/10.1007/JHEP04(2020)018}{\emph{JHEP} {\bfseries 04}
  (2020) 018}, [\href{https://arxiv.org/abs/1911.10174}{{\ttfamily
  1911.10174}}].

\bibitem{Henn:2019rmi}
J.~M. Henn, T.~Peraro, M.~Stahlhofen and P.~Wasser, \emph{{Matter dependence of
  the four-loop cusp anomalous dimension}},
  \href{http://dx.doi.org/10.1103/PhysRevLett.122.201602}{\emph{Phys. Rev.
  Lett.} {\bfseries 122} (2019) 201602},
  [\href{https://arxiv.org/abs/1901.03693}{{\ttfamily 1901.03693}}].

\bibitem{Bruser:2019auj}
R.~Br\"user, A.~Grozin, J.~M. Henn and M.~Stahlhofen, \emph{{Matter dependence
  of the four-loop QCD cusp anomalous dimension: from small angles to all
  angles}}, \href{http://dx.doi.org/10.1007/JHEP05(2019)186}{\emph{JHEP}
  {\bfseries 05} (2019) 186},
  [\href{https://arxiv.org/abs/1902.05076}{{\ttfamily 1902.05076}}].

\bibitem{Moch:2018wjh}
S.~Moch, B.~Ruijl, T.~Ueda, J.~A.~M. Vermaseren and A.~Vogt, \emph{{On quartic
  colour factors in splitting functions and the gluon cusp anomalous
  dimension}},
  \href{http://dx.doi.org/10.1016/j.physletb.2018.06.017}{\emph{Phys. Lett. B}
  {\bfseries 782} (2018) 627--632},
  [\href{https://arxiv.org/abs/1805.09638}{{\ttfamily 1805.09638}}].

\bibitem{Moch:2017uml}
S.~Moch, B.~Ruijl, T.~Ueda, J.~A.~M. Vermaseren and A.~Vogt, \emph{{Four-Loop
  Non-Singlet Splitting Functions in the Planar Limit and Beyond}},
  \href{http://dx.doi.org/10.1007/JHEP10(2017)041}{\emph{JHEP} {\bfseries 10}
  (2017) 041}, [\href{https://arxiv.org/abs/1707.08315}{{\ttfamily
  1707.08315}}].

\bibitem{Baranowski:2024vxg}
D.~Baranowski, M.~Delto, K.~Melnikov, A.~Pikelner and C.-Y. Wang,
  \emph{{Zero-Jettiness Soft Function to Third Order in Perturbative QCD}},
  \href{http://dx.doi.org/10.1103/PhysRevLett.134.191902}{\emph{Phys. Rev.
  Lett.} {\bfseries 134} (2025) 191902},
  [\href{https://arxiv.org/abs/2409.11042}{{\ttfamily 2409.11042}}].

\bibitem{Baranowski:2024ysi}
D.~Baranowski, M.~Delto, K.~Melnikov, A.~Pikelner and C.-Y. Wang, \emph{{Triple
  real-emission contribution to the zero-jettiness soft function at N$^3$LO in
  QCD}}, \href{http://dx.doi.org/10.1007/JHEP04(2025)084}{\emph{JHEP}
  {\bfseries 04} (2025) 084},
  [\href{https://arxiv.org/abs/2412.14001}{{\ttfamily 2412.14001}}].

\bibitem{Hoang:2008fs}
A.~H. Hoang and S.~Kluth, \emph{{Hemisphere Soft Function at $O(\alpha_s^2)$
  for Dijet Production in $e^+e^-$ Annihilation}},
  \href{https://arxiv.org/abs/0806.3852}{{\ttfamily 0806.3852}}.

\bibitem{Hoang:2021fhn}
A.~H. Hoang, C.~Lepenik and V.~Mateu, \emph{{REvolver: Automated running and
  matching of couplings and masses in QCD}},
  \href{http://dx.doi.org/10.1016/j.cpc.2021.108145}{\emph{Comput. Phys.
  Commun.} {\bfseries 270} (2022) 108145},
  [\href{https://arxiv.org/abs/2102.01085}{{\ttfamily 2102.01085}}].

\bibitem{Clavero:2024yav}
A.~M. Clavero, R.~Br\"user, V.~Mateu and M.~Stahlhofen, \emph{{Three-loop jet
  function for boosted heavy quarks}},
  \href{http://dx.doi.org/10.1007/JHEP04(2025)040}{\emph{JHEP} {\bfseries 04}
  (2025) 040}, [\href{https://arxiv.org/abs/2412.06881}{{\ttfamily
  2412.06881}}].

\bibitem{Mateu:2017hlz}
V.~Mateu and P.~G. Ortega, \emph{{Bottom and Charm Mass determinations from
  global fits to $Q\bar{Q}$ bound states at N$^3$LO}},
  \href{http://dx.doi.org/10.1007/JHEP01(2018)122}{\emph{JHEP} {\bfseries 01}
  (2018) 122}, [\href{https://arxiv.org/abs/1711.05755}{{\ttfamily
  1711.05755}}].

\bibitem{GehrmannDeRidder:2009dp}
A.~{Gehrmann-De Ridder}, T.~Gehrmann, E.~Glover and G.~Heinrich, \emph{{NNLO
  moments of event shapes in $e^+e^-$ annihilation}},
  \href{http://dx.doi.org/10.1088/1126-6708}{\emph{JHEP} {\bfseries 0905}
  (2009) 106}, [\href{https://arxiv.org/abs/0903.4658}{{\ttfamily 0903.4658}}].

\bibitem{Aveleira:2025svg}
B.~C. Aveleira, A.~Gehrmann-De~Ridder, T.~Gehrmann, N.~Glover, G.~Heinrich and
  C.~T. Preuss, \emph{{EERAD3 version 2: QCD corrections in hadronic
  colour-singlet decays}},  \href{https://arxiv.org/abs/2503.20610}{{\ttfamily
  2503.20610}}.

\bibitem{Benitez:2024nav}
M.~A. Benitez, A.~H. Hoang, V.~Mateu, I.~W. Stewart and G.~Vita, \emph{{On
  Determining $\alpha_s(m_Z)$ from Dijets in $e^+e^-$ Thrust}},
  \href{http://dx.doi.org/10.1007/JHEP07(2025)249}{\emph{JHEP} {\bfseries 07}
  (2025) 249}, [\href{https://arxiv.org/abs/2412.15164}{{\ttfamily
  2412.15164}}].

\bibitem{Dehnadi:2023msm}
B.~Dehnadi, A.~H. Hoang, O.~L. Jin and V.~Mateu, \emph{{Top quark mass
  calibration for Monte Carlo event generators \textemdash{} an update}},
  \href{http://dx.doi.org/10.1007/JHEP12(2023)065}{\emph{JHEP} {\bfseries 12}
  (2023) 065}, [\href{https://arxiv.org/abs/2309.00547}{{\ttfamily
  2309.00547}}].

\bibitem{Hoang:2008yj}
A.~H. Hoang, A.~Jain, I.~Scimemi and I.~W. Stewart, \emph{{Infrared
  Renormalization Group Flow for Heavy Quark Masses}},
  \href{http://dx.doi.org/10.1103/PhysRevLett.101.151602}{\emph{Phys. Rev.
  Lett.} {\bfseries 101} (2008) 151602},
  [\href{https://arxiv.org/abs/0803.4214}{{\ttfamily 0803.4214}}].

\bibitem{Hoang:2009yr}
A.~H. Hoang, A.~Jain, I.~Scimemi and I.~W. Stewart, \emph{{R-evolution:
  Improving perturbative QCD}},
  \href{http://dx.doi.org/10.1103/PhysRevD.82.011501}{\emph{Phys. Rev.}
  {\bfseries D82} (2010) 011501},
  [\href{https://arxiv.org/abs/0908.3189}{{\ttfamily 0908.3189}}].

\bibitem{Hoang:2017suc}
A.~H. Hoang, A.~Jain, C.~Lepenik, V.~Mateu, M.~Preisser, I.~Scimemi et~al.,
  \emph{{The MSR mass and the
  $\mathcal{O}\left({\Lambda}_{\mathrm{QCD}}\right)$ renormalon sum rule}},
  \href{http://dx.doi.org/10.1007/JHEP04(2018)003}{\emph{JHEP} {\bfseries 04}
  (2018) 003}, [\href{https://arxiv.org/abs/1704.01580}{{\ttfamily
  1704.01580}}].

\bibitem{Benitez:2025vsp}
M.~A. Benitez, A.~Bhattacharya, A.~H. Hoang, V.~Mateu, M.~D. Schwartz, I.~W.
  Stewart et~al., \emph{{A Precise Determination of $\alpha_s$ from the Heavy
  Jet Mass Distribution}},  \href{https://arxiv.org/abs/2502.12253}{{\ttfamily
  2502.12253}}.

\bibitem{Stewart:2014nna}
I.~W. Stewart, F.~J. Tackmann and W.~J. Waalewijn, \emph{{Dissecting Soft
  Radiation with Factorization}},
  \href{http://dx.doi.org/10.1103/PhysRevLett.114.092001}{\emph{Phys. Rev.
  Lett.} {\bfseries 114} (2015) 092001},
  [\href{https://arxiv.org/abs/1405.6722}{{\ttfamily 1405.6722}}].

\bibitem{Butenschoen:2016lpz}
M.~Butenschoen, B.~Dehnadi, A.~H. Hoang, V.~Mateu, M.~Preisser and I.~W.
  Stewart, \emph{{Top Quark Mass Calibration for Monte Carlo Event
  Generators}},
  \href{http://dx.doi.org/10.1103/PhysRevLett.117.232001}{\emph{Phys. Rev.
  Lett.} {\bfseries 117} (2016) 232001},
  [\href{https://arxiv.org/abs/1608.01318}{{\ttfamily 1608.01318}}].

\bibitem{Almeida:2014uva}
L.~G. Almeida, S.~D. Ellis, C.~Lee, G.~Sterman, I.~Sung et~al.,
  \emph{{Comparing and counting logs in direct and effective methods of QCD
  resummation}}, \href{http://dx.doi.org/10.1007/JHEP04(2014)174}{\emph{JHEP}
  {\bfseries 1404} (2014) 174},
  [\href{https://arxiv.org/abs/1401.4460}{{\ttfamily 1401.4460}}].

\bibitem{Catani:1998sf}
S.~Catani and B.~Webber, \emph{{Resummed C parameter distribution in $e^+\,
  e^-$ annihilation}},
  \href{http://dx.doi.org/10.1016/S0370-2693(98)00359-1}{\emph{Phys. Lett.}
  {\bfseries B427} (1998) 377--384},
  [\href{https://arxiv.org/abs/hep-ph/9801350}{{\ttfamily hep-ph/9801350}}].

\bibitem{Catani:1991bd}
S.~Catani, G.~Turnock and B.~Webber, \emph{{Heavy jet mass distribution in $e^+
  e^-$ annihilation}},
  \href{http://dx.doi.org/10.1016/0370-2693(91)91845-M}{\emph{Phys.Lett.}
  {\bfseries B272} (1991) 368--372}.

\end{thebibliography}\endgroup
\bibliographystyle{JHEP}

\end{document}